\documentclass[a4paper,11pt]{article}
\usepackage[utf8]{inputenc}
\pdfoutput=1

\usepackage{jcappub}
\usepackage{natbib}
\setcitestyle{square,numbers}
\usepackage{xcolor}
\usepackage{float}
\usepackage{multirow,array,longtable}
\usepackage{arydshln} 
\usepackage[makeroom]{cancel}
\usepackage{amsfonts}

\renewcommand\eqref[1]{Eq.~(\ref{#1})}
\newcommand\eqrefs[2]{Eqs.~(\ref{#1})-(\ref{#2})}
\newcommand\figref[1]{Fig.~\ref{#1}}
\newcommand\figrefs[2]{Figs.~\ref{#1}-\ref{#2}}
\newcommand\tabref[1]{Table~\ref{#1}}
\newcommand\tabrefs[2]{Tables~\ref{#1}-\ref{#2}}
\newcommand\secref[1]{Section~\ref{#1}}
\newcommand\appref[1]{Appendix~\ref{#1}}
\newcommand{\be}{\begin{equation}}
\newcommand{\ee}{\end{equation}}
\newcommand{\bear}{\begin{eqnarray}}
\newcommand{\eear}{\end{eqnarray}}
\newcommand{\nn}{\nonumber}

\def\cosw{c_{\rm w}}
\def\sinw{s_{\rm w}}
\def\tanw{t_{\rm w}}
\def\mh{m_{\rm H}}
\def\mw{m_{\rm W}}
\def\mz{m_{\rm Z}}
\def\gY{g'}
\def\gYc{g^{'2}}
\def\cone{c_{H\gamma\gamma}}
\def\ctwo{c_{H\gamma Z}}
\def\cHBB{c_{HBB}}
\def\cHWW{c_{HWW}}
\def\cHBW{c_{HBW}}
\def\vev{{\it v}}
\def\xiw{\xi_{\rm W}}
\def\xiz{\xi_{\rm Z}}
\def\xia{\xi_{\rm A}}

\def\Zf{{\cal Z}}
\def\div{\Delta_\epsilon}

\def\xq{\frac{4\mw^2}{q^2}}
\def\xh{\frac{4\mw^2}{\mh^2}}
\def\xz{\frac{4\mw^2}{\mz^2}}
\def\xixq{\frac{4\xi\mw^2}{q^2}}

\def\xixz{\frac{4\xi\mw^2}{\mz^2}}

\def\halfxixq{\frac{2\xi\mw^2}{q^2}}

\def\1loop{one-loop}
\def\decayboth{H\to\gamma V}
\def\greenfR{\Gamma^{\rm R}}
\def\greenfT{\Gamma^{\rm T}}
\def\greenfL{\Gamma^{\rm L}}
\def\greenfC{\Gamma^{\rm C}}
\def\amp{\mathcal{M}}
\def\ampLO{\mathcal{M}_{\rm LO}}
\def\ampNLO{\mathcal{M}_{\rm NLO}}
\def\ampEChL{\mathcal{M}}
\newcommand{\overbar}[1]{\mkern 3.5mu\overline{\mkern-6.mu#1\mkern-1.5mu}\mkern 3.5mu}
\def\ampSM{\overbar{\mathcal{M}}}
\def\ampone{\mathcal{M}_{\gamma\gamma}}
\def\amptwo{\mathcal{M}_{\gamma Z}}
\def\ampboth{\mathcal{M}_{\gamma V}}
\def\amponeSM{\overbar{\mathcal{M}}_{\gamma\gamma}}
\def\amptwoSM{\overbar{\mathcal{M}}_{\gamma Z}}
\def\ampbothSM{\overbar{\mathcal{M}}_{\gamma V}}
\def\amponeLoop{\mathcal{M}_{\gamma\gamma}^{\rm L}}
\def\amptwoLoop{\mathcal{M}_{\gamma Z}^{\rm L}}
\def\ampbothLoop{\mathcal{M}_{\gamma V}^{\rm L}}

\def\amponechiral{\mathcal{M}_{\gamma\gamma}^{\rm GB}}
\def\amptwochiral{\mathcal{M}_{\gamma Z}^{\rm GB}}
\def\ampbothchiral{\mathcal{M}_{\gamma V}^{\rm GB}}
\def\amponeSMchiral{\overbar{\mathcal{M}}_{\gamma\gamma}^{\rm GB}}
\def\amptwoSMchiral{\overbar{\mathcal{M}}_{\gamma Z}^{\rm GB}}
\def\ampbothSMchiral{\overbar{\mathcal{M}}_{\gamma V}^{\rm GB}}
\def\Veffone{V_{H\gamma\gamma}}
\def\Vefftwo{V_{H\gamma Z}}
\def\Veffboth{V_{H\gamma V}}
\newcommand{\overbarV}[1]{\mkern .5mu\overline{\mkern-1.mu#1\mkern-1.mu}\mkern .5mu}
\def\VeffoneSM{\overbarV{V}_{H\gamma\gamma}}
\def\VefftwoSM{\overbarV{V}_{H\gamma Z}}

\def\VFone{{V}_{H\gamma\gamma}^{\rm F}}
\def\VFtwo{{V}_{H\gamma Z}^{\rm F}}

\def\Veffonechiral{V_{H\gamma\gamma}^{\rm GB}}
\def\Vefftwochiral{V_{H\gamma Z}^{\rm GB}}

\def\VeffoneSMchiral{\overbarV{V}_{H\gamma\gamma}^{\rm GB}}
\def\VefftwoSMchiral{\overbarV{V}_{H\gamma Z}^{\rm GB}}

\newcommand{\overbarC}[1]{\mkern .5mu\overline{\mkern-.5mu#1\mkern-.5mu}\mkern .5mu}
\def\coupsEChL{\mathcal{C}}
\def\coupsSM{\overbarC{\mathcal{C}}}
\newcommand{\mL}{\mathcal{L}}
\newcommand{\mLtwo}{\mathcal{L}_2}
\newcommand{\mLfour}{\mathcal{L}_4}

\newcommand{\mO}{\mathcal{O}}

\newcommand\prop[3]{P_{#1#2}(#3)}
\newcommand\propG[1]{\frac{1}{(#1)^2-\xi\mw^2}}
\def\propGp{\frac{1}{p^2-\xi\mw^2}}
\usepackage{soul}
\usepackage{lineno} 

\title{Anatomy of Higgs decays into $\boldsymbol{\gamma \gamma}$ and $\boldsymbol{\gamma Z}$ within the EChL in the $R_\xi$ gauges}
\author[a]{Maria Herrero,}
\author[a]{and Roberto A. Morales}
\affiliation[a]{Departamento de F\'{\i}sica Te\'orica and Instituto de F\'{\i}sica Te\'orica, IFT-UAM/CSIC,\\
Universidad Aut\'onoma de Madrid, Cantoblanco, 28049 Madrid, Spain}
\emailAdd{maria.herrero@uam.es}
\emailAdd{robertoa.morales@uam.es}

\abstract{
In this work we study the Higgs boson decays into two photons and into one photon and one $Z$ gauge boson within the context of the non-linear Effective Field Theory called the Electroweak Chiral Lagrangian. We present a detailed computation of the corresponding amplitudes to \1loop level in the covariant $R_\xi$ gauges. We assume that the fermionic loop contributions are as in the Standard Model and focus here just in the computation of the bosonic loop contributions. Our renormalization program and the anatomy of the various contributions participating in the $R_\xi$ gauges are fully explored. With this present computation we demonstrate the gauge invariance of the EChL result, not only for the case of on-shell Higgs boson, but also for the most general and interesting case of off-shell Higgs boson. We finally analyse and conclude on the special relevance of the Goldstone boson loops, in good agreement with the expected chiral loops behaviour in Chiral Lagrangians. We perform a systematic comparison with the corresponding computation of the Standard Model in the $R_\xi$ gauges and with the previous EChL results in the unitary gauge.
This work represents the first computation within the EChL of these Higgs observables to \1loop in the most general $R_\xi$ gauges and with a full renormalization program description, not yet fully explored in the previous literature and which is different to the most frequently used in the linear Effective Field Theory (SMEFT).
}

\begin{document}
\begin{flushright}
	IFT-UAM/CSIC-20-75  \\	
	FTUAM-20-6
\end{flushright}
\maketitle

\section{Introduction}
\label{intro}

After the discovery of the Higgs boson particle~\cite{Aad:2012tfa,Chatrchyan:2012ufa}, a great effort has been done in exploring new aspects of the Higgs physics beyond the Standard Model (SM) with the tools of Effective Field Theories (EFTs), both in the linear and in the non-linear approach (for a  review, see for instance~\cite{Brivio:2017vri}). 
Within the non-linear approach, which we follow in this work, the EFT that has become more popular is the one based on the Electroweak Chiral Lagrangian (EChL), also named Higgs Effective Field Theory (HEFT) in the literature. 
This non-linear EFT is the most appropriate one in the case that the new ultraviolet physics beyond the SM be strongly interacting like, e.g., in composite models, since in that cases the dynamics of the Goldstone bosons (GBs) is well described by the non-linear effective Chiral Lagrangians (for a review, see for instance \cite{Dobado:2019fxe}). 
The inspiring example, predecessor of these type of non-linear EFTs, is the one provided by the Chiral Lagrangian of pions in Quantum Chromodynamics (QCD) and the Chiral Perturbation Theory (ChPT)~\cite{Gasser:1983yg,Manohar:1983md,Weinberg:1978kz}. Nowadays, it is generally  accepted that ChPT describes successfully the dynamics of the pions, that are the GBs associated to the spontaneous breaking of the chiral symmetry in QCD. The proposal of using a non-linear EFT within the context of the electroweak (EW) interactions based on the $SU(2)_L\times SU(2)_R$ chiral symmetry of the SM scalar sector that is spontaneously broken down to the subgroup $SU(2)_{L+R}=SU(2)_C$, called the custodial symmetry group, was done long ago~\cite{Appelquist:1980vg,Longhitano:1980iz,Chanowitz:1985hj,Cheyette:1987jf, Dobado:1989ax,Dobado:1989ue,Dobado:1990zh,Espriu:1991vm,Feruglio:1992wf,Herrero:1993nc,Herrero:1994iu}.
It is in the last years, after the discovery of the Higgs boson particle, that the EChL has been renewed with the incorporation of the Higgs field as an extra explicit light scalar field in addition to the three GBs. Consequently, the new version of the EChL contains more effective operators that now include also the Higgs field~\cite{Alonso:2012px,Brivio:2013pma,Espriu:2012ih,Espriu:2013fia,Delgado:2013loa,Delgado:2013hxa,Delgado:2014jna,Buchalla:2012qq,Buchalla:2013rka,Buchalla:2013eza,Buchalla:2015wfa,Buchalla:2015qju}.  
 
The key issue in this non-linear EChL is that the GBs are in a non-linear representation of the $SU(2)$ group whereas the Higgs boson is a singlet. This is in contrast to the linear EFT case  (like the SMEFT) where both the GBs and the Higgs boson are placed together in a linear representation, given by the usual SM Higgs doublet.  
The GBs in the non-linear case transform non-linearly under the EW chiral symmetry and, more interestingly, have derivative self-couplings, therefore growing with energy, which are not present in the linear case. This basically leads to a very different counting when extracting the relevant contributions in a practical computation of a given observable in both approaches, the linear and the non-linear one. The linear approach uses the canonical counting, whereas the non-linear one uses the so-called chiral counting~\cite{Buchalla:2012qq, Buchalla:2013eza,Buchalla:2013rka}. In consequence, this later leads to the corresponding chiral expansion being ordered by powers of momentum and soft masses and, therefore, giving rise to predictions for the observables that behave with momentum and masses very differently than in the linear case.  Furthermore, precisely due to this different counting, the renormalization programs in both approaches are also very different. 

Our focus here is in the non-linear case given by the EChL where, according to the usual rules with Chiral Lagrangians, the renormalization program is done perturbatively in the chiral expansion. Namely, terms of a given chiral order in the Chiral Lagrangian act as counter-terms to renormalize the generated divergences beyond tree level from the terms with lower chiral dimension. In particular, the generated divergences at the \1loop level and the corresponding renormalization program in the EChL have been explored in the literature, both at the effective action or effective Lagrangian level~\cite{Guo_2015,Buchalla:2017jlu,Buchalla:2020kdh} and for specific observables, like scattering processes involving EW gauge bosons. 
These later include $ \gamma \gamma \to VV$ with $VV=WW, ZZ$~\cite{Delgado:2014jna}, and vector boson scattering $VV \to VV$ with various options for the EW bosons $V$'s  being $W^\pm$ and $Z$~\cite{Espriu:2013fia, Delgado:2013hxa, Delgado:2013loa}. In all these scattering cases the \1loop computations with the EChL were simplified by considering just the loops containing only GBs and the Higgs boson and by neglecting all the other loops including the gauge bosons. However, there is not yet, to our knowledge a complete computation of these scattering processes to \1loop level within the EChL. 
The other interesting processes studied in the literature within the EChL at the \1loop level are the Higgs boson decays. In particular, $H \to \gamma \gamma$ and $H \to \gamma Z$ have been computed in~\cite{losBuchalla} within the unitary gauge, where only gauge bosons are involved in the loops by construction. But, to our knowledge there is not available computation of these decays in the covariant gauges and, therefore, the specific contributions in these $R_\xi$ gauges from GBs loops and the rest of loops have not being explored yet within this EChL non-linear context.

In this paper, we present an explicit and detailed computation of the \1loop amplitudes for the $H \to \gamma \gamma$ and $H \to \gamma Z$ decays within the EChL in the covariant $R_\xi$ gauges for the first time. Our purpose is to present this new computation in a didactic and illustrative way, showing in detail the various steps to follow, starting with the chiral expansion and ending with the renormalization program in order to get the final UV-finite and gauge invariant result for the total amplitude, within this non-linear EFT.
Of course our purpose is also to compare our results of the $R_\xi$ gauges with the previous results in the unitary gauge of~\cite{losBuchalla}. For the \1loop computation in the $H \to \gamma \gamma$ case and for the explicit demonstration of the gauge invariance of the EChL result we have followed very closely the method of~\cite{Marciano-paper} which was applied for the SM case, that we find very useful. 
We should also mention that other computations in the literature of these two Higgs boson decays within the context of EFT were done within the linear approach which, as we have said, involve different techniques, renormalization program and counting rules than the in non-linear case which we deal with here. Concretely,  in the SMEFT to \1loop with $R_\xi$ gauges,  $H \to \gamma \gamma$ has been computed in~\cite{Hartmann:2015oia,Dedes:2018seb}, and $H \to \gamma Z$ in~\cite{Dawson:2018pyl,Dedes:2019bew}. 

In our study of the anatomy of the various loop contributions participating in these Higgs boson decays, we also wish to explore the special role played by the GBs loops. Thus, we also include in our study an important part where we discuss on the role played by the GBs loops, differentiating both cases: on-shell versus off-shell Higgs boson. The results for the off-shell case involving high energies for the virtual Higgs particle, say at the TeV range, will tell us about the relevance on these GB loops (also called chiral loops) in high energy colliders like the LHC, the linear colliders as ILC or CLIC, etc., and then access to the EW symmetry breaking sector (EWSBS).

The paper is organized as follows. In \secref{sec-EChL} we review the main features of the EChL and the relevant operators participating in both $H\to\gamma\gamma,\,\gamma Z$ decays.
\secref{sec-renorm} is devoted to the renormalization program and to present the analytical results for the \1loop amplitudes of the EChL in the $R_\xi$ gauges. We also show in that section how the gauge invariance is established, and include a comparison with the corresponding SM results. 
The results for the associated \1loop vertex functions to these decays, differentiating the two cases with off-shell/on-shell Higgs boson, are discussed in \secref{sec-Veff}. 
The numerical results for the partial decay widths and the study of the behaviour of these vertex functions with the Higgs boson momentum are presented in \secref{sec-plots}. 
Finally, we conclude in \secref{sec-conclu}. The supplementary material needed for the computation, as the relevant Feynman rules, the chosen conventions for the loop integrals and the preparation of the \1loop diagrams for the automated computation, are summarized in the appendices.

\section{The Electroweak Chiral Lagrangian: relevant interactions}
\label{sec-EChL}

The EChL is a gauged non-linear EFT based on the $SU(2)_L\times SU(2)_R$ chiral symmetry of the EWSBS. 
This symmetry is spontaneously broken down to the subgroup $SU(2)_{L+R}=SU(2)_C$, called the custodial symmetry group.
Also, the EChL is Lorentz, CP and $SU(2)_L\times U(1)_Y$ gauge invariant.
In the present work, we focus on the bosonic part of the EChL assuming that the fermionic contributions to the observables of interest here, $H \to \gamma \gamma$ and $H \to \gamma Z$, are the same as in the SM.
For the present work, we consider only effective operators within the EChL that preserve custodial symmetry. With this assumption, the only source of custodial symmetry breaking in the bosonic sector of the EChL is, as in the SM case, the small non-zero hypercharge gauge coupling $\gY$ corresponding to $U(1)_Y$.
As dynamical fields, the EChL contains the EW gauge bosons $B_\mu$ and $W^a_\mu$ (with $a=1,2,3$), and the SM-like Higgs boson $H$. In contrast to the SM case, and other EFTs for BSM Higgs physics, where the Higgs field is implemented together with the would-be GBs inside a linear representation of $SU(2)$, given by the doublet 
$\Phi^T=(i\pi^+ \,\,,\,\, ((H+v)-i\pi^3)/\sqrt{2})$ with $\pi^\pm =(\pi^1\mp i \pi^2)/\sqrt{2}$, in the EChL the Higgs field and the GB fields are introduced separately. The $H$ field is a singlet of the EW chiral symmetry and the EW gauge symmetry and, consequently, there are not limitations from symmetry arguments on the implementation of this field and its interactions into the Lagrangian. Usually, in the EChL,  the interactions of $H$ with the other fields are introduced via multiplicative generic polynomials. Regarding the three GBs, $\pi^a$, they are introduced in the EChL in a non-linear representation of $SU(2)$. We use here the exponential parametrization given by:
\be 
U(\pi^a) = e^{i \pi^a \tau^a/\vev}
\label{expo}
\ee
where, $\tau^a$, $a=1,2,3$,  are the Pauli matrices and $v=246$ GeV.  
Under a EW chiral transformation of  $SU(2)_L \times SU(2)_R$, given by $L \in SU(2)_L$ and $R \in SU(2)_R$, the field $U$ transforms linearly as $L U R^\dagger$, whereas the GBs $\pi^ a$ transform non-linearly. In this sense, these EW GBs of the EChL behave similarly to the pions in low energy QCD, which are identified with the GBs of the Chiral symmetry breaking. Consequently, the building of this non-linear EFT for low energy EW interactions as given by the EChL is clearly inspired in the 
well known Chiral Perturbation Theory for low energy strong interactions given by the Chiral Lagrangian of QCD~\cite{Gasser:1983yg,Manohar:1983md,Weinberg:1978kz}.
On the other hand, the EW gauge bosons are introduced in the EChL via the $U(1)_Y$ and $SU(2)_L$ field strength tensors and the covariant derivative of the $U$ matrix by:
\bear
\hat{B}_{\mu\nu} &=& \partial_\mu \hat{B}_\nu -\partial_\nu \hat{B}_\mu \,,\quad
\hat{W}_{\mu\nu} = \partial_\mu \hat{W}_\nu - \partial_\nu \hat{W}_\mu + i  [\hat{W}_\mu,\hat{W}_\nu ] \,,  \nn\\
D_\mu U &=& \partial_\mu U + i\hat{W}_\mu U - i U\hat{B}_\mu \,,
\eear
where $\hat{B}_\mu = \gY B_\mu \tau^3/2$ and $\hat{W}_\mu = g W^a_\mu \tau^a/2$. In order to introduce the physical fields, we use the usual definitions:
\be
W_{\mu}^\pm = \frac{1}{\sqrt{2}}(W_{\mu}^1 \mp i W_{\mu}^2) \,,\quad
Z_{\mu} = \cosw W_{\mu}^3 - \sinw B_{\mu} \,,\quad
A_{\mu} = \sinw W_{\mu}^3 + \cosw B_{\mu} \,,
\ee
where we use the short notation $s_W=\sin \theta_W$ and $c_W=\cos \theta_W$.

The EChL structure is based on a momentum expansion following the usual counting rules of the Chiral Lagrangian approach, and the effective operators in the  EChL are organized through their chiral dimension. 
This chiral dimension is established by the scaling with the momentum $p$ of the various contributing building blocks.
In this context, derivatives and masses are {\it soft} scales and they count with the same power of the momentum:
\be
\partial_\mu \,,\,\mw \,,\,\mz \,,\,\mh \,,\,g\vev \,,\,\gY\vev \sim \mO(p) \,.
\ee
The corresponding counting rules for the gauge fields and the field strength tensors can then be obtained from the previous ones if we rewrite them in terms of the $g\vev$ and $\gY\vev$ combinations:
\bear
\hat{B}_\mu= (\gY\vev) (B_\mu/\vev) \tau^3/2 \,,\, \hat{W}_\mu = (g\vev) (W^a_\mu/\vev) \tau^a/2 &\sim & \mO(p) \,,  \nn\\
\hat{B}_{\mu\nu} = (\gY\vev) \left(\partial_\mu (B_\nu/\vev) -\partial_\nu (B_\mu/\vev)\right)\tau^3/2 &\sim & \mO(p^2) \,,  \nn\\
\hat{W}_{\mu\nu} = (g\vev) \left(\partial_\mu (W^a_\nu/\vev) - \partial_\nu (W^a_\mu/\vev) -(g\vev)\epsilon^{abc} (W^b_\mu/\vev) (W^c_\nu/\vev)\right)\tau^a/2 &\sim & \mO(p^2) \,.
\eear
Consequently, an effective operator containing the block 
$(1/g^2)(\hat{W}_{\mu\nu}\hat{W}^{\mu\nu})$ scales as ${\cal O}(p^2)$, etc..
With the above counting rules, the Chiral Lagrangian $\mL_{\rm EChL}$ is given by the sum of the various contributions $\mL_{d}$ with increasing chiral dimension $d$, i.e., of $\mO(p^d)$. The dimensionless parameter controling the convergence of this chiral expansion is $p/(4\pi\vev)$ with $4\pi\vev\approx 3$ TeV, in close analogy with the Chiral Lagrangian of QCD where the corresponding dimensionless parameter is $p/(4\pi f_\pi)$ with  $4\pi f_\pi\approx 1$ GeV. Thus, in absence of any other resonance appearing in the spectrum, the EChL is expected to be an EFT valid below this ${\cal O}(3 \,{\rm TeV})$ typical energy. 
 
We consider here just the two leading contributions, $\mL_{2}$ and $\mL_{4}$ in this chiral expansion of
the EChL, and then we select from them the effective operators which are relevant for the present computation of the Higgs boson decays into $\gamma \gamma$ and $\gamma Z$. 
Thus, we write:
\be 
\mL_{\rm EChL}=\mL_2 + \mL_4+...
\label{EChL}
\ee
where $\mL_2$  and $\mL_4$ are given, respectively, by:
 \bear
\mL_2 &=& \frac{\vev^2}{4}\Big[
1+2a\frac{H}{\vev}+b \left(\frac{H}{\vev}\right)^2+...\Big] {\rm Tr} \Big[D^\mu U^\dagger D_\mu U \Big] + \frac{1}{2} \partial^\mu H \, \partial_\mu H -{\it V}(H) \nn\\
&&
-\frac{1}{2 \gYc}
{\rm Tr} \Big[\hat{B}_{\mu\nu} \hat{B}^{\mu\nu}\Big] -\frac{1}{2 g^2} {\rm Tr}\Big[\hat{W}_{\mu\nu}\hat{W}^{\mu\nu}\Big] + \mL_{\rm GF} + \mL_{\rm FP}, 
\label{eq-L2}
\eear
\bear
\mLfour &=& a_1 {\rm Tr}\Big[ U \hat{B}_{\mu\nu} U^\dagger \hat{W}^{\mu\nu}\Big]  +i a_2  {\rm Tr}\Big[ U \hat{B}_{\mu\nu} U^\dagger [{\cal V}^\mu, {\cal V}^\nu ]\Big]  -i a_3 {\rm Tr}\Big[\hat{W}_{\mu\nu}[{\cal V}^\mu, {\cal V}^\nu]\Big] \nn\\
&&+ a_4 {\rm Tr}({\cal V}_\mu {\cal V}_\nu ){\rm Tr}({\cal V}^\mu {\cal V}^\nu )
+ a_5 {\rm Tr}({\cal V}_\mu {\cal V}^\mu ){\rm Tr}({\cal V}_\nu {\cal V}^\nu)
+\ldots \nn\\
&&-\cHBB \frac{H}{v}\, {\rm Tr} \Big[\hat{B}_{\mu\nu} \hat{B}^{\mu\nu} \Big]
   -\cHWW \frac{H}{v} {\rm Tr}\Big[\hat{W}_{\mu\nu} \hat{W}^{\mu\nu}\Big] +\cHBW \frac{H}{v} {\rm Tr}\Big[ U \hat{B}_{\mu\nu} U^\dagger \hat{W}^{\mu\nu}\Big] +\ldots  \nn\\
\label{eq-L4}
\eear
where the dots mean other terms in the bosonic sector of the EChL that we do not consider because they do not preserve custodial symmetry or because they are not relevant for the present work.
In $\mL_2$ above, the terms $\mL_{\rm GF}$ and $\mL_{\rm FP}$, denote the gauge-fixing and Faddeev-Popov Lagrangian, respectively, and we have assumed the same Higgs boson potential as in the SM:
\be
V(H) 
= \frac{1}{2}\mh^2 H^2 +\lambda \vev H^3 +\frac{\lambda}{4}H^4 =
\frac{1}{2}\mh^2 v^2
\left(\left(\frac{H}{\vev}\right)^2+\left(\frac{H}{\vev}\right)^3+\frac{1}{4}\left(\frac{H}{\vev}\right)^4\right) \,,
\ee
with $m_H^2=2\lambda v^ 2$, and $\lambda$ being the Higgs self-coupling. Notice that the second equality shows explicitely the chiral dimension of $V(H)$ belonging to $\mL_2$ since, as we have said, the Higgs mass $m_H$ counts as another soft mass of ${\cal O}(p)$. Notice also that there is not linear term in $H$ as in the SM. Regarding the relevant EChL parameters in $\mL_2$, it is just the $a$ paramater what enters in the present computation and not $b$ since this latter involves two Higgs fields. For $a=1$, the SM couplings of $H$ to two gauge bosons $WW$ and $ZZ$ are recovered, whereas for $a \neq 1$ the $HWW$ and $HZZ$ couplings differ from their SM values. Notice also that there are not terms corresponding to $HAA$ and $HZA$ interactions in $\mLtwo$ (as in the SM).
 
In $\mLfour$ above,  we have introduced the chiral vector ${\cal V}_\mu= (D_\mu U) U^\dagger$ (with chiral dimension 1) to write some of the effective operators in $\mLfour$ in a compact form. For the EChL parameters in front of the effective operators in the first two lines of $\mLfour$ we use the usual notation given by $a_i$'s (corresponding to the $\alpha_i$'s in the original formulation of the EChL~\cite{Longhitano:1980iz}, prior to the Higgs discovery). The operators with $a_{i}$'s contribute to anomalous vertices with three and four gauge bosons and are relevant for other observables like EW vector boson scattering, photon-photon scattering and others, but are not relevant for the present work. The parameter $a_1$ enters in the EW precision observables, concretely the oblique $S$ paramater, and also affects the two-point function $\gamma Z$ that enters {\it a priori} in the present computation of $H \to \gamma Z$. The most relevant effective operators in $\mLfour$ for the present work are the three last ones, with coefficients $c_{HBB}$, $c_{HWW}$ and $c_{HBW}$. These operators can be easily written in terms of the physical basis, $\gamma$, $W^\pm$ and $Z$, and if we just select those contributing to $H$ decays into $\gamma \gamma$ and $\gamma Z$ we get:
\bear
&&
-\cHBB \frac{H}{v}\, {\rm Tr} \Big[\hat{B}_{\mu\nu} \hat{B}^{\mu\nu} \Big] 
   -\cHWW \frac{H}{v} {\rm Tr}\Big[\hat{W}_{\mu\nu} \hat{W}^{\mu\nu}\Big] 
   +\cHBW \frac{H}{v} {\rm Tr}\Big[ U \hat{B}_{\mu\nu} U^\dagger \hat{W}^{\mu\nu}\Big] =  \nn\\
 &&  
-\frac{1}{2}\frac{e^2}{v}\cone H
(\partial_\mu A_\nu -\partial_\nu A_\mu)
(\partial^\mu A^\nu -\partial^\nu A^\mu)
-\frac{eg\cosw}{v}\ctwo H
(\partial_\mu A_\nu -\partial_\nu A_\mu)
(\partial^\mu Z^\nu -\partial^\nu Z^\mu)
\nn\\
 &&+\ldots
\label{eq-L4relevant}
\eear
where:
\bear
\cone &=& \cHBB+\cHWW-\cHBW \,,  \label{chgg}\\
\ctwo &=& \frac{1}{\cosw^2}(-\cHBB\sinw^2+\cHWW\cosw^2-\frac{1}{2}\cHBW(\cosw^2-\sinw^2)) \,.
\label{chgz}
\eear

Finally, regarding the quantization of the EChL we choose here to use the same gauge-fixing Lagrangian, $\mL_{\rm GF}$, as in the SM for the $R_\xi$ gauges~\cite{Fujikawa:1972fe}. The issues of $R_\xi$ gauge-fixing and renormalization within the context of the EChL were already studied long ago in~\cite{Herrero:1993nc,Herrero:1994iu} when the Higgs particle was not included explicitly in the Lagrangian. Generically, the quantizacion of the EChL requires the insertion of appropriate gauge-fixing functions $F_j$ involving the EW gauge bosons and the GBs. This gauge-fixing Lagrangian can be written in terms of the physical basis as:
\be
\mL_{\rm GF} = -\frac{1}{\xiw}F_+F_- -\frac{1}{2\xiz}F_{\rm Z}^2 -\frac{1}{2\xia}F_{\rm A}^2 \,,
\ee
where the gauge-fixing functions are:
\bear
F_\pm=\partial^{\mu}W_{\mu}^\pm-\xiw \mw \pi^\pm \,,\quad F_Z=\partial^{\mu}Z_{\mu}-\xiz \mz \pi^{3} \,,\quad F_A=\partial^{\mu}A_{\mu} \,.
\eear
and $\xi_W$, $\xi_Z$, $\xi_A$  are the typical gauge-fixing parameters of the $R_\xi$ gauges.

From the above gauge-fixing functions, $F_\pm$, $F_Z$ and $F_A$,  we derive the corresponding Faddeev-Popov Lagrangian~\cite{Faddeev:1967fc} (see also,~\cite{Herrero:1993nc}), by: 
\be
\mL_{\rm FP} = \sum_{i,j=+,-,Z,A} c^{i\,\dagger} \frac{\delta F_i}{\delta \alpha_j} c^j \,,
\ee
where $c^j$ are the ghost fields and $\alpha_j$ are the corresponding gauge transformation parameters ($j=+,-, Z,A$).
For the present computation of Higgs decays $\decayboth$, with $V=\gamma, Z$ only charged particles ($W^\pm$, $\pi^\pm$ and $c^\pm$) enter in the loops, therefore the only relevant gauge-fixing parameter for this work is $\xi_W$. We will use a short notation for this $\xi_W$ parameter from now on and call it simply $\xi$.

Once the relevant parts of the EChL have been set, we are ready to present the relevant interactions and Feynmam rules (FRs) for the present work. Specifically, the relevant vertices entering in the present computation are: those involving pure gauge bosons, those with pure scalars (GBs and Higgs boson), with mixed gauge-scalar bosons and vertices involving ghosts (ghosts with gauge bosons and ghosts with scalars). Regarding the interactions involving GBs, it is illustrative to remind that one has to perform first the expansion of the exponential matrix of \eqref{expo} in powers of the GB fields, $\pi^a$, and the non-linearity of the EFT is clearly manifest:
\bear
U (\pi^a)&=
& I_2 +i\frac{\pi^a}{\vev}\tau^a -\frac{2\pi^+\pi^-+\pi^3\pi^3}{2\vev^2}I_2 -i\frac{(2\pi^+\pi^-+\pi^3\pi^3)\pi^a}{6\vev^3}\tau^a +\ldots
\eear
where $I_2$ is the unity matrix and the dots stand for terms with four or more GBs. Only interactions with two GBs at most are relevant for this work.    

Then, following the standard procedure with Chiral Lagrangians, we distinguish between tree level interactions from $\mL_2$ and tree level interactions from $\mLfour$. Indeed, to differenciate the corresponding FRs, we use a different notation with a shaded box to mark the vertices from $\mLfour$.

The summary of all the relevant FRs for the present work is presented in \appref{App-FRules}. \tabrefs{relevant-FR1}{relevant-FR2} collect the relevant FRs from $\mL_2$, and \figref{diags-L4-FRules}, summarizes the relevant FRs from $\mLfour$. In \tabrefs{relevant-FR1}{relevant-FR2} (column on the right) we have also added the corresponding FRs within the SM, for a clear comparison. Some comments on these relevant EChL Feynman rules are in order. 
Firstly, from $\mL_2$, we see that the FRs for pure gauge boson interactions and for gauge boson interactions with ghosts are in the EChL as in the SM, as expected. 
Second, the  Feynman rules for the vertex $HWW$ and $HZZ$ within the EChL differ from that of the SM  by the coefficient $a$ in front. Only by setting $a=1$ one recovers the SM Feynman rules. This modification of the SM vertex is well known and, in fact, it is the main responsible for the most notably difference, respect to the SM result, from the loop contributions to the Higgs decays in the unitary gauge~\cite{PhysRevLett.27.1688}, as it has been already established \cite{losBuchalla}. 
Third, there are some interactions among gauge bosons and GBs that remain the same in the EChL as in the SM. Concretely, the FRs for $\gamma\pi\pi$,  $Z\pi\pi$, $\gamma W \pi$, $Z W \pi$, $\gamma \gamma \pi\pi$ and $\gamma Z \pi\pi$ are equal to the SM ones. Fourth, all the remaining relevant FRs from $\mL_2$ involve one Higgs boson and are different than in the SM. Specifically these are: $H\pi\pi$, $HW\pi$, $Hcc$, $H\gamma W\pi$, $H Z W\pi$, $H\gamma \gamma\pi\pi$, $H\gamma Z\pi\pi$. None of them recover the SM value by fixing $a=1$. In particular, the vertex $H\pi\pi$, in addition to the $a$ parameter involved, has a very different structure in momentum, typical from the non-linearity of the GBs and momentum expansion in Chiral Lagrangians. The vertex $HW\pi$ has also a different momentum dependence than in the SM.
Finally, there are vertices that are not present in the EChL but they are present in the SM, and viceversa. Particularly interesting is the absence of the interaction vertex of a Higgs boson with two ghosts, thus we have introduced a vanishing FR for $Hcc$ in the table. We also see that vertices with four and five legs, concretely, $H\gamma\pi\pi$, $HZ\pi\pi$, $H\gamma \gamma\pi\pi$ and $H\gamma Z\pi\pi$ are non-vanishing in the EChL, but they are not present in the SM (accordingly, we put in \tabref{relevant-FR2} a zero value for these FRs in the SM). 

Finally, regarding the interaction vertices from $\mLfour$, we see in \figref{diags-L4-FRules}, that there are two parameters involved which modify the Lorentz structure of the $H\gamma V$ vertices.  Concretely, $c_{H\gamma \gamma}$ that enters in  $H\gamma \gamma$,  and $c_{H\gamma Z}$ that enters in $H\gamma Z$ following \eqrefs{chgg}{chgz}. 

We will see in the following sections how all these interactions from the EChL enter in our computation of the Higgs decays into $\gamma \gamma$ and $\gamma Z$ within the $R_\xi$ gauges, and we will illustrate as well the comparison with respect to the SM computation.

\section{Computation of the amplitudes: Renormalization and loops}
\label{sec-renorm}
In this section we present the computation of the \1loop amplitudes for the two processes of our interest in this work, 
$\amp( H \to \gamma \gamma)$ and $ \amp(H \to \gamma Z)$, which from now on when commented together will be referred jointly as  $\amp( H \to \gamma V)$.  This computation will be done within the EFT given by the EChL, in the $R_\xi$ gauges, and at the \1loop level approximation. Contributions from two and more loops, are assumed to be negligible when compared with the \1loop contributions.

\subsection{Description of the computational  procedure}
First of all, we remind the main rules in doing a computation within the EChL to \1loop level. Generically, any amplitude of a process receives two types of contributions that are separated as Leading Order (LO) and Next to Leading Order (NLO),
\be
\amp=\ampLO+\ampNLO \,.
\label{amp-expansion}
\ee
The LO term, $\ampLO$, is the result of the amplitude when using just the tree level Lagrangian with chiral dimension 2, or equivalently in diagrammatic terms, when using just the Feynman Rules from $\mLtwo$ in \eqref{eq-L2} to build tree level diagrams. The NLO amplitude, $\ampNLO$, comes from adding two types of contributions, the ones from tree level diagrams using the Feynman Rules from $\mLfour$ in \eqref{eq-L4}, and the ones from \1loop diagrams using the Feynman Rules from $\mLtwo$ in \eqref{eq-L2}. This procedure is similar to the one in ChPT, where loop contributions from $\mLtwo$, are known to contribute to the same order in the chiral expansion, i.e. $\mO(p^4)$, as tree level contributions from the chiral dimension 4 Lagrangian, $\mLfour$. In fact, the operators in  
$\mLfour$ have a two-fold role in Chiral Lagrangians. On one hand they provide contributions to the amplitudes of next order in the momentum expansion respect to $\mLtwo$, and on the other hand they may act also as additional counter-terms of new divergences generated to \1loop by $\mLtwo$, which cannot be absorbed by just the renormalization of $\mLtwo$. 

For the two cases under study here, it is clear that the LO contribution to the amplitude $\amp( H \to \gamma V)_{\rm LO}$ vanishes, as it happens indeed within the SM where there are not tree level vertices $H \gamma \gamma$ and $H \gamma Z$. Thus $\amp( H \to \gamma V)$ only receives contributions from NLO within the EChL.

Regarding the regularization procedure of the loop contributions,  we use here dimensional regularization~\cite{tHooft:1972tcz} as usual. This method has the advantage that preserves all the relevant symmetries, including chiral invariance. 

Concerning the renormalization procedure,  we use here the standard method of generating counter-terms
for all parameters and fields appearing in the tree level Lagrangian,   $\mLtwo +\mLfour $,  according to the usual prescription that relates the bare quantities and the renormalized ones. In our present case, the relevant relations are summarised as follows: 
\bear
H_0 &=& \Zf_H^{1/2}H\,,\,\, B_{0\,\mu} = \Zf_B^{1/2}B_{\mu}\,,\,\, W_{0\,\mu}^{\pm,3} = \Zf_W^{1/2}W_{\mu}^{\pm,3} ,  \nn\\
\gY_0 &=& \Zf_B^{-1/2}(\gY +\delta \gY)\,,\,\,
g_0 = \Zf_W^{-1/2}(g +\delta g)\,,  \nn\\
a_i^0 &=& a_i+\delta a_i \,,\,\,\,\, c_i^0 = c_i+\delta c_i \,,
\label{counter}
\eear
where we have denoted with a $0$ script the bare quantities and with no script the renormalized quantities, to abbreviate the notation. The multiplicative renormalization constants $\Zf$ are split as usual:
 \bear
 \Zf_{H,B,W}  &=&1 +\delta\Zf_{H,B,W},
 \eear
and we have written generic counter-terms for both types of parameters in $\mLfour$, namely, the $a_i$  and the $c_i$ coefficients in \eqref{eq-L4}. The divergences in the EChL coefficients have been studied in the literature~\cite{Buchalla:2012qq,Espriu:2013fia,Delgado:2013hxa,Delgado:2013loa,Delgado:2014jna,Guo_2015,Buchalla:2017jlu,Buchalla:2020kdh} and they have also been used to renormalize some scattering processes to \1loop within the EChL, as $\gamma \gamma \to WW, ZZ$~\cite{Delgado:2014jna}, and vector boson scattering processes~\cite{Espriu:2013fia,Delgado:2013hxa,Delgado:2013loa}. However, in the present work, only the  $c_i$ from  \eqref{eq-L4} enter. Specifically, the involved counter-terms from  \eqref{eq-L4} are those of the derived coefficients  $c_{H \gamma \gamma}$ in \eqref{chgg}, for $H \to \gamma \gamma$ and $c_{H \gamma Z}$ in \eqref{chgz}, for $H \to \gamma Z$. 
 
For a clear presentation of the forthcoming results, our systematic computation of the decay amplitude $\amp(\decayboth)$ at \1loop level is organized in terms of  the relevant one-particle irreducible (1PI)  renormalized Green functions, such that our renormalization program will be first addressed to these 1PI functions,  and later these will be inserted into the
 corresponding $\amp( H \to \gamma V)$ amplitude. Generically, this can be written as:
\be
\amp(\decayboth)=\amp_{1-pt}+\amp_{2-pt}+\amp_{3-pt} +\amp_{w.f.r.}\,,
\label{ampHVV-generic}
\ee
where $\amp_{n-pt}$ means the contribution from the n-point 1PI renormalized function to the amplitude, and we have separated explicitly the potential contribution from the wave-function renormalization of the external legs, $ \amp_{w.f.r.}$. All these 1PI renormalized functions,  called generically here  $\greenfR$, are derived using diagrammatic methods, following  the usual decomposition:
\be
\greenfR =\greenfT+\greenfL+\greenfC \,.
\label{fgreen}
\ee
Within the EChL, $\greenfT$ means contributions from tree level Lagrangian, $\mLtwo +\mLfour $, $\greenfL$ are the contributions from the \1loop diagrams using the interaction vertices of $\mLtwo$, and $\greenfC$ summarises the contributions from all the counter-terms generated by the prescription in \eqref{counter}. All these quantities are expressed in terms of renormalized parameters.

Finally, once the regularization and renormalization procedures have been fixed, the next step is setting the renormalization conditions that we have adopted here. These particular conditions will provide the specific values of the counter-terms involved in the present computation.
We adopt a hybrid prescription in which we impose the on-shell (OS) scheme for the physical sector in $\mLtwo$, i.e. for $W$, $Z$, $\gamma$ and $H$, and the $\overline{MS}$ scheme for the EW chiral parameters in $\mLfour$, i.e., for the $a_i$ and $c_i$ coefficients. Explicitly, our renormalization conditions are the following: 
\begin{itemize}
    \item Vanishing Tadpole:
        \be
        T^{\rm R}=0 \,.
        \ee
    \item The pole of the renormalized propagator of the Higgs boson lies at $\mh$ and the corresponding residue is equal to 1:
        \be
        {\rm Re}\left[ \Sigma^{\rm R}_{HH}(\mh^{2}) \right] =0 \,,\quad {\rm Re}\left[ \frac{d\Sigma^{\rm R}_{HH}}{dq^2}(\mh^{2}) \right] =0 \,.
        \ee
    \item Properties of the photon: residue equal one; no $A-Z$ mixing propagators for on-shell photons; and the electric charge defined like in QED, since there is a remnant $U(1)_{\rm EM}$ symmetry.
        \be
        {\rm Re}\left[ \frac{d\Sigma_{AA}^{\rm R}}{dq^2}(0) \right] =0 \,,\quad \Sigma_{ZA}^{\rm R}(0) =0  \,,\quad \Gamma^{{\rm R}\,\mu}_{\gamma e e}\vert_{\rm OS}=i e \gamma^\mu \,. 
        \label{sigma}
        \ee    
    \item The poles of the renormalized propagator of the $W$ and $Z$ bosons lie at $\mw$ and $\mz$
        \be
        {\rm Re}\left[ \Sigma_{WW}^{\rm R}(\mw^{2}) \right] =0 \,,\quad {\rm Re}\left[ \Sigma_{ZZ}^{\rm R}(\mz^{2}) \right] =0 \,.
        \ee
    \item $\overline{MS}$ scheme for $\cone$ and $\ctwo$.
\end{itemize}
In the previous expressions, we used the notation $\Sigma^{\rm R}_{HH}$ for the renormalized self-energy of the Higgs boson and $\Sigma^{\rm R}_{VV'}$ for the renormalized transverse self-energy connecting the gauge bosons $V$ and $V'$.
The renormalization of the unphysical sector in $\mLtwo$ is not relevant for the present computation, then we do not need to fix the counter-terms participating in the gauge-fixing, in the longitudinal part of the self-energy of the gauge bosons, in the self-energy of the GBs and in the mixing propagators of photon-ghost or $Z$-ghost.

\begin{figure}[h!]
\begin{center}
\includegraphics[width=0.7\textwidth]{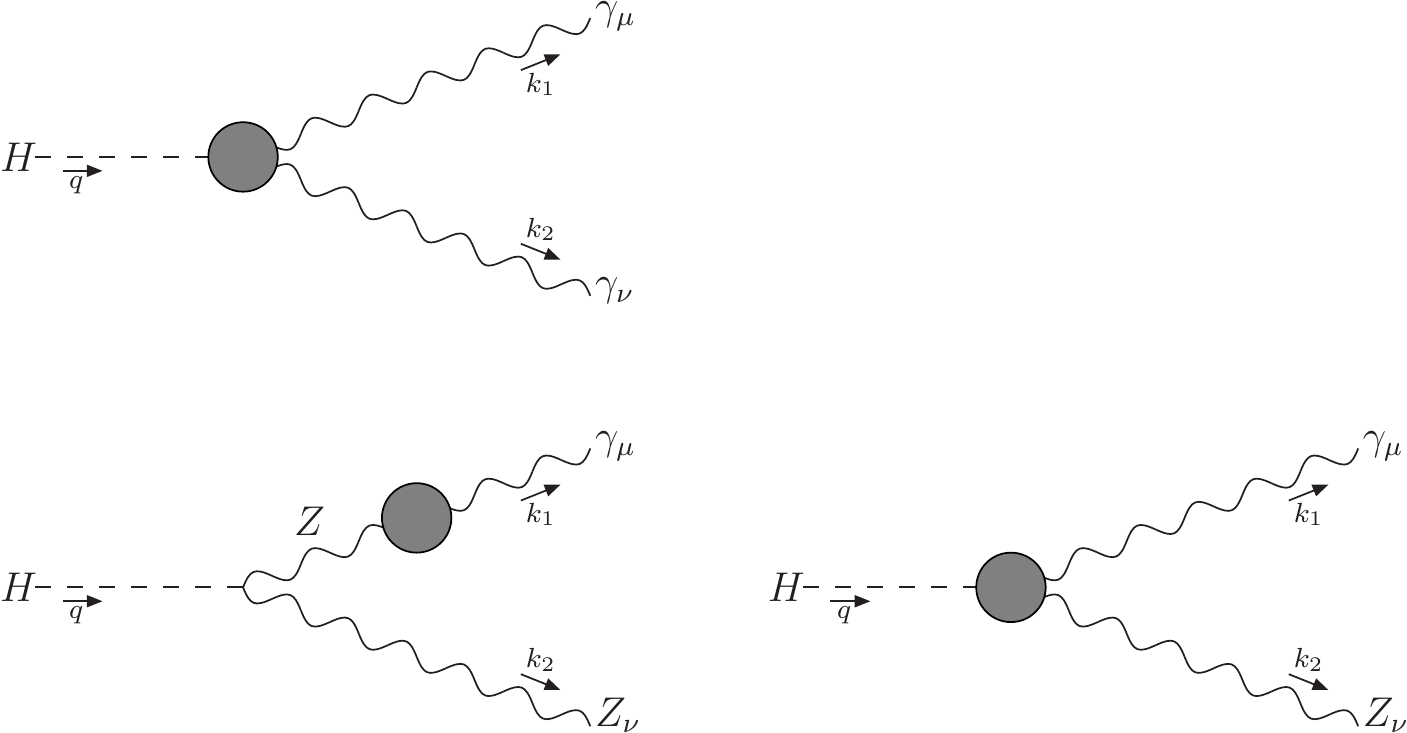}
\caption{Contributions of the two and three-point renormalized Green functions (denoted by gray circles). Upper line corresponding to $H\to\gamma\gamma$ and the lower line to $H\to\gamma Z$.}
\label{generic-decays}
\end{center}
\end{figure}

With the above renormalization conditions we can already conclude on which n-point renormalized 1PI functions contribute in each decay:
\be
\ampone \equiv \amp( H \to \gamma \gamma)=\langle \gamma(k_1)\gamma (k_2)\vert \mathcal{S}\vert H(q)\rangle=\amp_{3-pt} \,,
\label{ampHgg}
\ee
and 
\be
\amptwo \equiv \amp( H \to \gamma Z)=\langle \gamma(k_1) Z (k_2)\vert \mathcal{S}\vert H(q)\rangle =\amp_{2-pt}+\amp_{3-pt} \,.
\label{ampHgZ}
\ee
The corresponding graphical representations of  the above contributions are shown in \figref{generic-decays}. We have included in these graphs
our momentum convention for the $H(q)\to A_\mu(k_1)V_\nu(k_2)$ decays: $q$ is the incoming Higgs boson momentum, and $k_{1,2}$ are the outgoing gauge bosons momenta.

Some comments are in order. First, notice that there are not contributions from the 1-point 1PI function. This is due to the above vanishing renormalized Tadpole condition which implies  $\amp_{1-pt}=0$. Second, $\amp_{w.f.r.}$ also vanishes since it is obtained as $\amp_{w.f.r.}=\ampLO (\mathcal{R}_H+\mathcal{R}_A+\mathcal{R}_V)/2$ from the finite residues $\mathcal{R}$ of the external legs that generically could be different from 1, depending on the renormalization conditions. However, here  $\amp_{w.f.r.}=0$ because there is not LO contribution as we have said, i.e., due to $\ampLO=0$. Third, it is important noticing that for the $H\to\gamma Z$ decay, the renormalized self-energy connecting a photon with a $Z$ boson is present as an external leg correction (lower left diagram of \figref{generic-decays}). But our renormalization condition for this Green function evaluated at an on-shell photon implies that $\amp_{2-pt}=0$ (notice that no such diagram is present in the \1loop computation of $H\to\gamma\gamma$). Therefore, the unique contributions to both decay amplitudes are $\amp_{3-pt}$ and these
come exclusively from the corresponding renormalized three-point functions: $\greenfR_{HAA}$ (upper diagram of \figref{generic-decays}) and $\greenfR_{HAZ}$ (lower right diagram of \figref{generic-decays}), respectively.

Finally,  since the renormalization procedure implemented here preserves the gauge symmetry of the EChL, then the corresponding Ward identity for the $U(1)_{\rm EM}$ subgroup is also preserved .
This $U(1)_{\rm EM}$ Ward identity for the external on-shell photon implies a very precise structure for the decay amplitude of $H(q)\to\gamma(k_1)V(k_2)$ which can be written as:
\be
\ampboth=\ampboth^{\mu\nu}\epsilon_\mu(k_1)\epsilon_\nu(k_2) \quad{\rm with}\quad \ampboth^{\mu\nu}=\Veffboth(q,k_1,k_2)\left(g^{\mu\nu} -\frac{k_2^\mu k_1^\nu}{k_1\cdot k_2 } \right) \,,
\label{ward-structure}
\ee
where $\epsilon_\mu(k_1)$ and $\epsilon_\nu(k_2)$ are the polarization vectors of the outgoing gauge bosons and $\Veffboth$ is a complex function of the three external momenta ($q$, $k_1$ and $k_2$), with $q=k_1+k_2$ by momentum conservation.
Written the amplitude in this way, it is easy to check that it vanishes if the photon polarization vector is replaced by its corresponding momentum.
Notice that $\ampboth^{\mu\nu}=\Gamma_{HAV}^{\rm{R}\,\mu\nu}$ in the 1PI notation described before. Thus, 
our next aim is to provide the results for the two relevant functions  $\Gamma_{HAA}^{\rm{R}\,\mu\nu}$ and $\Gamma_{HAZ}^{\rm{R}\,\mu\nu}$ or, equivalently,  $\amp^{\mu\nu}_{\gamma \gamma}$ and $\amp^{\mu\nu}_{\gamma Z}$.

\subsection{Analytical results}
\label{analytical-results}
In this section we present the analytical results for the total \1loop amplitudes $\ampone$ and $\amptwo$ within the EChL  in the $R_\xi$ gauges.  In the following we assume that the final gauge bosons ($\gamma$ and $Z$) are on-shell and we present the results for the two interesting cases corresponding to: 1) the Higgs boson is off-shell, and 2) the Higgs boson is on-shell. We will also compare all these results with the corresponding ones in the SM case, which we have recomputed here in the $R_\xi$ gauges for illustrative purposes. From now on, and to distinguish clearly the EChL  from the SM results, we continue using the simple notation $\amp$ and $V$ for the EChL's amplitudes and vertex functions, but we use instead the `bar' notation for the corresponding SM results, $\ampSM$ and 
$\overbarV{V}$.

Next, we split all the results into three contributions, tree level, counter-terms and loops, following \eqref{fgreen}, which we compute with diagrammatic methods.
The contributions from tree level diagrams lead to: 
\bear
\amp_{\gamma\gamma}^{\rm{T}\,\mu\nu} &=& \frac{e^2 g}{\mw}\cone \left(k_1\cdot k_2g^{\mu\nu} -k_2^\mu k_1^\nu \right) \,, \label{treelevel-decayone}\\
\amp_{\gamma Z}^{\rm{T}\,\mu\nu} &=& \frac{eg^2\cosw}{\mw}\ctwo \left(k_1\cdot k_2g^{\mu\nu} -k_2^\mu k_1^\nu \right) \,.
\label{treelevel-decaytwo}
\eear
They are written in terms of the renormalized parameters, $e$, $g$, $\cone$, $\ctwo$ and $\mw$.
The corresponding Feynman rules for the relevant interaction vertices, which come from $\mLfour$,  are given in \appref{App-FRules}.
These tree level results  in the EChL  are clearly in contrast to the SM case where there are not tree level contributions.

\begin{figure}[h!]
\begin{center}
\includegraphics[width=0.6\textwidth]{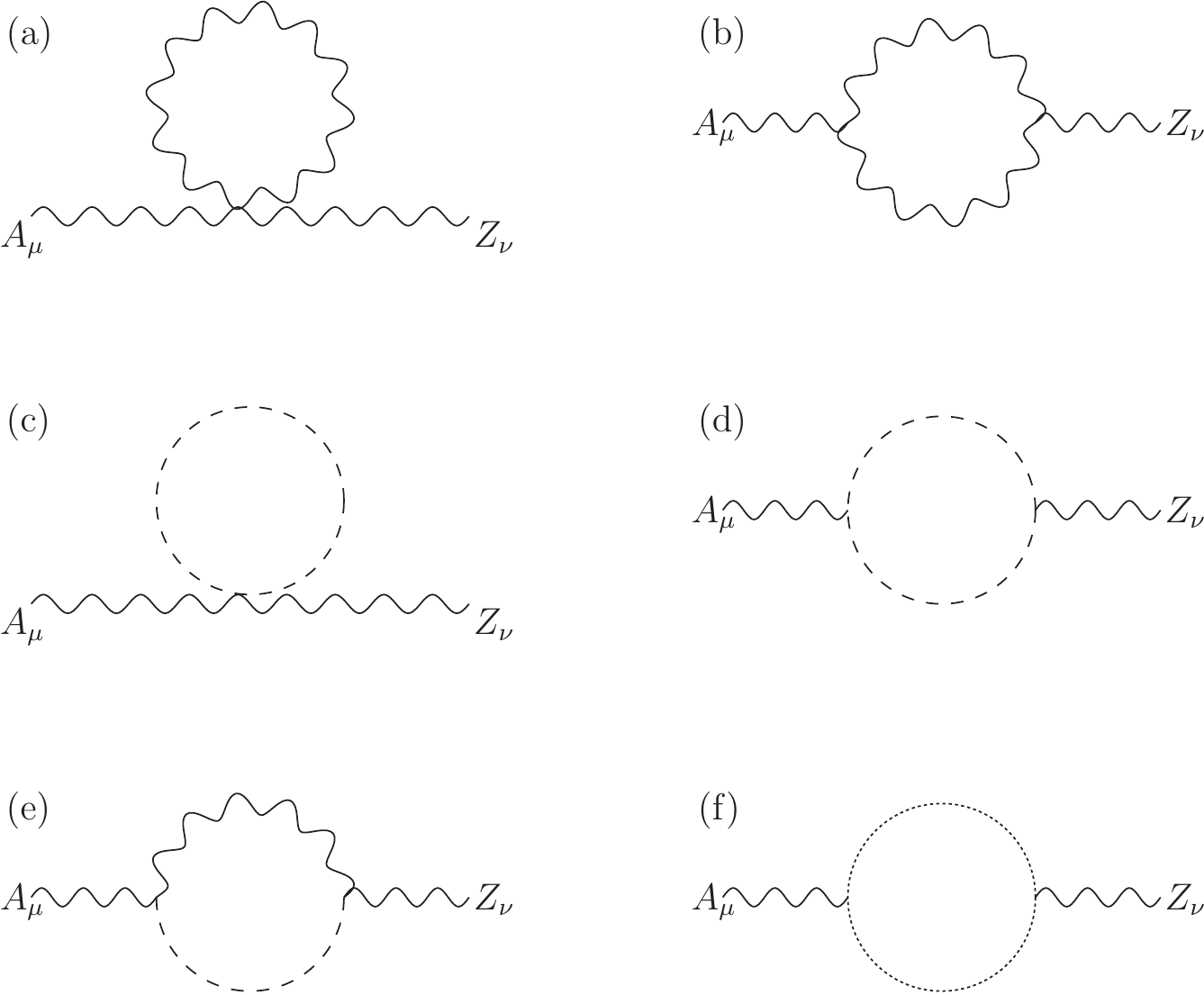}
\caption{Bosonic loop diagrams contributing to $\Sigma^{\rm L}_{ZA}(q^2)$ in the $R_\xi$ gauges. We denote the internal $W$ bosons, charged GBs and charged ghosts by wavy, dashed and dotted lines, respectively.}
\label{diags-SE}
\end{center}
\end{figure}

On the other hand, the counter-term contributions in the EChL lead to the following result: 
\bear
\ampone^{\rm{C}\,\mu\nu} &=& \frac{e^2 g}{\mw}\delta\cone \left(k_1\cdot k_2g^{\mu\nu} -k_2^\mu k_1^\nu \right) \,,  \label{counterterm-decayone}\\
\amptwo^{\rm{C}\,\mu\nu} &=& \frac{eg^2\cosw}{\mw}\delta\ctwo \left(k_1\cdot k_2g^{\mu\nu} -k_2^\mu k_1^\nu \right) +\frac{a g}{\mw}\mz^{2}\sinw\cosw\left( \frac{\delta g}{g} -\frac{\delta \gY}{\gY} \right)g^{\mu\nu} \,.
\label{counterterm-decaytwo}
\eear
The corresponding Feynman rules for the involved counter-terms can also be found in \appref{App-FRules}.
The previous result is also in contrast to the SM case, where there is not counter-term for the $H\to\gamma\gamma$ decay and the corresponding one for the $H\to\gamma Z$ decay is equal to just the second term with $a=1$ in the previous equation. 
This second term is derived from $\mLtwo$ with the prescription in \eqref{counter}, and its value is fixed from the renormalization condition on the self-energy of the photon-$Z$ mixing in \eqref{sigma}. 
Concretely, one starts with the renormalized two-point transverse self-energy of the photon-$Z$ mixing as given by:
\be
\Sigma_{ZA}^{\rm R} (q^2) = \Sigma_{ZA}^{\rm L} (q^2) -q^2\left( \delta\Zf_{ZA}-g\gY(\cosw^2-\sinw^2) (a_1+\delta a_1) \right) +\mz^{2}\sinw\cosw\left( \frac{\delta g}{g} -\frac{\delta \gY}{\gY} \right) \,.
\ee
where  $\Sigma_{ZA}^{\rm L}$ is the \1loop contribution to this self-energy and  $\delta\Zf_{ZA}$ is the corresponding counter-term from the wave-function renormalization.
Then, from the OS renormalization condition, $\Sigma_{ZA}^{\rm R}(0) =0$, and the explicit bosonic loop computation of the self-energy in the $R_\xi$ gauges of the diagrams in \figref{diags-SE}, we get:
\bear
\mz^{2}\sinw\cosw\left( \frac{\delta g}{g} -\frac{\delta \gY}{\gY} \right)&=&-\Sigma^{\rm L}_{ZA}(0)  \nn\\
&=&-\frac{e g}{16\pi^2}\frac{\mw^2}{4\cosw}\left(2(3+\xi)\div +5+\xi+\frac{2\xi(2+\xi)\log(\xi)}{1-\xi}\right)\,,
\label{sigmaaz}
\eear
where the divergence of dimensional regularization $\div$ is given in \eqref{div-definition}. Notice that the above result of $\Sigma^{\rm L}_{ZA}(0)$ in the $R_\xi$ gauges is valid for both the EChL and the SM, since the contributing diagrams in \figref{diags-SE} are the same, and also the FRs involved coincide in both cases.

Therefore, all the involved counter-terms in \eqrefs{counterterm-decayone}{counterterm-decaytwo} are fixed in terms of loop computations. The involved combination of gauge couplings counter-terms is fixed from the loops of the mixing two-point function, and there are just two left counter-terms $\delta\cone$ and $\delta\ctwo$ that need to be fixed from the loops of the three-point functions, $\Gamma_{HAA}^L$ and $\Gamma_{HAZ}^L$, respectively. Indeed, we can already anticipate our result that, for the $H\to\gamma\gamma$ decay, the sum of all the \1loop contributions in the $R_\xi$ gauges is UV-finite and, therefore we find $\delta\cone=0$. This result is in agreement with the result in the literature of the unitary gauge \cite{losBuchalla}, which is known to be finite without the need of renormalization. For the $H\to\gamma Z$ decay, the divergence of the sum of all the \1loop contributions in the $R_\xi$ gauges of the three-point function cancels out with the divergence from the \1loop contributions in the photon-$Z$ mixing self-energy, via the \eqref{sigmaaz}. Thus, we find in the $R_\xi$ gauges that $\delta\ctwo=0$. Again, this is in agreement with the result in the literature of the unitary gauge \cite{losBuchalla}, which is known to be finite without the need of renormalization. This is also in concordance with our finding that in the unitary gauge we get $\Sigma^{\rm L}_{ZA}(0)\vert^{\rm U}=0$, as expected.

Regarding the contributions from loop diagrams to $\ampbothLoop$, we classify them generically into four categories accordingly to the particles in the loops: i) loops with only gauge bosons (called `gauge' in short), ii) loops with both gauge and GBs (called `mix' in short reference to mixed loops), iii) loops with just GBs (called `GB' in short), and iv) loops with only ghosts (called `ghost' in short). Then we write the loop amplitude in the $R_\xi$ gauges as
\be
\ampbothLoop=\ampboth^{\rm{gauge}}+\ampboth^{\rm{mix}}+\ampbothchiral+\ampboth^{\rm{ghost}} \,.
\label{amp-classification}
\ee

Notice that, as we explained in \secref{sec-EChL}, the Higgs boson does not couple to the ghosts, thus there are not diagrams contributing to $\ampboth^{\rm{ghost}}$ in the EChL, but they are present in the SM computation.  Through this forthcoming loop computation we keep in parallel the comparison respect to the SM case which we find very illustrative. In fact, for the present loops computation and for the demonstration of gauge invariance of the result we follow closely the procedure and notation of diagrams in the reference \cite{Marciano-paper} that was devoted to the SM case. 

\begin{figure}[h!]
\begin{center}
\includegraphics[width=0.6\textwidth]{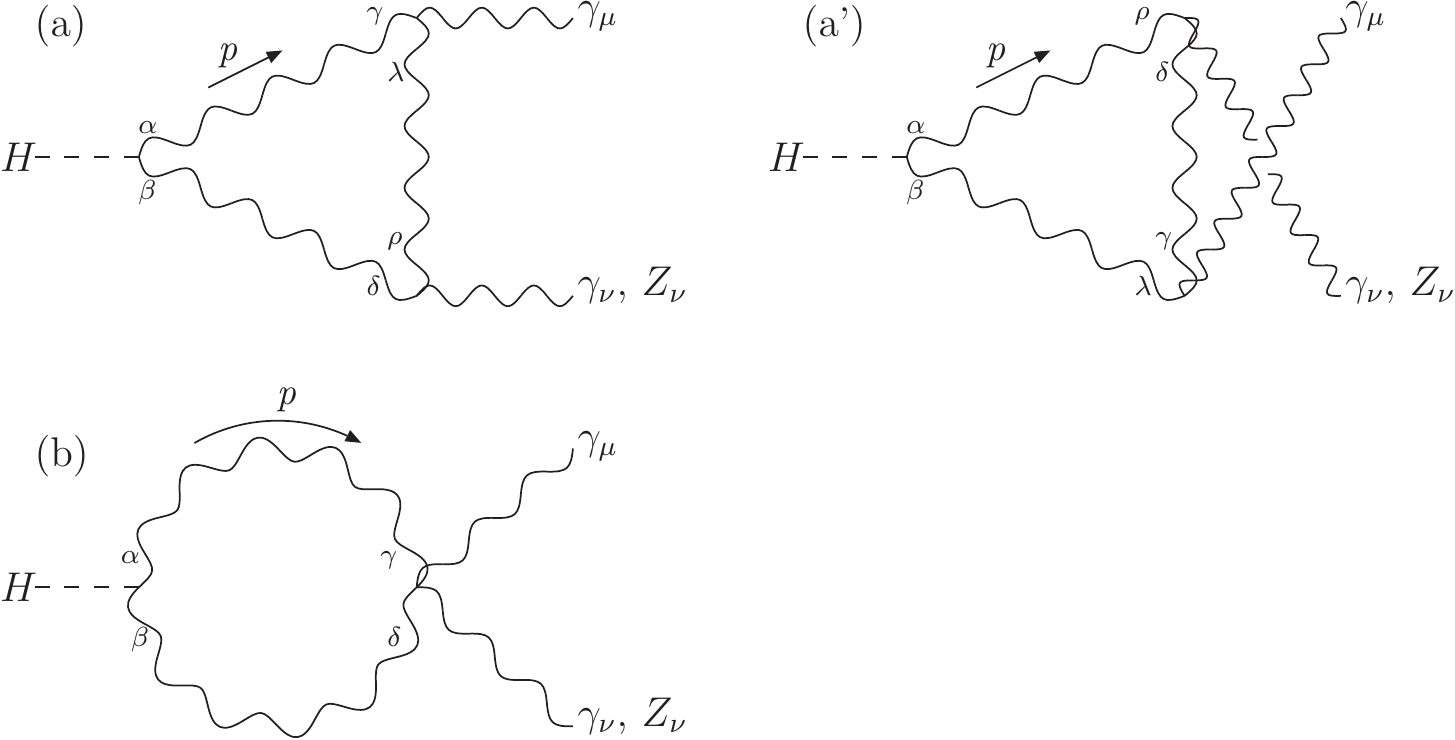}
\caption{One-loop diagrams contributing to $\ampboth^{\rm{gauge}}$ in both the EChL and the SM cases. Negative-charge flows in the same direction of the momentum $p$.}
\label{diags-puregauge}
\end{center}
\end{figure}

In \figref{diags-puregauge}, we show the diagrams contributing to $\ampboth^{\rm{gauge}}$. They are the same in both EChL and SM and coincide with the ones in the unitary gauge computation.
We explicitly draw the crossing diagrams (denoted by a prime ') because they give different results for the $H\to\gamma Z$ decay. For the $H\to\gamma\gamma$ case, they can be omitted by means of a factor 2 in the amplitude of the non-primed diagram. Our convention here is to show the diagrams in which the negative-charge flows in the same direction of the indicated momentum $p$.

\begin{figure}[h!]
\begin{center}
\includegraphics[width=0.6\textwidth]{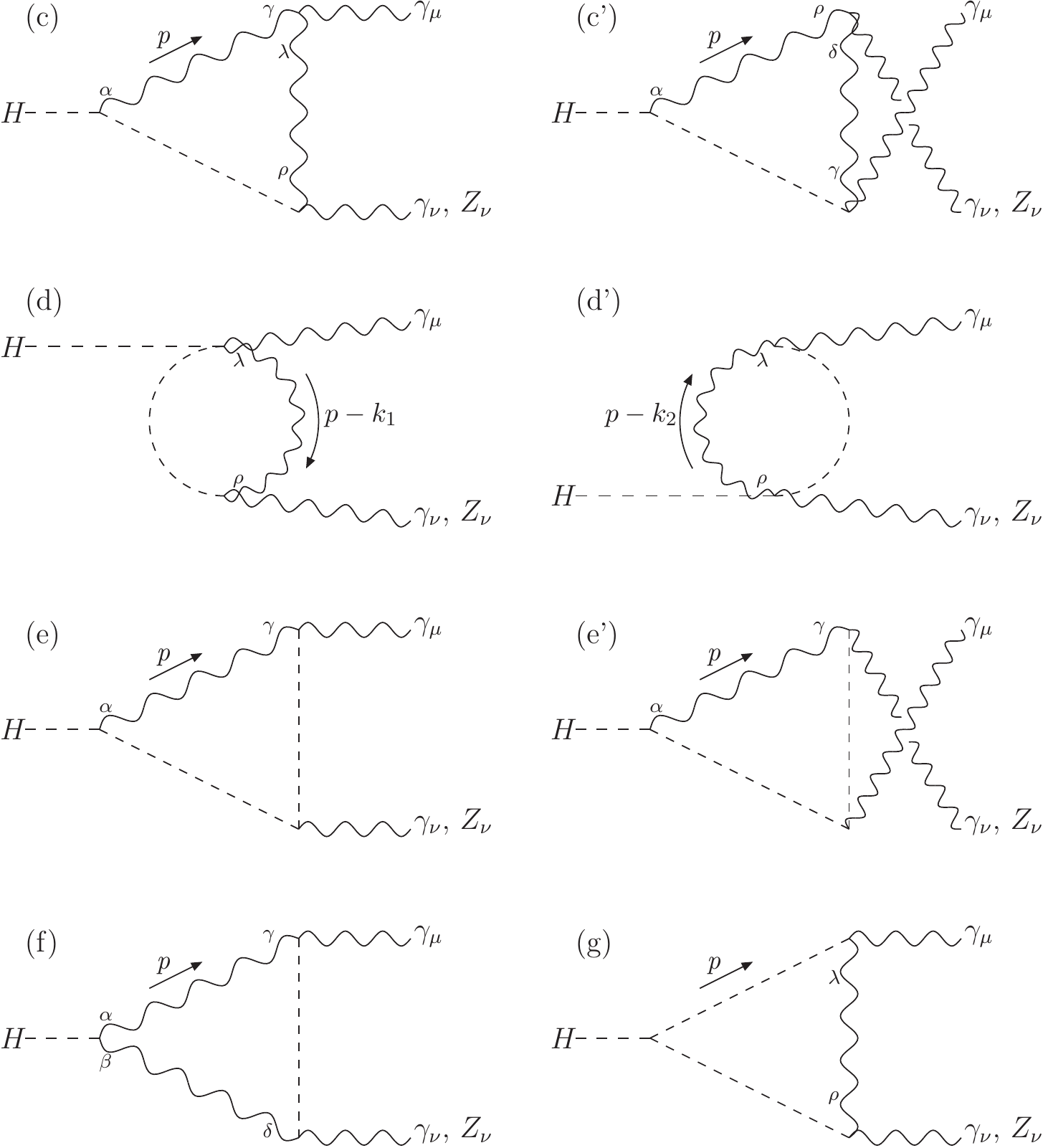}
\caption{One-loop diagrams contributing to $\ampboth^{\rm{mix}}$ in both the EChL and the SM cases. Negative-charge flows in the same direction of the momentum $p$.}
\label{diags-mix}
\end{center}
\end{figure}

In \figref{diags-mix}, we present the diagrams contributing to $\ampboth^{\rm{mix}}$. They are the same in both EChL and SM too.

\begin{figure}[h!]
\begin{center}
\includegraphics[width=0.6\textwidth]{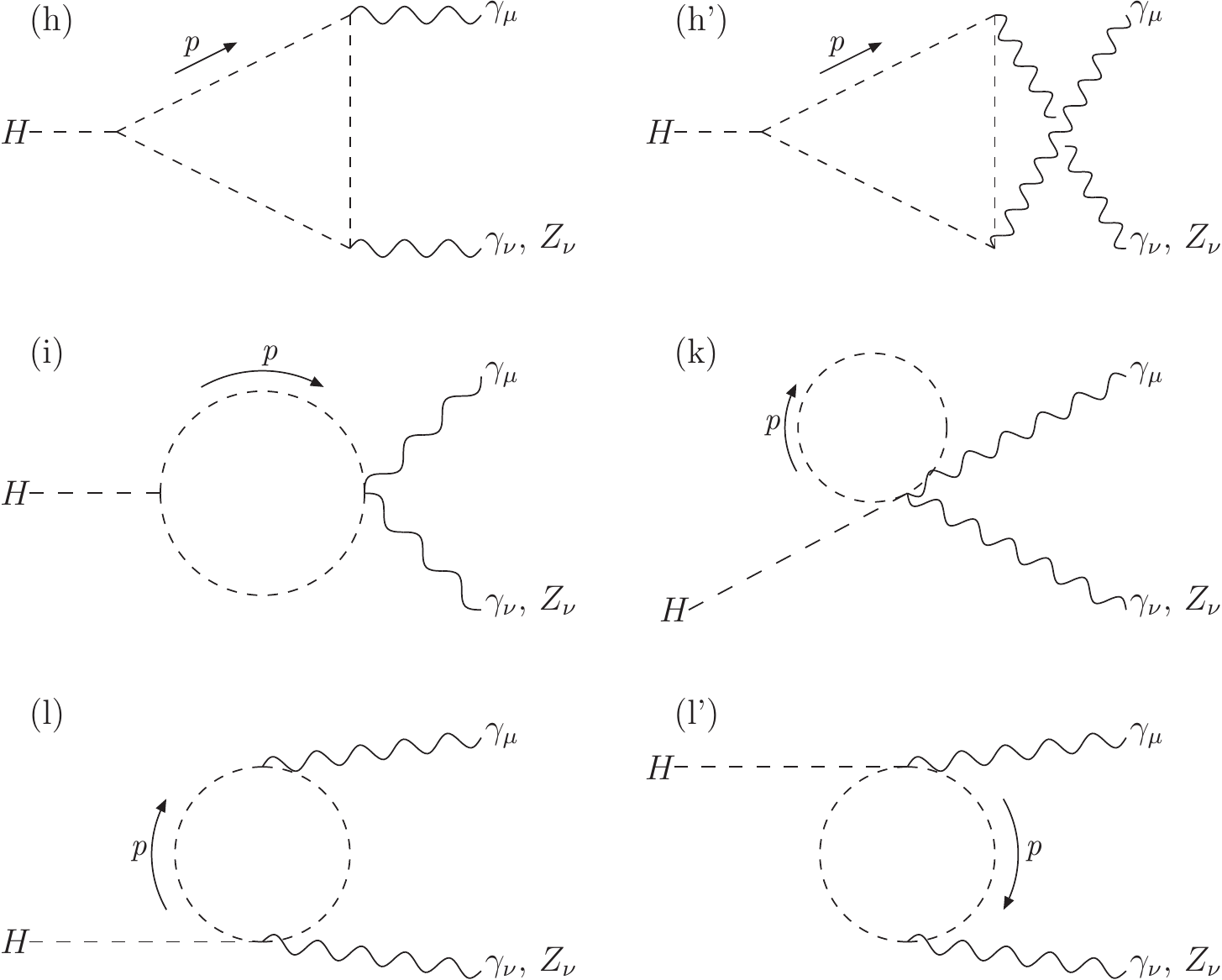}
\caption{One-loop diagrams contributing to $\ampboth^{\rm{GB}}$ in the EChL case. Only diagrams (h), (h') and (i) are present in the SM. Negative-charge flows in the same direction of the momentum $p$.}
\label{diags-GB}
\end{center}
\end{figure}

The \figref{diags-GB} shows the diagrams contributing to $\ampboth^{\rm{GB}}$. Now diagrams (k), (l) and (l') are only present in the EChL due to the multiple Goldstone boson interactions in this EFT (manifestation of its non-linearity).

\begin{figure}[h!]
\begin{center}
\includegraphics[width=0.6\textwidth]{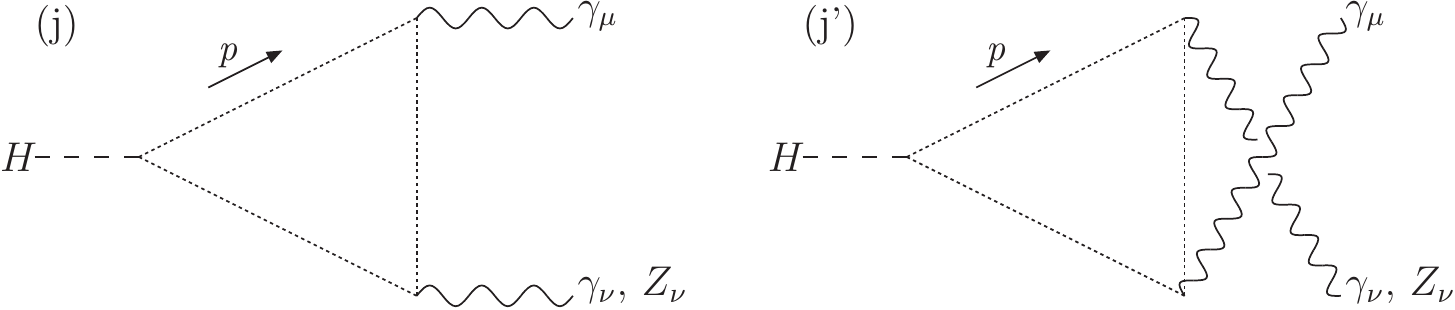}
\caption{One-loop diagrams contributing to $\ampboth^{\rm{ghost}}$ in the SM case. There are not such diagrams in the EChL. Negative-charge flows in the same direction of the momentum $p$.}
\label{diags-ghost}
\end{center}
\end{figure}

Finally, the \figref{diags-ghost} shows the diagrams contributing to $\ampboth^{\rm{ghost}}$ that are only present in the SM. 

Now we are ready to present the calculation of the previous diagrams and show how the gauge-fixing $\xi$-parameter cancellation proceeds in the EChL. The reading of each diagram in terms of the Feynman rules is provided separately in the \appref{App-LectureDiags} for both the EChL and the SM. The analytical \1loop computation was performed with Package-X \cite{packX} and we closely follow the presentation in \cite{Marciano-paper}. 
In particular, we split the $W$ propagator $-iP_{\mu\nu}$ into two pieces: i) the corresponding one to the unitary gauge $U_{\mu\nu}$, and ii) the remaining one $R_{\mu\nu}$ which contains the whole $\xi$-dependence of the $W$ propagator. The contributions to the amplitude from this latter part will be combined with the contributions from loops with GB and ghost propagators, that also provide  other $\xi$-dependent terms. 
In this way the final demonstration of the gauge invariance and $\xi$-independence of the total \1loop amplitude will be indeed manifest from the explicit cancellation of the $\xi$-dependent terms in doing the sum of all the above commented subset of contributions. Explicitly,
\be
\prop{\mu}{\nu}{p} = \frac{1}{p^2-\mw^2} \left( g_{\mu\nu} -(1-\xi)\frac{p_\mu p_\nu}{p^2-\xi\mw^2} \right) = U_{\mu\nu}(p) +R_{\mu\nu}(p)
\label{Wprop-spliting}
\ee
where
\be
U_{\mu\nu}(p)=\frac{g_{\mu\nu} -\frac{p_\mu p_\nu}{\mw^2}}{p^2-\mw^2} \quad\text{and}\quad R_{\mu\nu}(p)=\frac{\frac{p_\mu p_\nu}{\mw^2}}{p^2-\xi\mw^2}\,.
\label{Wprop-spliting2}
\ee

Firstly, it is interesting to notice that the `ghost' contributions in the EChL for both decays vanish:
\be
\ampboth^{\rm ghost}=0 ,
\label{EChLghost}
\ee
since, as we have said, the Higgs couplings to the ghost fields are zero within the EChL.
This is in contrast to the SM case where there are non-vanishing contributions from the diagrams in \figref{diags-ghost}, 
\be
\ampbothSM^{\rm ghost}=\ampSM_{\rm j}+\ampSM_{\rm j'}\,.
\label{SMghost}
\ee

On the other hand, the Goldstone boson loop contributions in the EChL are coming from diagrams in \figref{diags-GB}:
\be
\ampbothchiral=\amp_{\rm h}+\amp_{\rm h'}+\amp_{\rm i}+\amp_{\rm k}+\amp_{\rm l}+\amp_{\rm l'} \quad
\label{GB-decomposition}
\ee
Again, in contrast to the SM case, where the contributing loop diagrams from GB are:
\be
\ampbothSMchiral=\ampSM_{\rm h}+\ampSM_{\rm h'}+\ampSM_{\rm i}\,.
\label{GB-decomposition-SM}
\ee

Starting with the $H\to\gamma\gamma$ decay, from the explicit \1loop computation in the EChL for an arbitrary off- shell Higgs boson momentum  ($q^2 \neq m_H^2$)  we get the following result:
\be
\amponechiral=\frac{e^2 g}{\mw}\frac{a}{16\pi^2} \left(1 -\halfxixq \right) \left(2 -2\xixq f\left(\xixq\right)\right) \left(\frac{q^2}{2} g^{\mu\nu} -k_2^\mu k_1^\nu \right)\epsilon_\mu(k_1)\epsilon_\nu(k_2)
\label{eq-amponeEChLchiral}
\ee
where the function $f$ is
\be
f(r)=-\frac{1}{4}\log ^2\left(-\frac{1-\sqrt{1-r}}{1+\sqrt{1-r}}\right) = \left\lbrace
\begin{array}{ll}
\arcsin^2\left(\frac{1}{\sqrt{r}} \right) & \quad r\geq 1\\
-\frac{1}{4}\left( \ln\left(\frac{1+\sqrt{1-r}}{1-\sqrt{1-r}}\right) -i\pi \right)^2 & \quad 0<r<1
\end{array}
\right.
\label{fpaper}
\ee

Several interesting comments are in order. 
First, the above GB's amplitude is $\xi$-dependent as it is expected in the $R_\xi$ gauges.
Second, it is surprising that only this subset of diagrams respects the Ward structure of \eqref{ward-structure}. This structure is not maintained separately for the other contributions of  \eqref{amp-classification}. However, it is recovered for the loop amplitude $\ampone^{\rm L}$, as it is expected.
Third, this GB contribution is UV-finite and it is linear on the $a$ parameter (that controls the EChL interactions involving one Higgs). 
Fourth, we find also interesting that this GB's amplitude in the EChL is independent of $m_H$. This is due to the fact that the interaction of the Higgs boson to two GBs is given by the product of the two GB momenta, differently to the SM case where this coupling is given in terms of $m_H$ (see the FRs in \tabref{relevant-FR2}). Then, this independence on $m_H$ of the EChL GB's amplitude contrasts with the corresponding amplitude in the SM, where we find the following $m_H$ dependent result:
\be
\amponeSMchiral=\frac{e^2 g}{16\pi^2 \mw}\frac{\mh^2}{q^2}\left(2 -2\xixq f\left(\xixq\right)\right)  \left(\frac{q^2}{2} g^{\mu\nu} -k_2^\mu k_1^\nu \right)\epsilon_\mu(k_1)\epsilon_\nu(k_2) \,.
\label{eq-amponeSMchiral}
\ee
In fact, the two amplitudes above, $\amponechiral$ and $\amponeSMchiral$ are clearly different even for $a=1$, in contrast to the total loop amplitude $\ampone^{\rm L}$ where the SM result coincides with the EChL one for $a=1$.  
Finally, it is also worth commenting on the coincidences. First, both quantities are UV-finite. We also see clearly that the two results for $\amponechiral$ and $\amponeSMchiral$ above coincide for $a=1$ when the Landau gauge is chosen and the Higgs boson momentum is fixed to be on-shell, i.e., for $\xi=0$ and $q^2=m_H^2$. Then, in the Landau gauge, the potential differences from the non-linearity of the GB interactions in the EChL are not emergent in the on-shell result for the GB loop contributions.

Regarding the remaining loop contributions, `gauge' and `mix', and following the splitting of  \eqref{Wprop-spliting}, each \1loop diagram is separated into several parts. 
We adopt the same notation for the various parts of the amplitudes as in \cite{Marciano-paper}: for a given diagram, a contribution with two subscripts means two propagators and a contribution with three subscripts means three propagators.  
Thus, a subscript `0' is assigned for each GB propagator, a subscript `1' for each unitary part $U_{\mu\nu}$ and a subscript `2' for the remaining part $R_{\mu\nu}$ of the $W$ propagator. 
The explicit separation into the above commented contributions, for each `gauge' and `mix' diagrams in the EChL, is the following:
\bear
\amp_{\rm a,a'}&=&\amp_{111}+\amp_{112}+\amp_{121}+\amp_{122}+\amp_{211}+\amp_{212}+\amp_{221}+\amp_{222}\,,  \nn\\
\amp_{\rm b}&=&\amp_{11}+\amp_{12}+\amp_{21}+\amp_{22}\,,\quad \amp_{\rm c,c'}=\amp_{110}+\amp_{120}+\amp_{210}+\amp_{220}\,,  \nn\\
\amp_{\rm d,d'}&=&\amp_{10}+\amp_{20}\,,\quad \amp_{\rm e,e'}=\amp_{100}+\amp_{200}\,,  \nn\\
\amp_{\rm f}&=&\amp_{101}+\amp_{102}+\amp_{201}+\amp_{202}\,,\quad \amp_{\rm g}=\amp_{010}+\amp_{020}\,.
\label{gauge-and-mix-decomposition}
\eear
And a similar decomposition can be done for the corresponding parts in the SM loop amplitudes  $\amponeSM^{\rm L}$.  

Following the $\xi$-cancellation demonstration in \cite{Marciano-paper} for the SM case, we arrive to a set of relations among the different parts of the diagrams in the EChL, but in this case for an off-shell Higgs boson. We find the following result for the EChL in the $R_\xi$ gauges for arbitrary $q^2$:
\bear
&&\hspace{-12mm}\ampEChL_{122}=\ampEChL_{221}=\ampEChL_{222}=\ampEChL_{220}=0 \,,
\label{EChL-xicancelation_1stline} \\
&&\hspace{-12mm}\ampEChL_{20}+\ampEChL_{200}+\ampEChL_{202}+\ampEChL_{020}=-\amponechiral-\left(\ampEChL_{22}+\ampEChL_{212}+\ampEChL_{210}+\ampEChL_{010}\right) \,,
\label{EChL-xicancelation_2ndline} \\
&&\hspace{-12mm}\ampEChL_{12}+\ampEChL_{21}+\ampEChL_{112}+\ampEChL_{211}+\ampEChL_{110}+\ampEChL_{10}+\ampEChL_{120}+\ampEChL_{100}= \nn\\
&&\hspace{-12mm}\quad=-\left(\ampEChL_{121}+\ampEChL_{101}+\ampEChL_{102}+\ampEChL_{201}\right) \,.
\label{EChL-xicancelation_3rdline}
\eear
These relations above can be compared with the corresponding relations  in the SM case, which were provided in \cite{Marciano-paper} for the particular case of on-shell Higgs boson momentum, $q^2=m_H^2$:
\bear
&&\hspace{-12mm}\ampSM_{122}=\ampSM_{221}=\ampSM_{222}=\ampSM_{220}=0 \,,
\label{Marciano-xicancellation_1stline}  \\
&&\hspace{-12mm}\ampSM_{20}+\ampSM_{200}+\ampSM_{202}+\ampSM_{020}+\amponeSM^{\rm ghost}=-\amponeSMchiral-\left(\ampSM_{22}+\ampSM_{212}+\ampSM_{210}+\ampSM_{010}\right) \,,
\label{Marciano-xicancellation_2ndline}   \\
&&\hspace{-12mm}\ampSM_{12}+\ampSM_{21}+\ampSM_{112}+\ampSM_{211}+\ampSM_{110}+\ampSM_{10}=-\left(\ampSM_{121}+\ampSM_{101}\right) \,,
\label{Marciano-xicancellation_3rdline}  \\
&&\hspace{-12mm}\ampSM_{120}+\ampSM_{100}=-\left(\ampSM_{102}+\ampSM_{201}\right) \,,
\label{Marciano-xicancellation_4thline}
\eear
Notice that adding \eqref{Marciano-xicancellation_3rdline} and \eqref{Marciano-xicancellation_4thline} we get a similar relation to \eqref{EChL-xicancelation_3rdline}, but for $q^2=m_H^2$ in the SM case.

Therefore, in summary, putting all together from \eqrefs{EChL-xicancelation_1stline}{EChL-xicancelation_3rdline}, we have for the $\xi$-dependent parts of all diagrams a vanishing contribution in the EChL for an arbitrary $q^2$:
\be
\sum_{n\neq111,11}\amp_n=0 \,.
\label{ampone-xidependence}
\ee
And, similarly for the SM, but fixing $q^2=m_H^2$ in this case:
\be
\sum_{n\neq111,11}\ampSM_n\vert_{q^2=m_H^2}=0 \,.
\label{amponeSM-xidependence}
\ee

Now, for the \1loop contribution of the `111' part of diagrams (a), (a') and the `11' part of (b) for an arbitrary off-shell Higgs momentum we get the following result for the EChL in the $R_\xi$ gauges:
\bear
\ampEChL_{111}+\ampEChL_{11}&=& \frac{e^2 g}{\mw} \frac{a}{16\pi^2}\left( 2 +3\xq +3\left(2-\xq\right)\xq f\left(\xq\right) \right) \nn\\
&&\left(\frac{q^2}{2} g^{\mu\nu} -k_2^\mu k_1^\nu \right)\epsilon_\mu(k_1)\epsilon_\nu(k_2) \,.
\label{eq-amponeEChLU}
\eear
Notice that it is $\xi$-independent (by construction), UV-finite and has the Ward structure.
And this is precisely the same result than adding the three unique diagrams contributing in the unitary gauge, (a), (a') and (b). 
Therefore, we get the interesting result that the loop amplitude in the $R_\xi$ gauges and the unitary gauge coincide in the EChL for an arbitrary off-shell $q^2$ Higgs boson momentum, and not only for the particular on-shell case:
\be
\amponeLoop=\ampEChL_{111}+\ampEChL_{11}=\amponeLoop\vert^{\rm U}\,.
\label{EChL-amponeallloop}
\ee
For the SM case, we get the following expression for an arbitrary $q^2$ momentum: 
\bear
\ampSM_{111}+\ampSM_{11}&=& \frac{e^2 g}{\mw} \frac{1}{16\pi^2}\left( 2 +3 \frac{4\mw^2}{q^2} +3\left(2-\frac{4\mw^2}{q^2}\right)\frac{4\mw^2}{q^2} f\left(\frac{4\mw^2}{q^2}\right) \right) \nn\\
&&\left(\frac{q^2}{2} g^{\mu\nu} -k_2^\mu k_1^\nu \right)\epsilon_\mu(k_1)\epsilon_\nu(k_2) \,.
\label{eq-amponeSMLU}
\eear
This expression is $\xi$-independent (by construction), UV-finite and has the Ward structure. Notice that when fixing $q^2=\mh^2$ in the previous \eqref{eq-amponeSMLU}, we get the same result as in \cite{Marciano-paper}.
Hence the equality of the $R_\xi$ gauges and unitary gauge results is obtained exclusively for  $q^2=\mh^2$, as expected: 
\be
\amponeSM^{\rm L}\vert_{q^2=m_H^2}=\ampSM_{111}\vert_{q^2=m_H^2}+\ampSM_{11}\vert_{q^2=m_H^2}=\amponeSM^{\rm L}\vert^{\rm U}_{q^2=m_H^2}\,.
\label{SM-amponeallloop}
\ee
Therefore, we conclude that the SM loop amplitude and the EChL loop amplitude coincide for $a=1$ if (and only if) the Higgs boson is on-shell, i.e., we find for $q^2=m_H^2$: 
\be
\amponeSM^{\rm L}\vert_{q^2=\mh^2}=\ampone^{\rm L}\vert_{q^2=\mh^2;\,a=1}\,.
\ee

A different situation occurs in the SM when the Higgs boson is not on-shell.
For the Higgs boson off-shell, we arrive to the same \eqrefs{Marciano-xicancellation_1stline}{Marciano-xicancellation_2ndline} but the \eqrefs{Marciano-xicancellation_3rdline}{Marciano-xicancellation_4thline} 
are no longer valid. Hence, the SM result in the $R_\xi$ gauges for an arbitrary Higgs boson momentum $q^2$ does not respect the Ward structure and the equivalence with the unitary gauge is not fulfiled. 
This is in remarkable contrast with our result from the EChL where even for off-shell $q^2$ the Ward structure is respected and the equivalence  between  the $R_\xi$ and unitary gauge predictions is hold. In particular, we find the following $\xi$-dependent result for the sum of all the other terms of the SM \1loop amplitude for off-shell $q^2$ momentum: 
\bear
\hspace{-8mm}\sum_{n\neq111,11}\ampSM_n&=&\frac{e^2 g}{16\pi^2 \mw}\left(-\frac{q^2-\mh^2}{2q^2}\right) \epsilon_\mu(k_1)\epsilon_\nu(k_2) \nn\\
&& \left(g^{\mu\nu}\left( 2q^2+(1-\xi+\ln(1/\xi))\mw^2 \right.\right. \nn\\
&&\left.\left.\hspace{9mm}-2(1-\xi)\mw^2g(4\xi\mw^2/q^2)+3\xi\mw^2q^2f(4\xi\mw^2/q^2) \right.\right. \nn\\
&&\left.\left.\hspace{9mm} +(-q^2+(1-\xi)^2\mw^2)\mw^2C_0(0,q,\sqrt{\xi}\mw,\mw,\sqrt{\xi}\mw) \right)\right. \nn\\
&& \left.-\frac{k_2^\mu k_1^\nu}{k_1\cdot k_2} \left( 2q^2+2(1-\xi+\ln(1/\xi))\mw^2 \right.\right. \nn\\
&&\left.\left.\hspace{16mm}-4(1-\xi)\mw^2g(4\xi\mw^2/q^2)+3\xi\mw^2q^2f(4\xi\mw^2/q^2) \right.\right. \nn\\
&&\left.\left.\hspace{16mm} +(q^2+2(1-\xi)^2\mw^2)\mw^2C_0(0,q,\sqrt{\xi}\mw,\mw,\sqrt{\xi}\mw) \right)\right) 
\, ,
\label{eq-amponeSMRximinusU}
\eear
where the function $f$ was defined in \eqref{fpaper} and the function $g$ is defined as:
\be
g(r) = -\frac{1}{2}\sqrt{1-r}\log\left(-\frac{1-\sqrt{1-r}}{1+\sqrt{1-r}}\right) = \left\lbrace
\begin{array}{ll}
\sqrt{r-1}\arcsin\left(\frac{1}{\sqrt{r}} \right) & \quad r\geq 1\\
\frac{1}{2}\sqrt{1-r}\left( \ln\left(\frac{1+\sqrt{1-r}}{1-\sqrt{1-r}}\right) -i\pi \right) & \quad 0<r<1 \, .
\end{array}
\right.
\label{gpaper}
\ee
Here, $C_0$ is the scalar three-point \1loop integral function in the Passarino-Veltman notation (see \appref{App-LectureDiags} for details). Notice that the result in \eqref{eq-amponeSMRximinusU} is UV-finite.

In summary, the SM loop amplitude in the $R_\xi$ gauges for an arbitrary Higgs boson momentum is UV-finite and is given by:
\be
\amponeSM^{\rm L}=\ampSM_{111}+\ampSM_{11}+\sum_{n\neq111,11}\ampSM_n \,,
\label{SMtot}
\ee
where $\ampSM_{111}+\ampSM_{11}$ is displayed in \eqref{eq-amponeSMLU}, and the remaining contributions which are $\xi$-dependent are collected in \eqref{eq-amponeSMRximinusU}.
In particular, this explicitly shows that the $\xi$-independence and the Ward structure of the total SM loop amplitude in the $R_\xi$ gauges, is only obtained  for the case of on-shell Higgs boson. We believe that the above results in \eqref{eq-amponeSMLU}, and \eqref{eq-amponeSMRximinusU} and, therefore, the total SM result for the off-shell case in \eqref{SMtot}, are novel results that we have not found in the previous literature.

Finally, putting all contributions together, tree-level of  \eqref{treelevel-decayone}, loop of  \eqref{eq-amponeEChLU} and counter-term of  \eqref{counterterm-decayone}, we find the total \1loop amplitude corresponding to the $H\to\gamma\gamma$ decay in the EChL.  In fact,  we demonstrate that $\delta\cone=0$ since $\ampone^{\rm L}$ is UV-finite and no renormalization for the $\cone$ parameter is needed. We then conclude with our final EChL result in the $R_\xi$ gauges for an arbitrary $q^2$ off-shell Higgs boson momentum:
\bear
\ampone&=& \frac{e^2 g}{\mw} \left(\cone + \frac{a}{16\pi^2}\left( 2 +3\xq +3\left(2-\xq\right)\xq f\left(\xq\right) \right) \right) \nn\\
&&\left(\frac{q^2}{2} g^{\mu\nu} -k_2^\mu k_1^\nu \right)\epsilon_\mu(k_1)\epsilon_\nu(k_2) \,.
\label{eq-amponeEChL}
\eear
And we enphasize, that for the particular case of Higgs boson on-shell, as can be deduced from the previous equation by setting $q^2=m_H^2$, our result of the $R_\xi$ gauges is in agreement with the unitary gauge result in \cite{losBuchalla} and the corresponding SM total \1loop amplitude is recovered for $a=1$ and $\cone=0$, as expected.
\newline

Now, we move to the $H\to\gamma Z$ decay. The systematic of this computation is the same as before but now some interesting cancellations of the divergences and of the $\xi$-dependent terms among the loops and counter-term contributions take place in the $R_\xi$ gauges. 
First, for the Goldstone boson loops contribution in the EChL we find the following result for arbitrary off-shell $q^2$ momentum:
\bear
\amptwochiral&=&\frac{e g^2\cosw(1-\tanw^2)}{\mw}\frac{a}{16\pi^2} \frac{q^2-2\xi\mw^2}{q^2-\mz^2}\left(1 -\frac{4\xi\mw^2}{q^2-\mz^2} \left(f\left(\xixq\right)-f\left(\xixz\right)\right) \right.\nn\\
&&\left. -\frac{2\mz^2}{q^2-\mz^2}\left(g\left(\xixq\right)-g\left(\xixz\right)\right)\right) \left(\frac{q^2-\mz^2}{2} g^{\mu\nu} -k_2^\mu k_1^\nu \right)\epsilon_\mu(k_1)\epsilon_\nu(k_2) \,,  \nn\\
\label{eq-amptwoEChLchiral}
\eear
where the functions $f$ and $g$ were defined in \eqref{fpaper} and \eqref{gpaper}, respectively.
As for the $H\to\gamma\gamma$ case, this contribution is $\xi$-dependent, UV-finite, has the Ward structure and it does not depend on the Higgs boson mass. 
This $\mh$-independence is in contrast to the corresponding SM contribution, that we find as follows:
\bear
\amptwoSMchiral&=&\frac{e g^2\cosw(1-\tanw^2)}{16\pi^2 \mw} \frac{\mh^2}{q^2-\mz^2}\left(1 -\frac{4\xi\mw^2}{q^2-\mz^2} \left(f\left(\xixq\right)-f\left(\xixz\right)\right) \right.\nn\\
&&\left. -\frac{2\mz^2}{q^2-\mz^2}\left(g\left(\xixq\right)-g\left(\xixz\right)\right)\right) \left(\frac{q^2-\mz^2}{2} g^{\mu\nu} -k_2^\mu k_1^\nu \right)\epsilon_\mu(k_1)\epsilon_\nu(k_2) \,,  \nn\\
\label{eq-amptwoSMchiral}
\eear
and this difference comes from the different interaction $H\pi\pi$ in the EChL and in the SM, as it was discussed before. Again we find the same pattern of differences and coincidences between the two previous GB contributions as we found for the $H\to\gamma\gamma$ decay. In summary: $\amptwochiral $ and $\amptwoSMchiral$ are both UV-finite but do not coincide for $a=1$, nor for on-shell $q^2=\mh^2$. They, however, do coincide if the Landau gauge is chosen, together with the fixing of $a=1$ and $q^2=\mh^2$.

Concerning the `gauge' and `mix' contributions we follow the same notation for the separation of the various parts as in \eqref{gauge-and-mix-decomposition}, which also holds for the $\amptwoLoop$ case. 
However, now the \eqrefs{EChL-xicancelation_1stline}{EChL-xicancelation_3rdline} are no longer valid and we found for their sum the following result:
\be
\sum_{n\neq111,11}\amp_n = \frac{a}{16\pi^2}\frac{eg^2\mw}{4\cosw}\left(2(3+\xi)\div +5+\xi+\frac{2\xi(2+\xi)\log(\xi)}{1-\xi}\right) \epsilon(k_1)\cdot\epsilon(k_2) \,,
\label{amptwo-xidependence}
\ee
which does not respect the Ward structure and it is UV-divergent and $\xi$-dependent. Interestingly, as it can be seen from \eqref{counterterm-decaytwo} and \eqref{sigmaaz}, it precisely cancels the full counter-term contribution coming from the second term of the \eqref{counterterm-decaytwo}, and both the divergence and the $\xi$-dependence disappear in this sum. 

Regarding the remaining terms,  $\ampEChL_{111}$ and $\ampEChL_{11}$, our explicit computation leads to the following result:
\bear
&&\ampEChL_{111}+\ampEChL_{11}= \nn\\
&=& \frac{e g^2\cosw}{\mw} \frac{a}{16\pi^2} \frac{1}{\mw^2(q^2-\mz^2)} \left( 2q^2\mw^2+12 \mw^4-q^2\mz^2-2\mw^2\mz^2 \right.\nn\\
&&\left. -\frac{4\mw^2}{q^2-\mz^2}(-6q^2\mw^2+12\mw^4+q^2\mz^2+6 \mw^2 \mz^2-2 \mz^4) \left(f\left(\xq\right)-f\left(\xz\right)\right) \right.\nn\\
&&\left. -\frac{2\mz^2}{q^2-\mz^2}(2q^2\mw^2+12 \mw^4-q^2\mz^2-2\mw^2\mz^2)\left(g\left(\xq\right)-g\left(\xz\right)\right)\right) \nn\\
&& \left(\frac{q^2-\mz^2}{2} g^{\mu\nu} -k_2^\mu k_1^\nu \right)\epsilon_\mu(k_1)\epsilon_\nu(k_2) \,,
\label{eq-amptwoEChLRxiLoop}
\eear
which, as in the previous case of $H\to \gamma \gamma$, is $\xi$-independent, UV-finite and it has the Ward structure. Again, this result coincides with the result from the three contributing diagrams in the unitary gauge, (a), (a') and (b) leading to $\amptwoLoop\vert^{\rm U}$. 

Therefore, we conclude that in the $R_\xi$ gauges, the Ward structure is recovered for the combination of the loop and counter-term amplitudes resulting in a UV-finite contribution and, as a consequence,  there is not need of the renormalization of  $\ctwo$. Thus, we find $\delta\ctwo=0$,  and: 
\be
\amptwoLoop+\amptwo^{\rm C}=\ampEChL_{111}+\ampEChL_{11}=\amptwoLoop\vert^{\rm U} \,.
\label{EChL-amptwoallloop}
\ee

Finally, by adding the tree-level contribution of \eqref{treelevel-decaytwo} to the previously reported parts we get the final result for the total \1loop amplitude, in the $R_\xi$ gauges, of the $H\to\gamma Z$ decay in the EChL and for the general case the Higgs boson momentum $q^2$  off-shell:
\bear
\amptwo&=& \frac{e g^2\cosw}{\mw}\left(\ctwo + \frac{a}{16\pi^2} \frac{1}{\mw^2(q^2-\mz^2)} \left( 2q^2\mw^2+12 \mw^4-q^2\mz^2-2\mw^2\mz^2 \right.\right.\nn\\
&&\left.\left. -\frac{4\mw^2}{q^2-\mz^2}(-6q^2\mw^2+12\mw^4+q^2\mz^2+6 \mw^2 \mz^2-2 \mz^4) \left(f\left(\xq\right)-f\left(\xz\right)\right) \right.\right.\nn\\
&&\left.\left. -\frac{2\mz^2}{q^2-\mz^2}(2q^2\mw^2+12 \mw^4-q^2\mz^2-2\mw^2\mz^2)\left(g\left(\xq\right)-g\left(\xz\right)\right)\right)\right) \nn\\
&& \left(\frac{q^2-\mz^2}{2} g^{\mu\nu} -k_2^\mu k_1^\nu \right)\epsilon_\mu(k_1)\epsilon_\nu(k_2) \,.
\label{eq-amptwoEChLRxi}
\eear
Now, by setting the Higgs boson momentum on-shell in the previous formula, i.e., by fixing $q^2=m_H^2$, we see that our result is in agreement with unitary gauge result in \cite{losBuchalla}, and again the corresponding SM total \1loop amplitude is recovered for $a=1$ and $\ctwo=0$, as expected. 

On the other hand, we have also found that
the corresponding SM result for an arbitrary Higgs boson momentum $q^2$ in the $R_\xi$ gauges does not respect the Ward structure and, again, the equivalence with the unitary gauge result is lost, as it also happened for the $H\to\gamma\gamma$ case. We believe this is a novel result, again. Besides, we derived an analog expression to \eqref{eq-amponeSMRximinusU} for the present case of $H \to \gamma Z$, but it is too long and not much illuminating for the main purpose of this work. Thus, we omit to present this long expression here. 
Nevertheless, it is worth mentioning our explicit check that this long expression is UV-finite, and this fact follows from the cancellation among the divergent parts of the different loop categories together with the counter-term contribution that can be derived from \eqref{sigmaaz}. Specifically:
\bear
\amptwoSM^{\rm{gauge}} &=& \frac{e g^2}{16\pi^2}\frac{3\mw}{2}\cosw(1+\xi)\Delta_\epsilon\,\epsilon(k_1)\cdot\epsilon(k_2) +\text{finite}\,,  \nn\\
\amptwoSM^{\rm{mix}} &=& \frac{e g^2}{16\pi^2}\frac{\mw}{2\cosw}(3+\xi-\cosw^2(3+2\xi))\Delta_\epsilon\,\epsilon(k_1)\cdot\epsilon(k_2) +\text{finite}\,,  \nn\\
\amptwoSM^{\rm{ghost}} &=& -\frac{e g^2}{16\pi^2}\frac{\mw}{2}\cosw\xi\Delta_\epsilon\,\epsilon(k_1)\cdot\epsilon(k_2)  +\text{finite}\,,  \nn\\
\amptwoSM^{\rm{C}} &=& -\frac{e g^2}{16\pi^2}\frac{\mw}{2\cosw}(3+\xi)\Delta_\epsilon\,\epsilon(k_1)\cdot\epsilon(k_2) +\text{finite}\,,
\label{div-SMtwo}
\eear
Remember that the GB contribution of the \eqref{eq-amptwoSMchiral} is UV-finite.

\section{Vertex functions $\Veffone$ and $\Vefftwo$: off-shell versus on-shell Higgs boson}
\label{sec-Veff}

In this section we study analytically the vertex functions $\Veffone$ and $\Vefftwo$ that describe the  \1loop level interactions of the Higgs boson with the gauge boson pairs $\gamma \gamma$ and $\gamma Z$ respectively. 
We also find illustrative to show here some approximate analytical results that help in understanding the behaviour with energy of the Higgs decays $H \to \gamma V$ in the case where the initial Higgs boson is off-shell but the final gauge bosons are on-shell. Thus, we explore here the behaviour of the vertex function $\Veffboth(q,k_1,k_2)$ in \eqref{ward-structure} as a function of  $q^2\neq m_H^2$, with $q=k_1+k_2$ and 1) $k_1^2=0$, $k_2^2=0$ for $H \to \gamma \gamma$, or 2)  $k_1^2=0$, $k_2^2=m_Z^2$ for $H \to \gamma Z$. We are using an EFT based on a momentum expansion, so it is clearly motivated to explore how the result from the EChL behaves with $q^2$, which provides the available energy for the decay, and how it compares with the SM result.

Let us start with the $H \to \gamma \gamma $ case.
The full result for $\Veffone(q,k_1,k_2)$ can be extracted from \eqref{eq-amponeEChL} and it is summarized by:
\be
\Veffone= \frac{e^2 g}{2\mw} \left( \frac{a}{16\pi^2}\left(2q^2+12\mw^2\left(1+\left(2-\xq\right)f\left(\xq\right)\right)\right)+\cone q^2 \right).
\label{eq-VeffoneEChL}
\ee
Now, this is a complex function and involves two different scales, $q^2$ and $m_W^2$.  Thus, to find an approximate formula we have to take into account the different branches of the function $f$, corresponding to $4\mw^2<q^2$ and $4\mw^2\ge q^2$ respectively (see \eqref{fpaper}). Next we use the approximate expressions of $f(r)$ for the two regimes of large $r$ ($r\gg1$) and small $r$ ($r\ll1$) given by:
\be
f(r)\sim\left\lbrace
\begin{array}{ll}
\frac{1}{r} +\frac{1}{3r^2} +\mO(r^{-3}) &\quad\text{large}\,\,r\\
-\frac{1}{4}\left( \ln^2\left(\frac{4}{r}\right)-\pi^2 -2i\pi\ln\left(\frac{4}{r}\right) \right) +\mO(r) &\quad\text{small}\,\,r
\end{array}
\right.
\label{fpaper-asymptotic}
\ee
and by using them in \eqref{eq-VeffoneEChL} we get the following approximate results for $\Veffone$ in the two regimes of small $q^2$ ($q^2\ll 4\mw^2$) and large $q^2$ ($q^2\gg 4\mw^2$): 
\be
\Veffone \sim \left\lbrace
\begin{array}{ll}
\frac{e^2 g}{2\mw}\left(\frac{7a}{16\pi^2}+\cone\right)q^2 & \quad\text{small}\,q^2\\
\frac{e^2 g}{2\mw}\left(\frac{a}{8\pi^2}\left(q^2-3\mw^2\ln^2(q^2/\mw^2)+3(\pi^2+2)\mw^2\right)+\cone q^2\right.&\\
\left.\hspace{10mm}+i\pi\frac{a}{8\pi^2}6\mw^2\ln(q^2/\mw^2)\right) & \quad\text{large}\,q^2
\end{array}
\right.
\label{eq-VeffoneEChLapprox}
\ee
From these results we see that, for small $q^2$,  $\Veffone$ is real and vanishes at $q^2=0$. At large $q^2$, it is in contrast complex. The real part dominates, growing as $q^2$, whereas the imaginary part grows just logarithmically.

One interesting exercise, is to compare the previous result $\Veffone$ with the contribution from just GB loops, i.e., from chiral loops in the usual terminology of Chiral Lagrangians. The complete result of this contribution is extracted from \eqref{eq-amponeEChLchiral}: 
\be
\Veffonechiral = \frac{e^2 g}{2\mw}\frac{a}{16\pi^2} \left(1 -\halfxixq\right) \left(2q^2 -8\xi\mw^2 f\left(\xixq\right)\right) \,.
\label{eq-VeffoneEChLchiral}
\ee
It is a complex function that depends again on two energy scales $q^2$ and $m_W^2$ but now it is $\xi$-dependent.
In particular, working in the Landau gauge, i.e., for $\xi=0$, we found that the vertex function is real for any arbitrary $q^2$ momentum and equal to: 
\be
\Veffonechiral|^{\rm{Landau}}=\frac{e^2 g}{2\mw}\frac{a}{8\pi^2}q^2 \,.
\label{eq-VeffoneEChLchiralLandau}
\ee

However, for $\xi\neq0$, the vertex function of  \eqref{eq-VeffoneEChLchiral} has a zero at $1=2\xi\mw^2/q^2$ and at $1=(4\xi\mw^2/q^2)f(4\xi\mw^2/q^2)$, and a branch point at $1=4\xi\mw^2/q^2$.
We can use again the simple expressions of $f$ in the two regimes of small $q^2$ ($q^2\ll 4\xi \mw^2$) and large $q^2$ ($q^2\gg 4\xi \mw^2$) and we get the following approximate result:
\bear
\Veffonechiral &\sim& \left\lbrace
\begin{array}{ll}
\frac{e^2 g}{2\mw}\frac{a}{8\pi^2}\frac{q^2}{6} & \quad\text{small}\,q^2\,\\
\frac{e^2 g}{2\mw}\frac{a}{8\pi^2}\left(q^2 +\xi\mw^2\ln^2(q^2/\xi\mw^2)-(\pi^2+2)\xi\mw^2\right.&\\
\left.\hspace{15mm}-i\pi2\xi\mw^2\ln(q^2/\xi\mw^2)\right)& \quad\text{large}\,q^2\,
\end{array}
\right.
\label{eq-VeffoneEChLchiralapprox}
\eear
We see that at small $q^2$ it is real, whereas at large $q^2$ it is complex, as for the total contribution. The leading term in this regime is also real, grows as $q^2$ and it is $\xi$-independent. This $\xi$-independent term is precisely the result in the Landau gauge of  \eqref{eq-VeffoneEChLchiralLandau}.
We find this an interesting result, since it demonstrates that the GB loops in the Landau gauge provide the polynomic $q^2$ contribution of the total loop result of  \eqref{eq-VeffoneEChLapprox} which is gauge-invariant and therefore has a physical meaning.
This fact is clearly related with the line of thinking in Chiral Lagrangians where the chiral loops provide the most relevant loops at large $q^2$.

On the other hand, it is also interesting to compare the GB contribution in the EChL with the SM case. From \eqref{eq-amponeSMchiral}, we deduce the corresponding expression in the Landau gauge for an arbitrary Higgs boson momentum $q^2$: 
\be
\VeffoneSMchiral|^{\rm{Landau}}=\frac{e^2 g}{2\mw}\frac{1}{8\pi^2}\mh^2 \,,
\label{eq-VeffoneSMchiralLandau}
\ee
which is constant in the whole $q^2$ range and it coincides with the EChL case just for the Higgs boson on-shell and $a=1$. Now, for $\xi\neq0$, the approximate result in the two regimes is 
\bear
\VeffoneSMchiral&\sim& \left\lbrace
\begin{array}{ll}
-\frac{e^2 g}{2\mw}\frac{1}{8\pi^2}\frac{\mh^2}{12\xi\mw^2}q^2 & \quad\text{small}\,q^2\\
\frac{e^2 g}{2\mw}\frac{1}{8\pi^2}\mh^2 & \quad\text{large}\,q^2
\end{array}
\right.
\label{eq-VeffoneSMchiralapprox}
\eear
Thus, the Landau contribution provides the leading loop contribution for large $q^2$, as in the EChL case.
Regarding the SM total \1loop amplitude for the Higgs boson off-shell, as we discussed along the \eqrefs{EChL-xicancelation_1stline}{SMtot}, it does not respect the Ward structure of the \eqref{ward-structure} and we can not define the corresponding vertex function in the $R_\xi$ gauges $\VeffoneSM(q,k_1,k_2)$ for an arbitrary $q^2$.
In that case, two vertex functions arise, corresponding to the coefficients of $g^{\mu\nu}$ and $k_2^\mu k_1^\nu$ in the amplitude of \eqref{eq-amponeSMRximinusU}, but the behaviour of these two functions with the Higgs boson momentum is beyond the scope of this work.
However, for the case of on-shell Higgs boson, we get the equivalence of the SM vertex in the 
$R_\xi$ gauges and  in the unitary gauge, and in addition we also get the equivalence with the corresponding EChL vertex for $a=1$ and $\cone=0$, as expected: 
\bear
\VeffoneSM\vert_{q^2=\mh^2}&=&\Veffone\vert_{q^2=\mh^2;\,a=1;\,\cone=0} \, \nn\\
 &=&\frac{e^2 g}{2\mw}  \frac{1}{16\pi^2}\left(2\mh^2+12\mw^2\left(1+\left(2-\xh\right)f\left(\xh\right)\right)\right)\,.
\label{eq-VeffoneSM}
\eear

\vspace{3mm}
We close this section presenting the $H \to \gamma Z$ decay, in which we proceed similarly as before.
A useful check of the following vertex functions is that we recover (leaving apart coupling factors) the corresponding expressions of the $H\to\gamma\gamma$ case in the limit $\mz^2\to0$, since this mass comes just from the kinematics of the interaction.
We start with the full result of the vertex function coming from \eqref{eq-amptwoEChLRxi}:
\bear
\Vefftwo&=&\frac{e g^2\cosw}{2\mw} \left(\frac{a}{16\pi^2} \left( 2q^2+12 \mw^2-\frac{\mz^2}{\mw^2}q^2-2\mz^2 \right.\right.\nn\\
&&\left.\left. \hspace{5mm}-\frac{4\mw^2}{q^2-\mz^2}\left(-6q^2+12\mw^2+\frac{\mz^2}{\mw^2}q^2+6\mz^2-\frac{2\mz^4}{\mw^2}\right) \left(f\left(\xq\right)-f\left(\xz\right)\right) \right.\right.\nn\\
&&\left.\left. \hspace{5mm}-\frac{2\mz^2}{q^2-\mz^2}\left(2q^2+12 \mw^2-\frac{\mz^2}{\mw^2}q^2-2\mz^2\right)\left(g\left(\xq\right)-g\left(\xz\right)\right)\right)\right. \nn\\
&&\left. \hspace{13mm}+\ctwo(q^2-\mz^2) \right) \,.
\label{eq-VefftwoEChL}
\eear
Taking into account the simple expression of $f$ in \eqref{fpaper-asymptotic} and the corresponding one for the function $g$:
\be
g(r) \sim \left\lbrace
\begin{array}{ll}
1-\frac{1}{3r}-\frac{2}{15r^2}+\mO(r^{-3}) & \quad\text{large}\,\,r\\
\frac{1}{2}\left(\ln\left(\frac{4}{r}\right) -i\pi\right) +\mO(r) & \quad\text{small}\,\,r
\end{array}
\right.
\label{gpaper-asymptotic}
\ee
we arrive to the approximate expression for the vertex function of $H\to\gamma Z$ decay in the two Higgs boson momentum regimes, including the effect of the $Z$ boson mass, i.e., small $q^2$ ($q^2\ll 4\mw^2$) and large $q^2$ ($q^2\gg 4\mw^2$):

\bear
\Vefftwo\vert_{\text{small}\,q^2} &\sim& \frac{e g^2\cosw}{2\mw} \left(\frac{a}{8\pi^2}\left(18\mw^2-3\mz^2-2(6\mw^2-\mz^2)g\left(\frac{4\mw^2}{\mz^2}\right)  \right.\right.\nn\\
&&\left.\left.\hspace{24mm}-2\left(\frac{12\mw^4}{\mz^2}+6\mw^2-2\mz^2\right)f\left(\frac{4\mw^2}{\mz^2}\right)\right)-\ctwo\mz^2  \right.\nn\\
&&\left.\hspace{8mm}+q^2\left(\frac{a}{24\pi^2}\left(\frac{54\mw^2}{\mz^2}+9-\frac{7\mz^2}{\mw^2}-\frac{36\mw^2}{\mz^2}g\left(\frac{4\mw^2}{\mz^2}\right)+\frac{3\mz^2}{\mw^2}g\left(\frac{4\mw^2}{\mz^2}\right)  \right.\right.\right.\nn\\
&&\left.\left.\left.\hspace{18mm}-\frac{72\mw^4}{\mz^4}f\left(\frac{4\mw^2}{\mz^2}\right)+6f\left(\frac{4\mw^2}{\mz^2}\right)\right)+\ctwo\right)\right) \,,  \nn\\
\Vefftwo\vert_{\text{large}\,q^2} &\sim& \frac{e g^2\cosw}{2\mw} \left(\frac{a}{8\pi^2}\left(\left(1-\frac{\mz^2}{2\mw^2}\right)\left(q^2-\mz^2\left(\ln\left(\frac{q^2}{\mw^2}\right)-2g\left(\frac{4\mw^2}{\mz^2}\right)\right)\right)  \right.\right.\nn\\
&&\left.\left.\hspace{2mm}+\left(3\mw^2-\mz^2/2\right)\left(-\ln^2\left(\frac{q^2}{\mw^2}\right)+\pi^2+2-4f\left(\frac{4\mw^2}{\mz^2}\right)\right)\right)+\ctwo (q^2-\mz^2)  \right.\nn\\
&&\left.\hspace{12mm}+i\pi\frac{a}{8\pi^2}\left((6\mw^2-\mz^2)\ln\left(\frac{q^2}{\mw^2}\right)+\mz^2\left(1-\frac{\mz^2}{2\mw^2}\right)\right)\right) \,.
\label{eq-VefftwoEChLapprox}
\eear
In view of the previous expression for small $q^2$, the $\Vefftwo$ is real and, for $q^2=0$, there is a remaining contribution which vanishes only at $\mz^2\to0$ (recovering the $H\to\gamma\gamma$ result in \eqref{eq-VeffoneEChLapprox}).
On the other hand, this vertex function is complex at large $q^2$. The real part dominates, growing as $q^2$, whereas the imaginary part grows just logarithmically. Also, \eqref{eq-VeffoneEChLapprox} is recovered when $\mz^2\to0$ in this regime.

The complete Goldstone boson contribution, coming from \eqref{eq-amptwoEChLchiral}, is:

\bear
\Vefftwochiral&=&\frac{e g^2\cosw(1-\tanw^2)}{2\mw}\frac{a}{16\pi^2} \frac{q^2-2\xi\mw^2}{q^2-\mz^2}\left(q^2-\mz^2 -4\xi\mw^2 \left(f\left(\xixq\right)-f\left(\xixz\right)\right) \right.\nn\\
&&\left. -2\mz^2\left(g\left(\xixq\right)-g\left(\xixz\right)\right)\right) \,, 
\label{eq-VefftwoEChLchiral}
\eear
which is a complex function that depends on the $\xi$-parameter. The Landau gauge result ($\xi=0$) for arbitrary Higgs boson momentum, $q^2$, then corresponds to:
\be
\Vefftwochiral\vert^{\rm Landau}=\frac{e g^2\cosw(1-\tanw^2)}{2\mw}\frac{a}{16\pi^2} q^2 \left(1 +\frac{\mz^2}{q^2-\mz^2}\ln\left(\frac{\mz^2}{q^2}\right)\right) \,.
\label{eq-VefftwoEChLchiralLandau}
\ee
If $\xi\neq0$, the \eqref{eq-VefftwoEChLchiral} has a zero at $1=2\xi\mw^2/q^2$, and a branch point at $1=4\xi\mw^2/q^2$.

Thus we find the approximate GB vertex functions in the two regimes, of small $q^2$ ($q^2\ll 4\xi \mw^2$) and large $q^2$ ($q^2\gg 4\xi \mw^2$): 

\bear
\Vefftwochiral\vert_{\text{small}\,q^2} &\sim& \frac{e g^2\cosw(1-\tanw^2)}{2\mw} \frac{a}{8\pi^2}\left(\xi\mw^2\left(-3+2g\left(\frac{4\xi\mw^2}{\mz^2}\right) +\frac{4\xi\mw^2}{\mz^2}f\left(\frac{4\xi\mw^2}{\mz^2}\right)\right)  \right.\nn\\
&&\left.\hspace{6mm}+q^2\left(\frac{1}{6}+\left(-\frac{1}{2}+\frac{\xi\mw^2}{\mz^2}\right)\left(-3+2g\left(\frac{4\xi\mw^2}{\mz^2}\right) +\frac{4\xi\mw^2}{\mz^2}f\left(\frac{4\xi\mw^2}{\mz^2}\right)\right)\right)\right) \,,  \nn\\
\Vefftwochiral\vert_{\text{large}\,q^2} &\sim& \frac{e g^2\cosw(1-\tanw^2)}{2\mw} \frac{a}{16\pi^2}\left(q^2-\mz^2\left(\ln\left(\frac{q^2}{\xi\mw^2}\right)-2g\left(\frac{4\xi\mw^2}{\mz^2}\right)\right)  \right.\nn\\
&&\left.\hspace{36mm}+\xi\mw^2\left(\ln^2\left(\frac{q^2}{\xi\mw^2}\right)-(\pi^2+2)+4f\left(\frac{4\xi\mw^2}{\mz^2}\right)\right)  \right.\nn\\
&&\left.\hspace{36mm}-i\pi\left(2\xi\mw^2\ln\left(\frac{q^2}{\xi\mw^2}\right)-\mz^2\right)\right) \,.
\label{eq-VefftwoEChLchiralapprox}
\eear
As for the $H\gamma\gamma$ vertex function of the GBs, at small $q^2$ the vertex function $\Vefftwochiral$ is real, whereas at large $q^2$ it is complex. The leading loop  contribution in this regime is also real, grows polynomicaly as $q^2$ and it is $\xi$-independent. This $\xi$-independent term is precisely the result of the polynomial term in the Landau gauge of  \eqref{eq-VefftwoEChLchiralLandau}. Therefore, we conclude again that the chiral loops provide in the EChL the most relevant loop contribution at large $q^2$ to the loop amplitude, also in the $H \to \gamma Z$ case. 

Finally, we compare with the GB contribution of the SM. From \eqref{eq-amptwoSMchiral}, the GB vertex function in the Landau gauge is: 
\be
\VefftwoSMchiral\vert^{\rm Landau}=\frac{e g^2\cosw(1-\tanw^2)}{32\pi^2\mw}\mh^2 \left(1 +\frac{\mz^2}{q^2-\mz^2}\ln\left(\frac{\mz^2}{q^2}\right)\right) \,,
\label{eq-VefftwoSMchiralLandau}
\ee
whose dominant contribution at large $q^2$ is constant and depends on the Higgs boson mass. On the other hand, for $\xi\neq0$, the approximate result in the two regimes is: 
\bear
\VefftwoSMchiral\vert_{\text{small}\,q^2} &\sim& -\frac{e g^2\cosw(1-\tanw^2)}{32\pi^2\mw}\mh^2 \left(-3+2g\left(\frac{4\xi\mw^2}{\mz^2}\right) +\frac{4\xi\mw^2}{\mz^2}f\left(\frac{4\xi\mw^2}{\mz^2}\right)  \right.\nn\\
&&\left.\hspace{4mm}+\frac{q^2}{\xi\mw^2}\left(\frac{1}{6}+\left(-\frac{1}{2}+\frac{\xi\mw^2}{\mz^2}\right)\left(-3+2g\left(\frac{4\xi\mw^2}{\mz^2}\right) +\frac{4\xi\mw^2}{\mz^2}f\left(\frac{4\xi\mw^2}{\mz^2}\right)\right)\right)\right) \,,  \nn\\
\VefftwoSMchiral\vert_{\text{large}\,q^2} &\sim& \frac{e g^2\cosw(1-\tanw^2)}{32\pi^2\mw}\mh^2 \,.
\label{eq-VefftwoSMchiralapprox}
\eear
Hence, the Landau contribution provides again the leading contribution for large $q^2$, as in the EChL case, but now it is constant with $q^2$.
Concerning the SM total \1loop amplitude for the Higgs boson off-shell, as we discussed in the $H\to\gamma\gamma$ case, it does not respect the Ward structure and we can not define the corresponding vertex function in the $R_\xi$ gauges $\VefftwoSM(q,k_1,k_2)$ for an arbitrary $q^2$.
However, for the case of on-shell Higgs boson, we get again the equivalence of the SM vertex in the 
$R_\xi$ gauges and  in the unitary gauge, and in addition we also get the equivalence with the corresponding EChL vertex for $a=1$ and $\ctwo=0$, as expected:

\bear
&&\hspace{-2mm}\VefftwoSM\vert_{q^2=\mh^2}=\Vefftwo\vert_{q^2=\mh^2;\,a=1;\,\ctwo=0} \nn \\
&&\quad= \frac{e g^2\cosw}{2\mw} \frac{1}{16\pi^2} \left( 2\mh^2+12\mw^2-\frac{\mz^2}{\mw^2}\mh^2-2\mz^2 \right.\nn\\
&&\left. \quad\hspace{5mm}-\frac{4\mw^2}{\mh^2-\mz^2}\left(-6\mh^2+12\mw^2+\frac{\mz^2}{\mw^2}\mh^2+6\mz^2-\frac{2\mz^4}{\mw^2}\right) \left(f\left(\xh\right)-f\left(\xz\right)\right) \right.\nn\\
&&\left. \quad\hspace{5mm}-\frac{2\mz^2}{\mh^2-\mz^2}\left(2\mh^2+12 \mw^2-\frac{\mz^2}{\mw^2}\mh^2-2\mz^2\right)\left(g\left(\xh\right)-g\left(\xz\right)\right)\right)
\label{eq-VefftwoSM}
\eear

\section{Numerical Results}
\label{sec-plots}

In this section we present the numerical predictions for some previously derived quantities. We start with the EChL partial widths in terms of the relevant parameters $a$ and $c_{H\gamma V}$. Next we analyse the various bosonic $\rm {R}_\xi$  contributions to these partial widths. Later we discuss on the differences found between the off-shell Higgs boson case versus the on-shell case, and focus in particular in the interesting features found for the vertex functions at large off-shell Higgs boson momentum.

In order to give realistic predictions for the decay widths, we must include the fermionic loop contributions. As we said in the introduction, we consider them as in the SM. In particular, they provide a contribution to the amplitude that is UV-finite, $\xi$-independent, has the Ward structure,  and does not depend on the EChL parameters $a$ and $c_{H\gamma V}$. In particular, they do not modify our conclusions on the $\xi$-independence of the bosonic contributions and on the renormalization of the $\mLfour$ operators.

The partial width of the $H\to\gamma\gamma$ decay in the EChL is then constructed from the tree-level `T', bosonic `B' and fermionic `F' loop contributions. Taking into account all the polarization final states and an extra $1/2$ identical particles factor in the phase space, the $H\to\gamma\gamma$ partial width in the EChL can then be written as:
\be
\Gamma^{\rm EChL}_{\gamma\gamma}=\frac{1}{16\pi\mh}\frac{1}{2}\sum_{\text{\{pol\}}}\Big\vert\ampone^{\rm T+B+F}\Big\vert^2 =\frac{1}{16\pi\mh}\Big\vert\Veffone\vert_{q^2=\mh^2}+\VFone\vert_{q^2=\mh^2}\Big\vert^2 \,,
\label{generic-partialwidth-one}
\ee
where the vertex function $\Veffone$, showed in \eqref{eq-VeffoneEChL}, contains the tree-level and bosonic loop contributions.
The fermionic vertex function $\VFone$~\cite{HHG} is well known in the SM case and it is dominated by the top quark loop, which is given by:
\be
\VFone\vert_{q^2=\mh^2}= \frac{e^2 g}{2\mw}\frac{\mh^2}{16\pi^2}\frac{4}{3}A_{\gamma\gamma}^{\rm F}\left(\frac{4m_{\rm top}^2}{\mh^2}\right) \,.
\label{eq-fermone}
\ee
We neglect the QCD corrections of $\mO(\alpha_s)$ due to their small impact in our numerical results and we provide the explicit function $A_{\gamma\gamma}^{\rm F}$ in \eqref{Amedios}.
It is important to stress that the polarization states $\{++,--\}$ give equal non-vanishing contributions and the $\{+-,-+\}$ ones are vanishing (due to the angular momentum conservation in the decay). 

Similarly, in the partial width of the $H\to\gamma Z$ decay only the pairs  with equal  transverse polarizations $\{++,--\}$ contribute, whereas the corresponding contributions from $\{+-,-+\}$ and $\{+0,-0\}$ are vanishing. Thus we have:
\be
\Gamma^{\rm EChL}_{\gamma Z}=\frac{\mh^2-\mz^2}{16\pi\mh^3}\sum_{\text{\{pol\}}}\Big\vert\amptwo^{\rm T+B+F+CT}\Big\vert^2 =\frac{\mh^2-\mz^2}{8\pi\mh^3}\Big\vert\Vefftwo\vert_{q^2=\mh^2}+\VFtwo\vert_{q^2=\mh^2}\Big\vert^2 \,,
\label{generic-partialwidth-two}
\ee
where now the vertex function $\Vefftwo$, showed in \eqref{eq-VefftwoEChL}, contains the tree-level, bosonic loop and counter-term `CT' contributions.
The fermionic vertex function $\VFtwo$ is as in the SM, and again it is dominated by the top quark loop, which is given by~\cite{HHG}:
\be
\VFtwo\vert_{q^2=\mh^2} = \frac{e g^2\cosw}{2\mw}\frac{\mh^2-\mz^2}{16\pi^2}\frac{4\left(1/2-4\sinw^2/3\right)}{\cosw^2}A_{\gamma Z}^{\rm F}\left(\frac{4m_{\rm top}^2}{\mh^2},\frac{4m_{\rm top}^2}{\mz^2}\right) \,.
\label{eq-fermtwo}
\ee
We neglect again the QCD corrections of $\mO(\alpha_s)$ and we provide the explicit function $A_{\gamma Z}^{\rm F}$ in \eqref{Amedios}.

\vspace{2mm}
From the previous analytical results we already see that both partial widths of \eqref{generic-partialwidth-one} and \eqref{generic-partialwidth-two} grow quadratically with each of the two relevant EChL  parameters,  $a$ and $c_{H\gamma V}$, since they are the square of  linear functions in these parameters. 
To be more concrete, it is illustrative to explore this growing numerically and to provide predictions on the size of the corresponding widths with these two EChL parameters, together with a comparison respect to the SM predictions. For this comparison, it is convenient to define the $a$ value respect to the SM via the difference $\Delta a=a-1$.

\begin{figure}[h!]
\begin{center}
\includegraphics[width=0.49\textwidth]{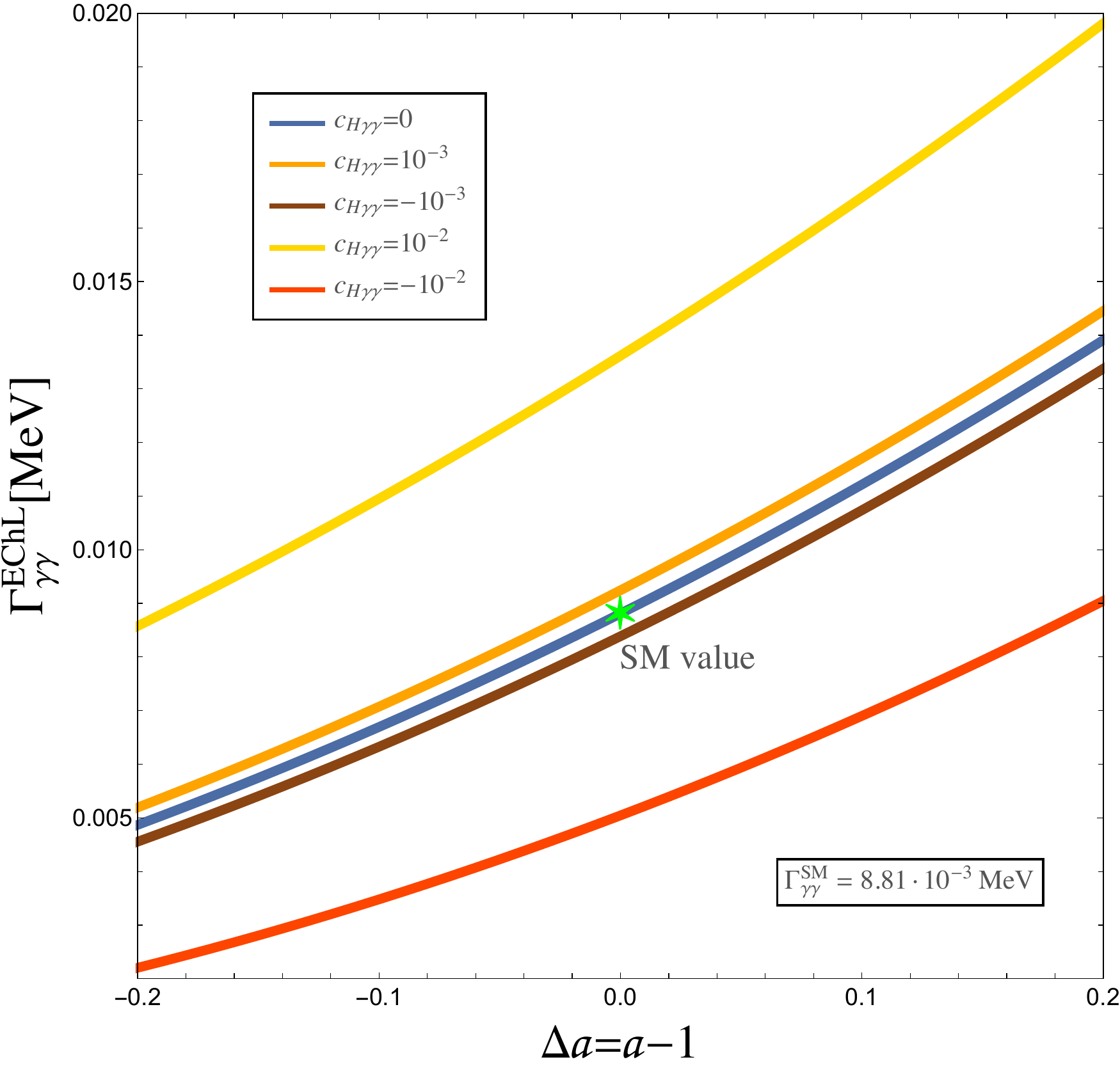}
\hspace{1mm}\includegraphics[width=0.49\textwidth]{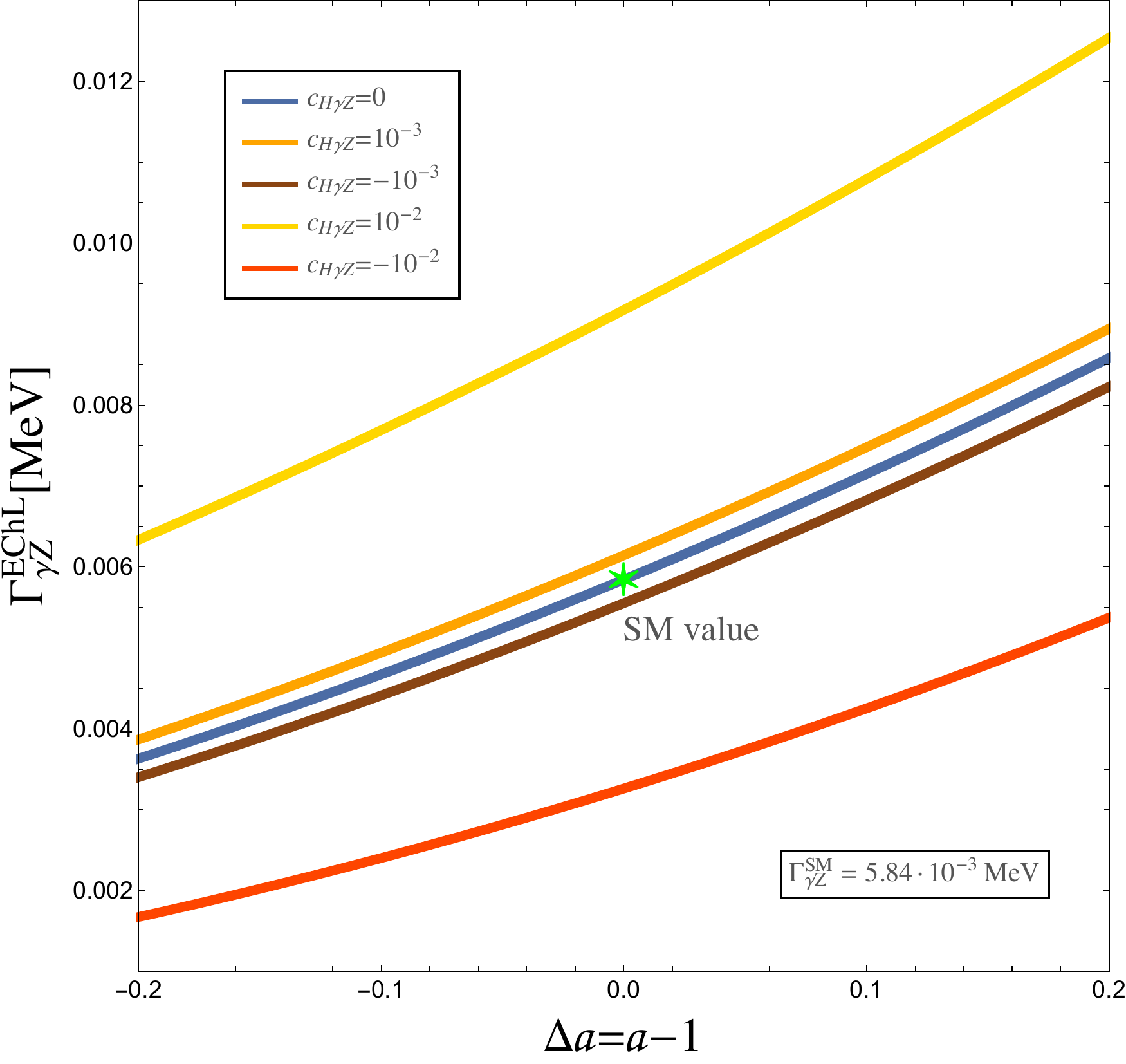}
\caption{EChL partial width (fermionic loop included) corresponding to the $H\to\gamma\gamma$ (left) and $H\to\gamma Z$ (right) decays as function of $\Delta a=a-1$ for different values of $c_{H\gamma V}$. The SM value is shown for reference (green star).}
\label{fig:allEChL}
\end{center}
\end{figure}

Our numerical results for these partial widths, as functions of $\Delta a$ for different values of $c_{H\gamma V}$, are shown in \figref{fig:allEChL}. 
We have explored the intervals on the EChL parameters  given by $-0.2\leq\Delta a\leq0.2$ and $-10^{-2}\leq c_{H\gamma V}\leq10^{-2}$. The reason for the different size of these two intervals is obvious given that $a$ enters via loop corrections, therefore with extra factors of ${\cal O}(1/(16\pi^2))$ at the amplitude level, whereas $c_{H\gamma V}$ enters to tree level, hence with no extra suppression factors.
The point corresponding to $\Delta a=0$ and $\cone=\ctwo=0$ is included in both plots for numerical comparison since it corresponds to the SM value, as we found in \eqref{eq-VeffoneSM} and in \eqref{eq-VefftwoSM}. 
In this figure we see clearly the behaviour of the EChL partial widths with these two parameters $\Delta a$ and $c_{H\gamma V}$ and the departures respect to the SM value. 
Setting $\cone=0$, the EChL partial witdth grows (decreases) with positive (negative) $\Delta a$ and separates from the SM prediction, reaching values well above (below) the SM value. The produced shift in 
$|\Gamma_{\gamma\gamma}^{\rm EChL}-\Gamma_{\gamma\gamma}^{\rm SM}|$ reaches values up to around $5 \times 10^{-3}$ MeV for the maximum explored values of $|\Delta a|=0.2$. 
Setting $\Delta a=0$, the behaviour with $\cone$ is similar, and the produced shift in this width difference also reaches values up to $5 \times 10^{-3}$ MeV for the maximum explored values of $|\cone|=10^{-2}$. 
The behaviour of the decay witdth in the other studied channel, $\gamma Z$,  with $\Delta a$ and  $\ctwo$ is similar, but in this case the predicted shift is of smaller size, reaching values of up to around $3 \times 10^{-3}$ MeV for the maximum explored values of $|\Delta a|=0.2$ or $|\ctwo|=10^{-2}$. 
Taking into account both non-vanishing parameters, the shift in both channels increases reaching values of up to around $10^{-2}$ MeV for the $\gamma \gamma$ channel and around $6 \times 10^{-3}$ for the $\gamma Z$ channel, at the extreme values considered of both EChL parameters.

\begin{figure}[h!]
\begin{center}
\includegraphics[width=\textwidth]{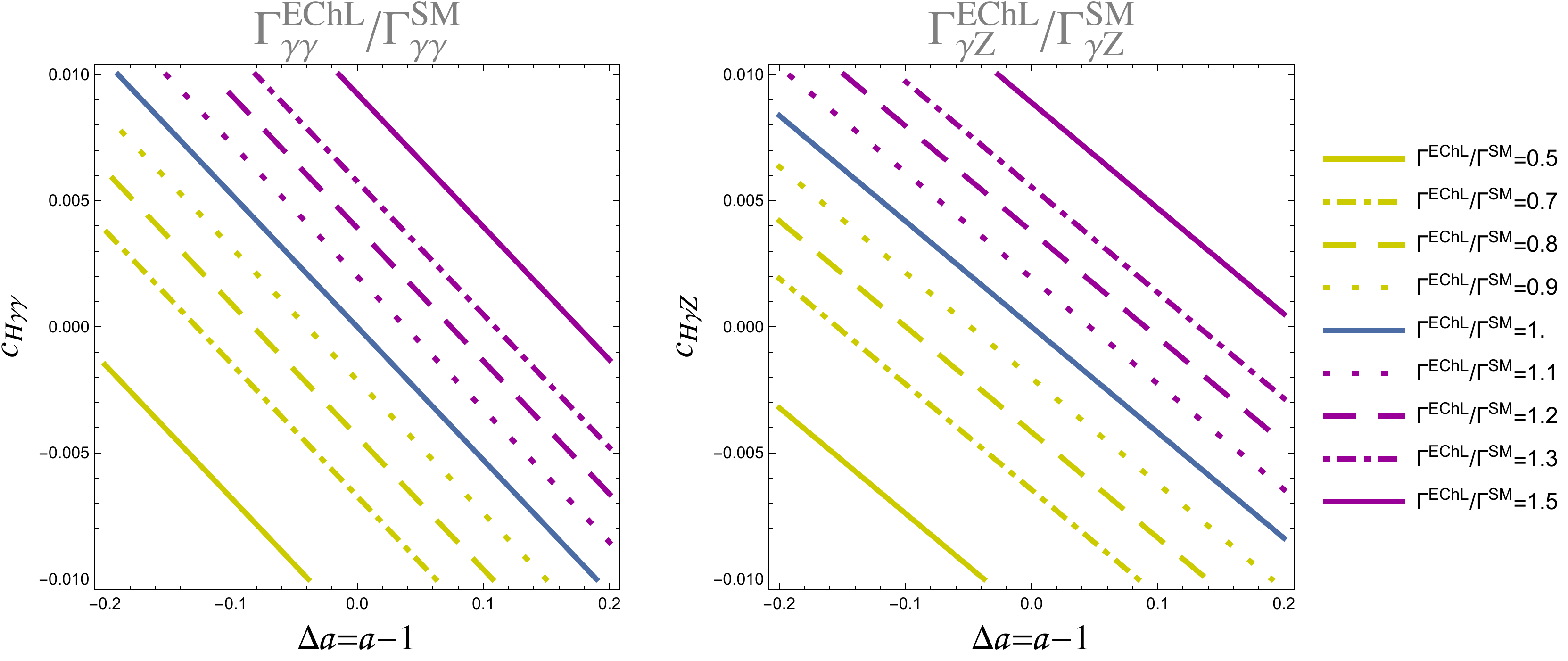}
\caption{Contour lines for the ratio of the EChL partial width respect to the SM one, corresponding to the $H\to\gamma\gamma$ (left) and $H\to\gamma Z$ (right) decays. The various lines correspond to specific settings for the deviations respect to the SM value, $(\Gamma^{\rm EChL}-\Gamma^{\rm SM})/ \Gamma^{\rm SM}$, of  $0\%$ (solid blue line), $\pm10\%$ (dotted), $\pm20\%$ (dashed), $\pm30\%$ (dot-dashed) and $\pm50\%$ (solid) purple(green), respectively.}
\label{fig:CL-bosfermEChL}
\end{center}
\end{figure}

\vspace{2mm}
Over the explored region of the parameter space, we also construct in \figref{fig:CL-bosfermEChL} the contour lines in the  $(\Delta a, c_{H\gamma V})$ plane for various fixed values of the ratio of the EChL partial width over the SM one, $\Gamma^{\rm EChL}/\Gamma^{\rm SM}$. Specifically, we show the contours for $\Gamma^{\rm EChL}/\Gamma^{\rm SM}$ equal to $1.$ , $1.\pm 0.1$, $1.\pm 0.2$, $1.\pm 0.3$ and $1. \pm 0.5$ , that correspond to deviations respect to the SM value, $(\Gamma^{\rm EChL}-\Gamma^{\rm SM})/ \Gamma^{\rm SM}$, of $0\%$, $\pm10\%$, $\pm20\%$, $\pm30\%$ and $\pm50\%$, respectively. 
The resulting contour lines are parallel straight lines, corresponding to the linear dependence of $\Veffboth\vert_{q^2=\mh^2}$ as function of $a$ (or equivalently $\Delta a=a-1$) and $c_{H\gamma V}$. 
In this figure we see clearly that the same numerical prediction for the ratio $\Gamma^{\rm EChL}/\Gamma^{\rm SM}$ can be reached for the infinite points $(\Delta a, c_{H\gamma V})$ on top of the corresponding  straight line. In particular, the same EChL prediction as in the SM, can be reached for all the points on top of the contour line with  $\Gamma^{\rm EChL}/\Gamma^{\rm SM}=1$ and not just  for $(\Delta a, c_{H\gamma V})=(0,0)$. Therefore, potential signatures from BSM physics via these Higgs decay channels get strongly reduced sensitivity at all these points and the corresponding ones on top of the closest parallel lines to these reference lines. The more distant lines to the blue line correspond to the larger sensitivities to BSM physics. Thus, the two farest contours summarise the $(\Delta a, c_{H\gamma V})$ points with largest deviations respect to the SM, and therefore the first ones to be constrained by data. The present status of experimental searches for the $H\to\gamma\gamma$ decay corresponds to the measured signal strengths relative to the SM expectation $\mu_{\rm ATLAS}=0.99^{+0.15}_{-0.14}$~\cite{Aaboud:2018xdt} and $\mu_{\rm CMS}=1.18^{+0.17}_{-0.14}$~\cite{Sirunyan:2018ouh}. However, the $H\to\gamma Z$ decay is not measured yet and we only have an upper limit on the production cross section times the branching ratio~\cite{Aaboud:2017uhw,Aad:2020plj,Sirunyan:2018tbk}. Also, the corresponding prospects to HL-LHC and HE-LHC are given in~\cite{Cepeda:2019klc}. A detailed analysis of the constraints on the EChL parameters entering in the Higgs observables is given in~\cite{losBuchalla}.

\vspace{2mm}
Next we explore the anatomy of the various loop contributions participating in the $R_\xi$ gauges. For simplicity, we fix $\cone=\ctwo=0$ from now on in our numerical estimates. 
As we said in the previous sections this splitting leads to contributions that are separately $\xi$-dependent. Thus, when comparing numerically the size of the various loop contributions  to the Higgs decays we have to set $\xi$ to a specific value. We choose in particular to make this numerical comparison for the two most popular $R_\xi$ gauges, the Landau gauge ($\xi=0$) and the Feynman-'t Hooft gauge ($\xi=1$). Since, as discussed in \secref{analytical-results}, it is only the GB contribution, among all the bosonic loop contributions, which respects the Ward structure of the amplitude and it is UV-finite by itself, we have done this separation in the total amplitude specifically as GB contribution plus the rest, meaning that in our numerical predictions we put together the other `gauge'+`mix'+`CT' contributions. 
On the other hand, our interest in showing the GB loop contributions separately is also because we know that they play a distinct role in non-linear EFTs based on chiral symmetries. The GB loops of these chiral theories, usually called chiral loops, play the most relevant  role in the chiral (momentum) expansion of physical quantities, like scattering amplitudes, decay amplitudes, etc., since they provide the largest contributions due to the GB derivative couplings. This is clearly the case of pion loops in the context of ChPT. Thus, we wish to explore here what is the role of the GB loops in our EChL case, and analyse if they are also relevant or not in the selected Higgs decays. We wish also to analyse if this relevance depends or not on the gauge-fixing $\xi$ parameter and on the particular setting of the Higgs momentum, i.e., being either on-shell or off-shell. Next we analyse these issues, first for the partial widths and later for the vertex functions.

\begin{figure}[h!]
\begin{center}
\includegraphics[width=0.49\textwidth]{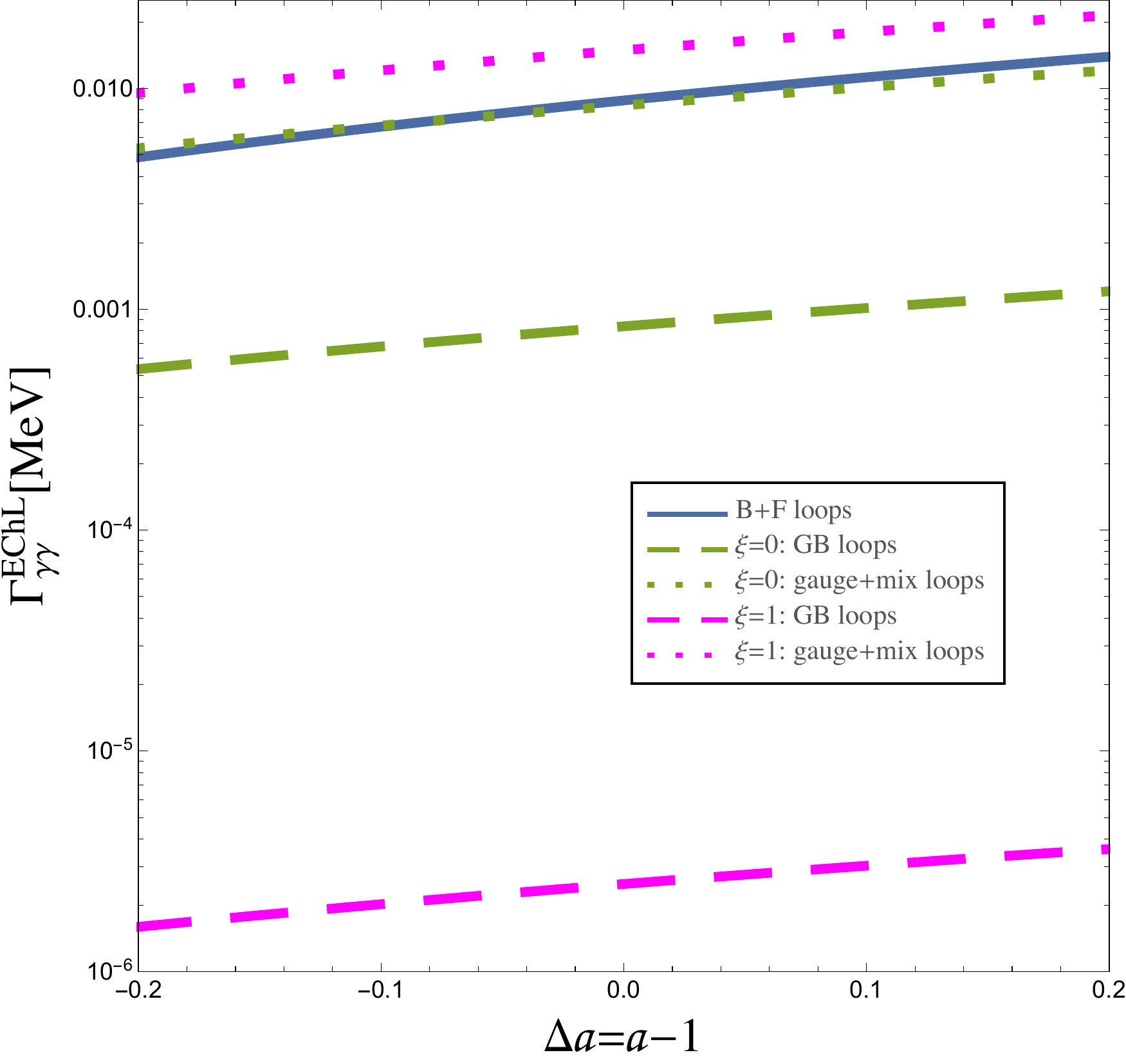}
\includegraphics[width=0.49\textwidth]{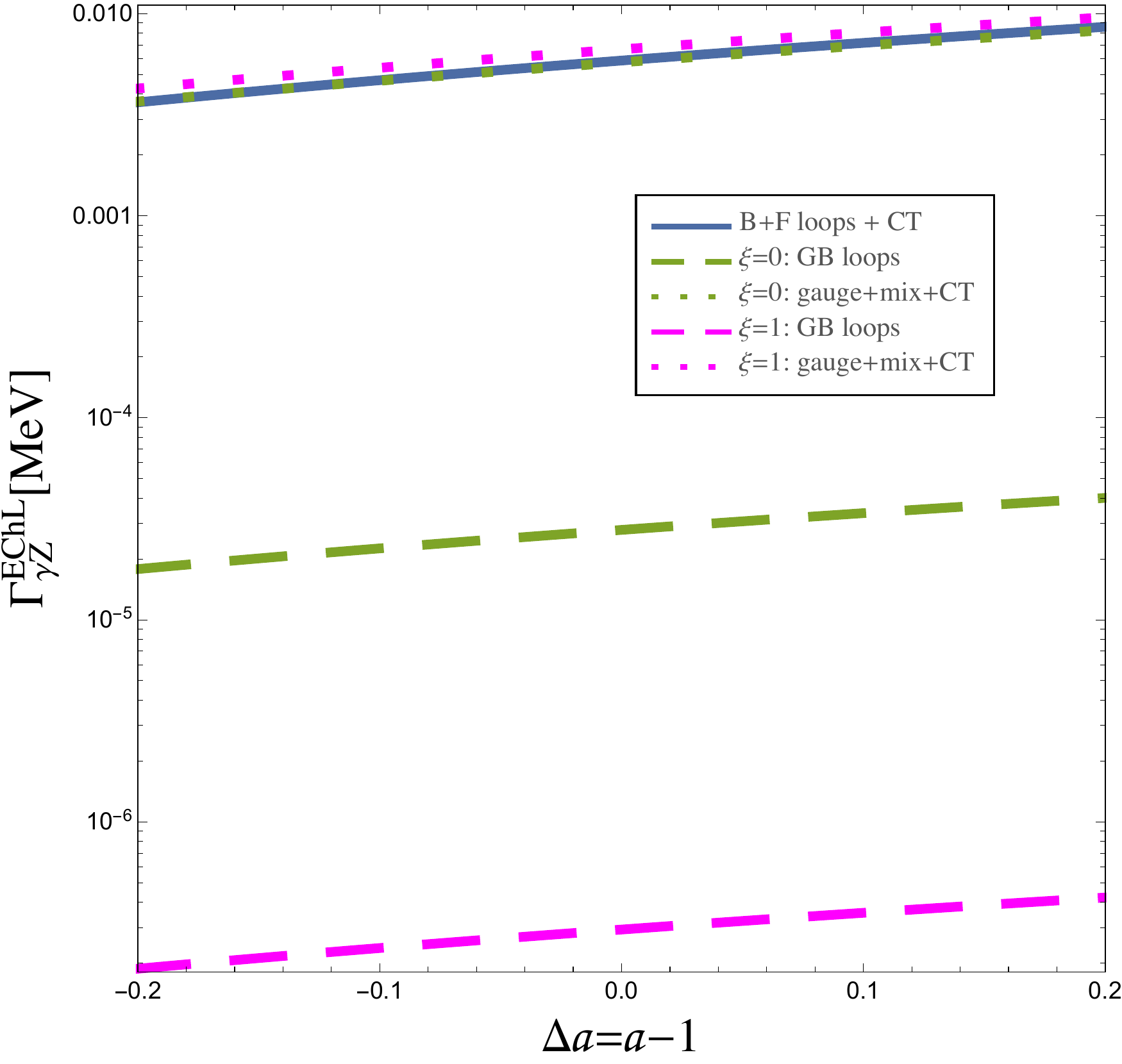}
\caption{Anatomy of the various contributions to the EChL partial widths as functions of $\Delta a=a-1$ in the Landau and Feynman gauges, for the $H\to\gamma\gamma$ (left) and $H\to\gamma Z$ (right) decays. The predictions for the complete partial width (solid), the GB loop contribution (dashed) and the contribution form the rest (dotted) are shown separately.}
\label{fig:anatomyEChLlandauFtH}
\end{center}
\end{figure}

\figref{fig:anatomyEChLlandauFtH} shows our separate predictions within the EChL in the $R_\xi$ gauges of the partial widths for both decays as function of $\Delta a$. The fermion loop contribution is added in all predictions of the complete partial widths shown, as explained above. In these plots, the contributions from each separate loops types to the partial widths means that the others have been set to zero.  
Specifically, we include in \figref{fig:anatomyEChLlandauFtH} the predictions for: the complete partial width,  the GB loops contributions, and the contributions form the rest, i.e., from `gauge'+`mix' in the $\gamma \gamma$  case and from `gauge'+`mix'+`CT' in the $\gamma Z$ case. As we can clearly  see from these plots, the GB loops contributions do not provide the leading contribution to the partial widths for any of the two chosen gauges in the case of the on-shell Higgs boson, i.e., with $q^2=\mh^2$. At this low momentum, the most relevant loops are indeed the remaining ones, i.e., the loops with only gauge bosons, and the loops with both gauge and GB loops.

\begin{figure}[h!]
\begin{center}
\includegraphics[width=0.49\textwidth]{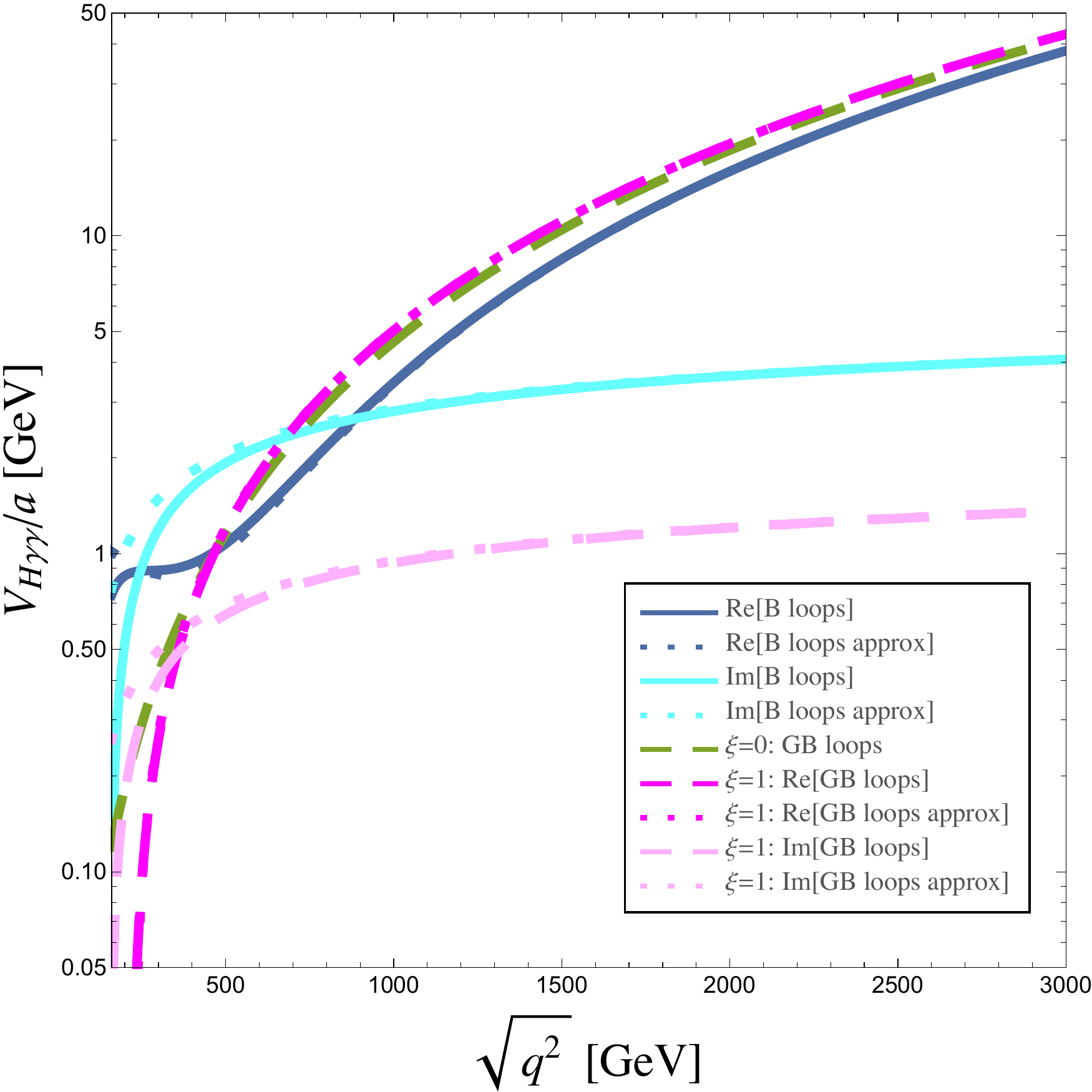}
\includegraphics[width=0.49\textwidth]{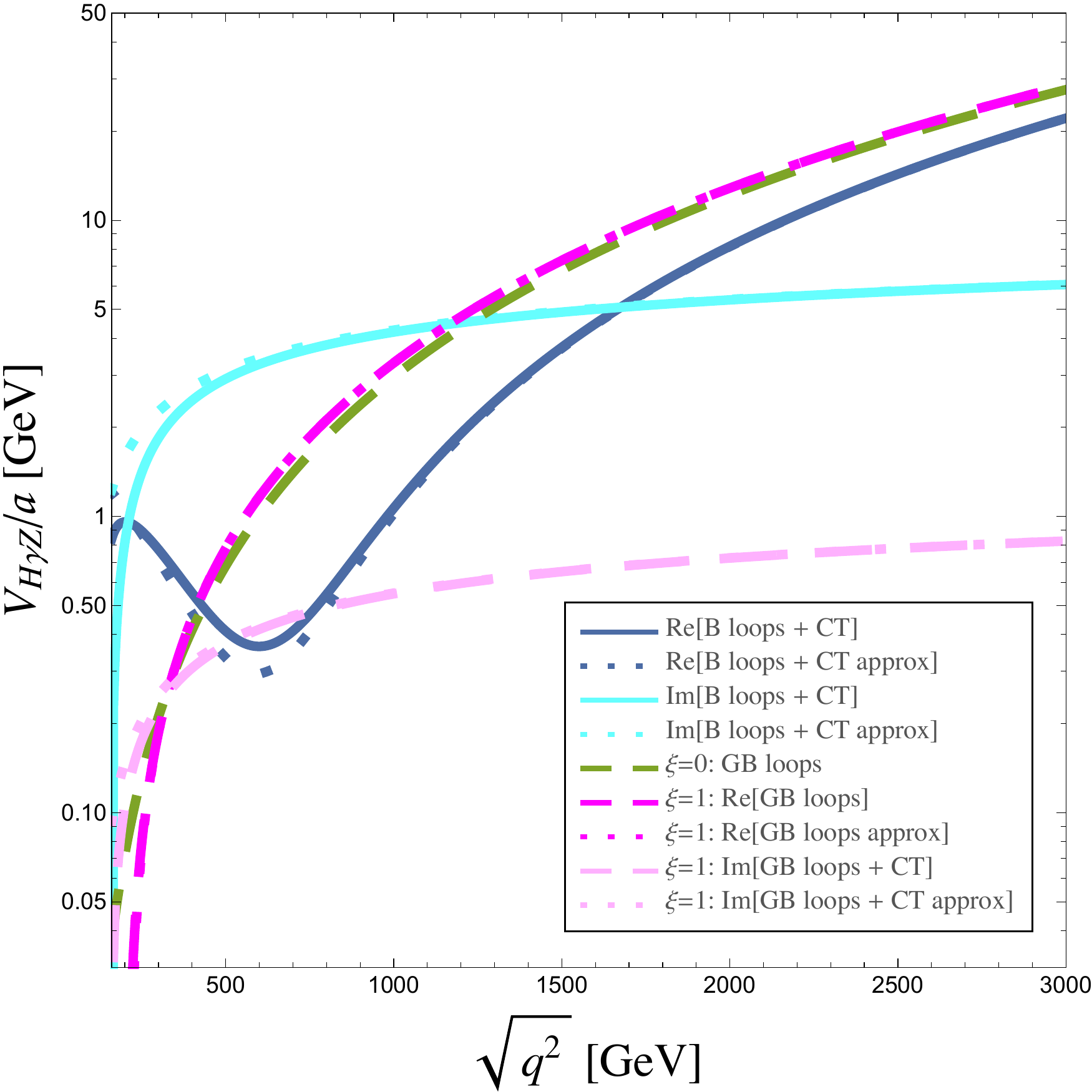}
\caption{Anatomy of the EChL vertex functions, $\Veffone$ (left) and $\Vefftwo$ (right), split into the various boson loop contributions corresponding to the total bosonic (solid lines) and GB loop contributions (dashed lines), as function of the Higgs boson momentum $\sqrt{q^2}$. We set $\cone=\ctwo=0$ and normalize the resulting vertex functions respect to the $a$ parameter. The real and imaginary parts (solid blue and light-blue lines respectively) of the total vertex functions are displayed separately. We include predictions for both gauge choices, $\xi=0$ (dashed green) and $\xi=1$ (dashed pink and light-pink for its real and imaginary parts). The predictions from the approximate vertex functions at large $q^2$ are also shown (dotted lines).}
\label{fig:anatomyVeff}
\end{center}
\end{figure}

\vspace{2mm}
Finally, in order to explore the off-shell case, we study the behavior of the EChL vertex functions in the $R_\xi$ gauges in terms of the off-shell Higgs boson momentum squared, $q^2$. We show in \figref{fig:anatomyVeff} our numerical predictions for the two vertex functions, $\Veffone$ and $\Vefftwo$, normalized by the $a$ parameter, as functions of $\sqrt{q^2}$, focusing this time on just the boson loops (i.e. we do not include here the fermion loops) and setting again $\cone=\ctwo=0$. Since these are complex functions we display separately their corresponding real and imaginary parts. 
The explored interval in $q^2$ goes up to $(3\, {\rm TeV})^2$ which corresponds typically to the maximum allowed energy for valid predictions within this non-linear EFT of the EChL given approximately by $4 \pi v$.  As in the previous plots, we include predictions for the two cases, $\xi=0$ and $\xi=1$.  
Specifically, we show predictions from the total boson loops and from just the GB loops.  Notice that we have also included the contribution from the counter-term in the case of $\Vefftwo$ since, as we said, it is crucial for the $\xi$-independence, UV-finiteness and Ward structure of the total vertex function result. Furthermore, for illustrative purposes and to help in extracting the conclusions from this plot, we also include in this figure the predictions from the simple approximate vertex functions found in the previous sections. Concretely those valid at large $q^2$. 

First of all, in regard the validity of our approximate formulas for $\Veffone$ and $\Vefftwo$ at large $q^2$, we compare the predictions in this figure with the predictions using the corresponding full formulas. We can see in \figref{fig:anatomyVeff} a very good agreement in all cases of the approximate results with the corresponding full results at large $q^2$. The agreement is indeed excellent for a Higgs boson momentum say above 1 TeV. 
Second, we see that at large energies the real part dominates over the imaginary part in the total amplitudes. In particular, this hierarchy is achieved roughly above 1 and 1.7 TeV for $H\to\gamma\gamma$ and $H\to\gamma Z$, respectively.
Third, comparing the real parts, the predictions of the GB loops for both $\xi=0$ and $\xi =1$ are very close to each other above 500 GeV for both decays. Remember that in the Landau gauge the GB loops are indeed real. 
Finally, and most importantly, we also see that these real parts from the GB loops approach clearly to the real part of the total prediction at large $q^2$. This occurs in $\Veffone$  already at intermediate energies, say above 500 GeV, and in $\Vefftwo$ at larger energies, say above 1500 GeV. 
The growing dependence with the energy of these predictions is very well approximated by the simple prediction from the Landau gauge, which is polynomic in $q^2$ at large momentum. This polynomic contribution is indeed UV-finite and gauge independent. Thus, we conclude from this plot that the GB loops (i.e. the chiral loops) do provide the most relevant loop contributions to the vertex functions in the case of the off-shell Higgs boson, at large energies, and their main effect can be described by the simple polynomic term that is gauge invariant. This is in concordance with the expectations in EFT described by Chiral Lagrangians.

\section{Conclusions}
\label{sec-conclu}
The use of EFTs containing the Higgs boson particle and its interactions with all the SM particles, in a $SU(3) \times SU(2) \times U(1)$ gauge invariant way, is nowadays with no doubt, the best tool we have at hand to describe in a model independent way the potential new Higgs physics from BSM dynamics. In this work we have focused in the bosonic sector of the non-linear EFT given by the EChL which is the most appropriate one if the dynamics behind the EWSBS is strongly interacting. In that case, the GBs of the electroweak symmetry breaking are also the GBs of the electroweak chiral symmetry breaking, $SU(2)_L \times SU(2)_R \to SU(2)_{L+R}$, and the non-linearity of the GBs under $SU(2)$ transformations leads to notable differences with respect to the SM case and also with respect to linear EFTs like the SMEFT. These include derivative GB self-interactions, different GB couplings to  the EW gauge bosons and to the Higgs boson, multiple particle interactions and others.  These differences, in turn,  may also affect in a relevant way to the Higgs physics itself,  particularly when going beyond the tree level within this EChL approach.    

In this paper we have worked out in full detail the computation, within the EChL to \1loop level, of two Higgs decays which are of particular relevance due to their phenomenological implications for collider physics. Concretely, $H \to \gamma \gamma$ and $H \to \gamma Z$. We have presented here, in a supposedly didactic and illustrative way,  the sequential steps to follow in this \1loop computation, according to the standard rules of Chiral Lagrangians and also including the proper renormalization program for this case. Our computation in the EChL with 
$R_\xi $ gauges is a novel one,  and we believe complements in an interesting way the computation available in the literature~\cite{losBuchalla}, which is performed in the unitary gauge instead.

Firstly, we have provided an explicit demonstration of the gauge invariance of the resulting EChL 
$R_\xi$-amplitudes, for both decays $H \to \gamma \gamma$ and $H \to \gamma Z$. Specifically, we demonstrate that both decay amplitudes are UV-finite, $\xi$-independent, preserve the structure of the Ward identity, and find out a final result that coincides with the unitary gauge result. 
One of the most  interesting features is that these findings are true for both assumptions on the external Higgs momentum, on-shell and off-shell. The two external gauge bosons are always assumed here to be on-shell. This situation is different than in the SM case, where it was proven in~\cite{Marciano-paper} the  gauge invariance of the decay amplitude $H \to \gamma \gamma$ for the on-shell Higgs case, but the off-shell case was not considered. Thus, we have also computed here, for comparison with our EChL computation, the SM Higgs amplitudes in both decay channels for the missing off-shell case in the $R_\xi$ gauges. For this part of the gauge invariance demonstration we have followed very closely the procedure described in~\cite{Marciano-paper}. 
Our overall outcome from that comparison is that whereas the \1loop EChL amplitudes preserve gauge invariance and the Ward structure in both Higgs cases, on-shell and off-shell, it is not the same for the SM case. For the SM $R_\xi$-amplitudes we have found that, for arbitrary Higgs boson momentum $q^2$, they are both UV-finite but they are $\xi$-dependent and do not preserve the Ward structure. And it is in the on-shell case, $q^2=m_H^2$, where the gauge invariance of the total SM amplitudes are recovered. 

In addition to the gauge invariance demonstration,  we have also discussed, both analytically and numerically, on the relevance of the various parts contributing to the total EChL \1loop decay amplitudes. These include the contributions from:  1) the tree-level  EChL coefficients entering in the chiral dimension 4 operators, $\cone$ and $\ctwo$,  2) the various types of loop diagrams built from the chiral dimension 2 operators (hence, providing the dependence on the EChL coefficient $a$), namely, those with only gauge bosons in the loops, those with only GBs, those with both gauge and GBs,  and those with ghosts and, finally, 3) the various counter-terms. In the analytical part, we have found that the contribution from GB loops is by itself UV-finite and preserves the Ward structure, although it is $\xi$-dependent.

The anatomy of the studied Higgs boson decays in terms of the mentioned contributions within the $R_\xi$ gauges has allowed us to unveil the role played by the GBs, usually called chiral loops in the Chiral Lagrangian approach. For that purpose, we consider the corresponding \1loop vertex functions, $\Veffone$ and $\Vefftwo$, and analyse them also numerically in the two different cases for the external Higgs boson momentum: on-shell and off-shell. When comparing the size of the various contributions (this is true for both the Higgs boson partial widths and the \1loop vertex functions) with the corresponding total result, with have found that the size of the GB loop contributions is very small (compared to the others) in the case that the Higgs boson is on-shell. We understand this smallness because the momentum transferred to the loop is small, being set by $q^2=m_H^2$. In contrast,  for the off-shell Higgs case, we have found that the GB contribution is indeed the most relevant one when the Higgs boson momentum reaches the TeV domain. The typical growing behaviour with energy of the Chiral Lagrangian amplitudes  is understood, in the present case, by performing an expansion at large momentum of the \1loop Higgs decay EChL amplitudes and the corresponding vertex functions. Here by large $\sqrt{q^2}$ we mean large compared with the soft masses involved, $m_Z$, $m_W$ and $m_H$, but below the maximum energy that is allowed for this EFT to be valid, which in  the present EChL case is set  approximately by $4\pi v \sim  3\, {\rm TeV}$. We have also checked numerically that, indeed,  the GB loops provide the most relevant contributions at these ${\cal O}({\rm TeV})$ large energies, and they approach clearly to the total result. Finally, with the help of our approximate formulas from the expansion at large $q^2$ of the \1loop vertex functions, we have understood that the dominant term in the total amplitude is polynomial in $q^2$ and it is reproduced by the $\xi$-independent part of the GB contributions, in concordance with the expectations in EFTs described by Chiral Lagrangians.

In conclusion and in summary, we believe that the results of the $R_\xi$ gauges presented in this paper complement nicely the previous ones in the literature of the unitary gauge,  and they may be useful in the colliders analysis of BSM Higgs signatures via these two channels. The \1loop vertex functions derived here, $\Veffone$ and $\Vefftwo$, can also have interesting applications for Higgs-mediated processes where the intermediate propagating Higgs boson is off-shell. 

\section*{Acknowledgments}
This work is supported by the European Union through the ITN ELUSIVES H2020-MSCA-ITN-2015//674896 and the RISE INVISIBLESPLUS H2020-MSCA-RISE-2015//690575, by the CICYT through the project FPA2016-78645-P and by the Spanish MINECO's ``Centro de Excelencia Severo Ochoa''  Programme under grant SEV-2016-0597. 
\newpage

\section*{Appendices}
\appendix

\section{Relevant Feynman rules}
\label{App-FRules}

In this appendix we summarise all the relevant Feynman rules participating in the computation.
Regarding the $W$ propagator $-iP_{\mu\nu}$, we split it into two pieces as in \eqrefs{Wprop-spliting}{Wprop-spliting2}, in order to demonstrate the $\xi$-independence of the total \1loop amplitude in the $R_\xi$ gauges computation.
In \tabrefs{relevant-FR1}{relevant-FR2} we collect these relevant Feynman rules for the EChL interaction vertices coming from $\mLtwo$ of \eqref{eq-L2} and the corresponding SM Feynman rules for a clear comparison between them. In particular,  \tabref{relevant-FR1} contains the FRs participating in both the unitary and the $R_\xi$ gauge computations, and we use the following short notation for the standard Lorentz tensors of the gauge boson self couplings:
\bear
\Gamma^{\mu\nu\rho}(p_-,p_+,p_0) &=& g^{\mu\nu}(p_--p_+)^\rho+g^{\nu\rho}(p_+-p_0)^\mu+g^{\rho\mu}(p_0-p_-)^\nu \,,  \nn\\
S^{\mu\nu,\rho\sigma} &=& 2g^{\mu\nu}g^{\rho\sigma} -g^{\mu\rho}g^{\nu\sigma} -g^{\mu\sigma}g^{\nu\rho} \,.
\label{self-gauge-vertex}
\eear
On the other hand, \tabref{relevant-FR2} contains the other relevant FRs involved in the $R_\xi$ gauges.
Notice that, to shorten the table, we present together the photon and $Z$ boson interaction vertices in most of the Feynman rules. In that cases, the first coupling corresponds to the photon and the second one to the $Z$.
\begin{table}[H]
\begin{tabular}{ >{\centering\arraybackslash}m{45mm}| >{\arraybackslash}c| >{\arraybackslash}c}
Interaction & EChL & SM \\
\hline 
\includegraphics[width=40mm]{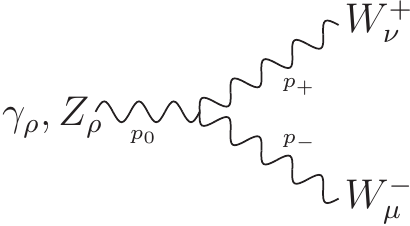} & $-i\{e,g\cosw\}\Gamma_{\mu\nu\rho}(p_-,p_+,p_0)$ & $-i\{e,g\cosw\}\Gamma_{\mu\nu\rho}(p_-,p_+,p_0)$  \\
\includegraphics[width=38mm]{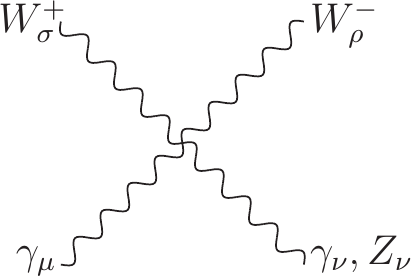} & $-i\{e^2,eg\cosw\}S_{\mu\nu,\rho\sigma}$ & $-i\{e^2,eg\cosw\}S_{\mu\nu,\rho\sigma}$ \\
\includegraphics[width=33mm]{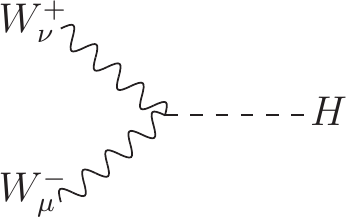} & $\frac{2i a \mw^2}{\vev}g^{\mu\nu}$ & $\frac{2i\mw^2}{\vev}g^{\mu\nu}$ \\
\includegraphics[width=33mm]{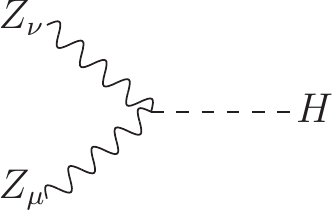} & $\frac{2i a \mz^2}{\vev}g^{\mu\nu}$ & $\frac{2i\mz^2}{\vev}g^{\mu\nu}$ \\
\end{tabular}
\caption{Relevant EChL Feynman rules from $\mLtwo$ participating in the unitary and $R_\xi$ gauge computations. All momenta are incoming. In the rules with `$\{\,,\}$', the first and second argument correspond to the photon and $Z$ boson, respectively. We also include the corresponding SM Feynman rules for comparison.}
\label{relevant-FR1}
\end{table}
\begin{table}[H]
\begin{tabular}{ >{\centering\arraybackslash}m{45mm}| >{\arraybackslash}c| >{\arraybackslash}c}
Interaction & EChL & SM \\
\hline 
\includegraphics[width=33mm]{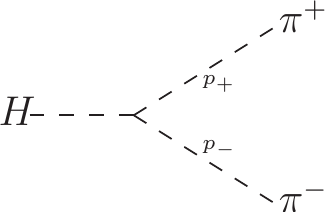} & $-\frac{2i a}{\vev}p_+ \cdot p_-$ & $-\frac{i}{2}g\frac{\mh^2}{\mw}$  \\
\includegraphics[width=33mm]{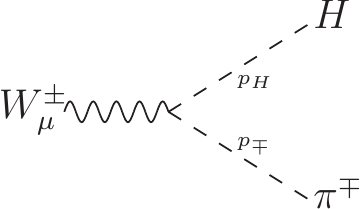} & $\frac{2a \mw}{\vev}p_\mp^\mu$ & $\frac{\mw}{\vev}(p_\mp-p_H)^\mu$ \\
\includegraphics[width=38mm]{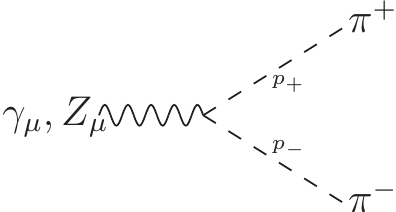} & $i \{e,\frac{g\cosw(1-\tanw^2)}{2}\}(p_- -p_+)^\mu$  & $i \{e,\frac{g\cosw(1-\tanw^2)}{2}\}(p_- -p_+)^\mu$  \\
\includegraphics[width=38mm]{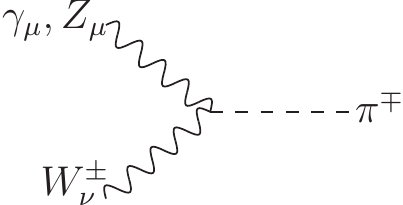} & $\{\pm e\mw,\mp e\tanw \mw\}g^{\mu\nu}$ & $\{\pm e\mw,\mp e\tanw \mw\}g^{\mu\nu}$  \\
\includegraphics[width=38mm]{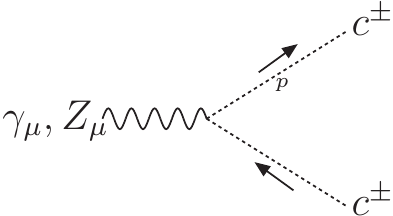} & $\mp i \{e,g\cosw\} p_\mu$ & $\mp i \{e,g\cosw\} p_\mu$  \\
\includegraphics[width=31mm]{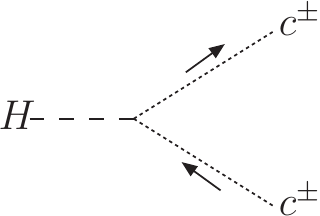} & 0 & $-\frac{ig}{2}\xi\mw$  \\
\includegraphics[width=38mm]{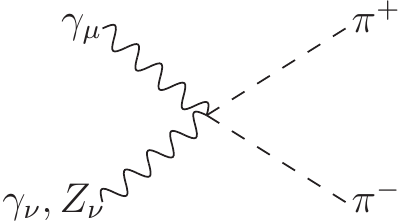} & $i\{2e^2,\frac{eg(\cosw^2-\sinw^2)}{\cosw}\} g^{\mu\nu}$ & $i\{2e^2,\frac{eg(\cosw^2-\sinw^2)}{\cosw}\} g^{\mu\nu}$  \\
\includegraphics[width=38mm]{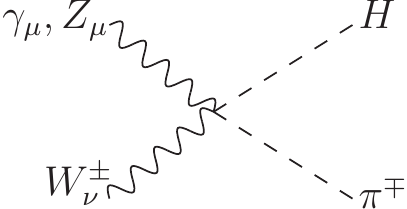} & $\{\pm \frac{2a e \mw}{\vev},\mp \frac{2a g\sinw^2\mw}{\vev\cosw}\} g^{\mu\nu}$ & $\{\pm \frac{e\mw}{\vev},\mp \frac{g\sinw^2\mw}{\vev\cosw}\} g^{\mu\nu}$ \\
\includegraphics[width=38mm]{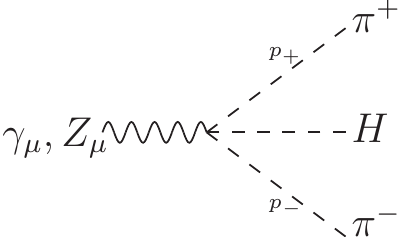} & $i\{\frac{2 e a}{\vev},\frac{a g (\cosw^2-\sinw^2)}{\vev\cosw}\}(p_- -p_+)^\mu$ & 0  \\
\includegraphics[width=38mm]{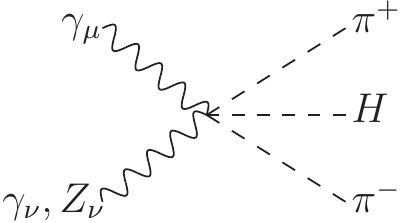} & $i\{\frac{4a e^2}{\vev},\frac{2a eg(\cosw^2-\sinw^2)}{\cosw\vev}\} g^{\mu\nu}$ & 0  \\
\end{tabular}
\caption{Additional relevant EChL Feynman rules from $\mLtwo$ participating in the $R_\xi$ gauge computation. All momenta are incoming (except for the $c$'s, given by the arrows). In the rules with `$\{\,,\}$', the first and second argument are for the photon and $Z$ boson, respectively. The corresponding SM rules are included for comparison.}
\label{relevant-FR2}
\end{table}

\begin{figure}[h!]
\begin{center}
\includegraphics[width=0.8\textwidth]{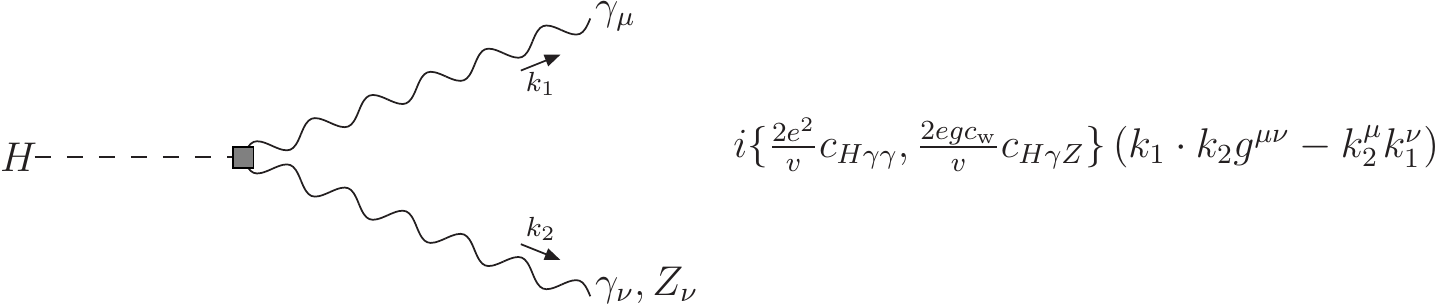}
\caption{Relevant EChL Feynman rules from $\mLfour$ (not present in the SM). The interaction vertex for $H \gamma V$  is denoted in this case by a shadowed little box. The first and second argument in `$\{\,,\}$' correspond to the photon and $Z$ boson, respectively.}
\label{diags-L4-FRules}
\end{center}
\end{figure}

The Feynman rules corresponding to the relevant $\mLfour$ operators of the \eqref{eq-L4relevant} are shown in \figref{diags-L4-FRules}. Notice that these are not present in the SM.

\begin{figure}[h!]
\begin{center}
\includegraphics[width=0.8\textwidth]{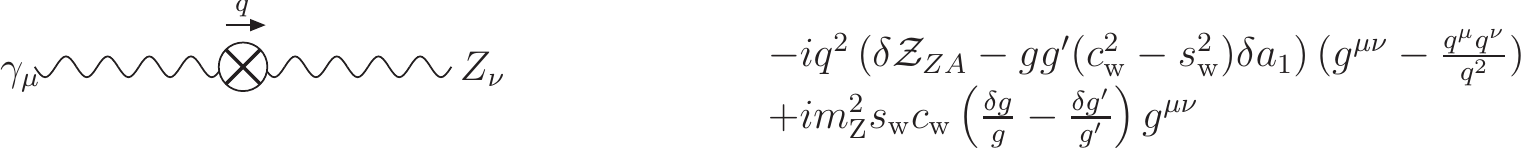}\\
\vspace{3mm}
\includegraphics[width=0.8\textwidth]{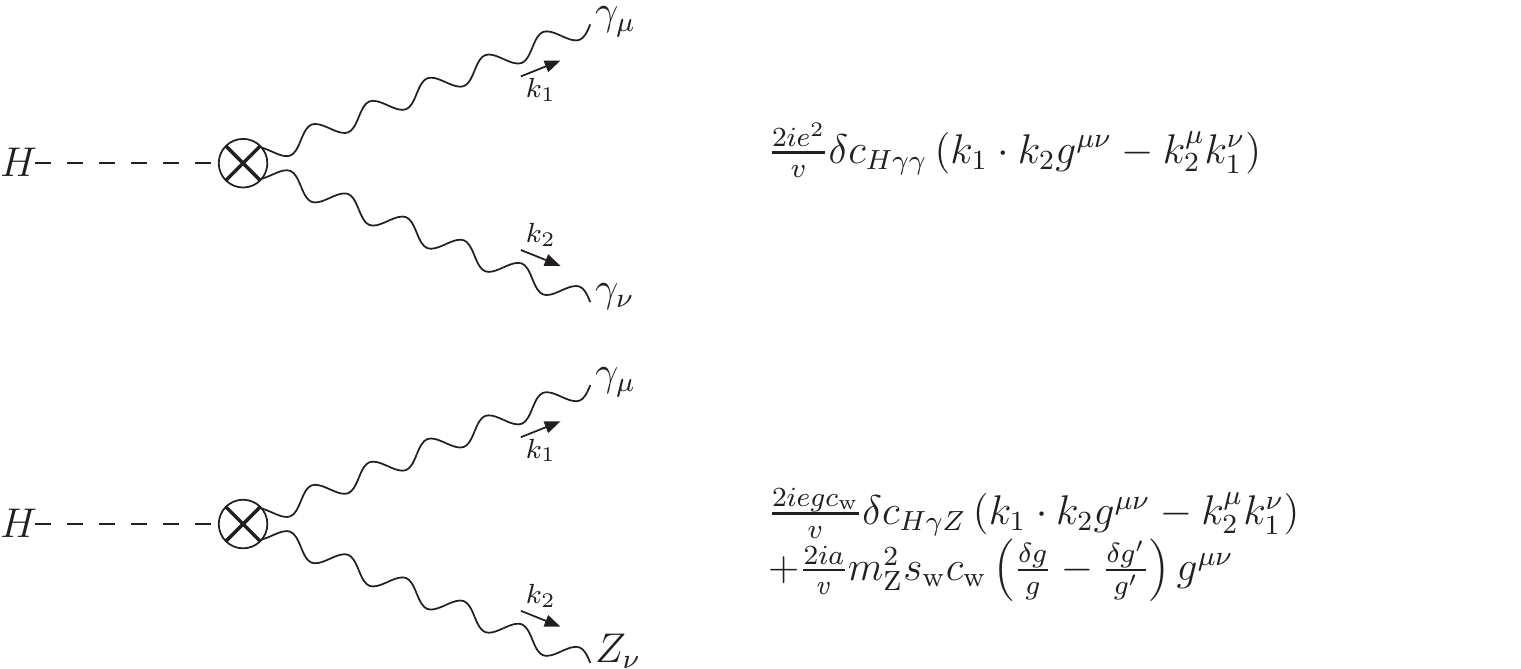}
\caption{Relevant EChL Feynman rules from all the counter-terms involved in the computation. A crossed circle is drawn to denote these counter-terms. We use the short notation $\delta\Zf_{ZA}=\sinw\cosw(\delta\Zf_W -\delta\Zf_B)$ for the contribution from the counter-term of the wave function renormalization of the mixing.}
\label{diags-counterterms}
\end{center}
\end{figure}

Finally, the Feynman rules for all the counter-terms involved in the present computation, according to the explanation presented in the text,  are collected  in \figref{diags-counterterms}.

\section{Preparing the \1loop diagrams for the automated computation}
\label{App-LectureDiags}
In this appendix we present the contributions to the Higgs decay amplitudes  from the various \1loop diagrams. For this computation we have used the automated procedure provided by the Package-X~\cite{packX} which requires to write the input diagrams in a given format. Thus, we prepare here the input diagrams to be ready for this automated computation.   

The  \1loop computation is performed  with dimensional regularization in $D=4-\epsilon$ dimensions and we use the standard definitions for the associated divergence:
\be
\div=\frac{2}{\epsilon}-\gamma_E+\log\left(\frac{4\pi\mu_0^2}{\mw^2}\right) \,,
\label{div-definition}
\ee
where $\mu_0$ is the usual scale and the presence of the $W$ mass comes from the fact that only charged particles run in the loops (we separate the $\xi$-dependence coming from the $R_{\mu\nu}$ part of the $W$ propagator and from the GB and ghost ones). We implement the compact notation for the momentum integral given by:
\be
\int_k=\mu_0^{4-D}\int\frac{d^D k}{(2\pi)^D} \,.
\ee
To display some results, we also use the scalar two and three-point \1loop integral functions in the Passarino-Veltman notation \cite{Passarino:1978jh}, with the following conventions:
\be
  \frac{i}{16\pi^2} B_0(q_1,m_1,m_2) = \int_k \frac{1}{[k^2 - m_1^2][(k+q_1)^2 - m_2^2]}\,,
\ee
\be
   \frac{i}{16\pi^2}C_0(q_1, q_2, m_1, m_2, m_3) = \int_k \frac{1}{[k^2 - m_1^2][(k + q_1)^2 - m_2^2][(k + q_1 + q_2)^2 - m_3^2]}.
\ee
The other functions $f$ and $g$ that we use in this work are defined in \eqref{fpaper} and \eqref{gpaper} respectively, and are particular cases of the previous two and three-point integral functions. Explicitly:
\be
C_0(0,q,M,M,M)=-\frac{2}{q^2}f\left(\frac{4M^2}{q^2}\right)
\quad\text{and}\quad
B_0(q,M,M)=\Delta_\epsilon+g\left(\frac{4M^2}{q^2}\right) \,.
\label{fg_as_floops}
\ee

With all these conventions above, we present next each diagram of  \figrefs{diags-puregauge}{diags-ghost} in terms of the Feynman rules of \tabrefs{relevant-FR1}{relevant-FR2}, in an explicit form such that, as we have said,  they are ready for the automated computation with the Package-X~\cite{packX}.
In all categories, only charged particles are present in the loops and we just show the diagrams with negative-charged particles running clockwise in the loops (along the momentum $p$). The possible additional diagrams with positive-charged particles running clockwise (or negative-charged in counter-clockwise) are taken into account by means of an extra factor 2 in the amplitude.
It is important to stress that the kinematics of each decay is different and the resulting $\mz$-dependence in the $H\to\gamma Z$ decay breaks the symmetry under the exchange of the external gauge bosons (present in $H\to\gamma\gamma$). Then we explicitly draw the crossing diagrams and denote them by adding a prime '.
However, the reading of the diagrams in both decays are the same and they just differ on the coupling factors of  \tabrefs{relevant-FR1}{relevant-FR2}. In particular, the results of the $H\to\gamma\gamma$ case are recovered in the limit $\mz\to0$ from the corresponding ones of the $H\to\gamma Z$ case, leaving apart these coupling factors.

The results for the diagrams are presented in the format of contributions to the rank-two tensor $\ampboth^{\mu\nu}$  from each diagram, following the definition of  \eqref{ward-structure}, inspired on the $U(1)_{\rm EM}$ Ward identity.
It is important to stress that this Ward identity structure does not arise on each individual diagram, and it is only present for the contributions from the GB loops and for the total \1loop amplitude (as we discussed along the text). We present the amplitudes for each diagram in both the EChL and the SM cases simultaneously, for comparison. We omit to include here the explicit output for each diagram from the Package-X computation, due to their extremely large size. However, when summing over the diagrams the outcome is shorten and it is indeed included explicitly in the text.

\vspace{2mm}
The EChL amplitudes of the diagrams type (a) and (b), in terms of the FRs, are:
\bear
&&\hspace{-3mm}\quad i\amp_{\rm a}^{\mu\nu}= \nn\\
&&\hspace{-3mm}\coupsEChL_{\rm a}\int_p \prop{\alpha}{\gamma}{p}\prop{\lambda}{\rho}{p-k_1}\prop{\delta}{\beta}{p-k_1-k_2}g^{\alpha\beta}\Gamma^{\gamma\lambda\mu}(p,-p+k_1,-k_1)\Gamma^{\rho\delta\nu}(p-k_1,-p+k_1+k_2,-k_2) \nn\\
&&\hspace{-3mm}\quad i\amp_{\rm a'}^{\mu\nu}= i\amp_{\rm a}^{\mu\nu}\vert_{k_1\leftrightarrow k_2;\,\mu\leftrightarrow\nu;\,\coupsEChL_{\rm a}\rightarrow\coupsEChL_{\rm a'}}\,,  \nn\\
&&\hspace{-3mm}\quad i\amp_{\rm b}^{\mu\nu}=\mathcal{C}_{\rm b}\int_p \prop{\alpha}{\gamma}{p}\prop{\delta}{\beta}{p-k_1-k_2}g^{\alpha\beta}S^{\mu\nu,\gamma\delta} \,,
\eear
with coupling factors equal to:
\begin{center}
    \begin{tabular}{c||c||c}
    & $\coupsEChL_{\rm a}$ = $\coupsEChL_{\rm a'}$ & $\coupsEChL_{\rm b}$ \\ \hline
    $H\to\gamma\gamma$ & $e^2 g \mw a$ & $-e g^2 \cosw \mw a$ \\
    $H\to\gamma Z$ & $e g^2 \cosw \mw a$ & $-e g^2 \cosw \mw a$
    \end{tabular}
\end{center}
The corresponding SM amplitudes to these diagrams are the same except for the strength of the $HWW$ vertex (where we must set $a=1$), as it can be seen from \tabref{relevant-FR1}:
\be
\ampSM_{\rm a}^{\mu\nu}=\amp_{\rm a}^{\mu\nu}\vert_{a=1} \,,\quad \ampSM_{\rm a'}^{\mu\nu}=\amp_{\rm a'}^{\mu\nu}\vert_{a=1} \quad\text{and}\quad \ampSM_{\rm b}^{\mu\nu}=\amp_{\rm b}^{\mu\nu}\vert_{a=1} \,.
\ee
The EChL amplitudes of the diagrams type (c), in terms of the Feynman rules, are:
\bear
&&i\amp_{\rm c}^{\mu\nu}=\coupsEChL_{\rm c}\int_p \prop{\alpha}{\gamma}{p}\prop{\lambda}{\rho}{p-k_1}\propG{p-k_1-k_2}(p-k_1-k_2)^\alpha\Gamma^{\gamma\lambda\mu}(p,-p+k_1,-k_1)g^{\rho\nu} \nn\\
&&i\amp_{\rm c'}^{\mu\nu}=i\amp_{\rm c}^{\mu\nu}\vert_{k_1\leftrightarrow k_2;\,\mu\leftrightarrow\nu;\,\coupsEChL_{\rm c}\rightarrow\coupsEChL_{\rm c'}} \,.
\eear
 The corresponding SM amplitudes differ on the $HW\pi$ vertex:
\bear
&&i\ampSM_{\rm c}^{\mu\nu}=\coupsSM_{\rm c}\int_p \prop{\alpha}{\gamma}{p}\prop{\lambda}{\rho}{p-k_1}\propG{p-k_1-k_2}(p-2k_1-2k_2)^\alpha\Gamma^{\gamma\lambda\mu}(p,-p+k_1,-k_1)g^{\rho\nu} \nn\\
&&i\ampSM_{\rm c'}^{\mu\nu}= i\ampSM_{\rm c}^{\mu\nu}\vert_{k_1\leftrightarrow k_2;\,\mu\leftrightarrow\nu;\,\coupsSM_{\rm c}\rightarrow\coupsSM_{\rm c'}}\,, 
\eear
with coupling factors equal to:
\begin{center}
    \begin{tabular}{c||c|c||c|c}
    & $\coupsEChL_{\rm c}$ & $\coupsEChL_{\rm c'}$ & $\coupsSM_{\rm c}$ & $\coupsSM_{\rm c'}$ \\ \hline
    $H\to\gamma\gamma$ & $2e^2 g \mw a$ & $2e^2 g \mw a$ & $2e^2 g\frac{1}{2} \mw$ & $2e^2 g\frac{1}{2} \mw$ \\
    $H\to\gamma Z$ & $-2e g^2 \frac{\sinw^2}{\cosw} \mw a$ & $-2e g^2 \cosw \mw a$ & $-2e g^2 \frac{\sinw^2}{2\cosw} \mw$ & $-2e g^2 \frac{\cosw}{2} \mw$
    \end{tabular}
\end{center}
The EChL amplitudes of the diagrams type (d), in terms of the Feynman rules, are:
\bear
&&i\amp_{\rm d}^{\mu\nu}=\coupsEChL_d\int_p \prop{\lambda}{\rho}{p-k_1}\propG{p-k_1-k_2}g^{\lambda\mu}g^{\rho\nu} \nn\\
&&i\amp_{\rm d'}^{\mu\nu}=i\amp_{\rm d}^{\mu\nu}\vert_{k_1\leftrightarrow k_2;\,\mu\leftrightarrow\nu;\,\coupsEChL_{\rm d}\rightarrow\coupsEChL_{\rm d'}}\,. 
\eear
The corresponding SM amplitudes differ on the $H\gamma W\pi$ and $HZW\pi$ vertices:
\bear
&&i\ampSM_{\rm d}^{\mu\nu}=\coupsSM_{\rm d}\int_p \prop{\lambda}{\rho}{p-k_1}\propG{p-k_1-k_2}g^{\lambda\mu}g^{\rho\nu} \nn\\
&&i\ampSM_{\rm d'}^{\mu\nu}=i\ampSM_{\rm d}^{\mu\nu}\vert_{k_1\leftrightarrow k_2;\,\mu\leftrightarrow\nu;\,\coupsSM_{\rm d}\rightarrow\coupsSM_{\rm d'}}\,, 
\eear
with coupling factors equal to:
\begin{center}
    \begin{tabular}{c||c||c}
    & $\coupsEChL_{\rm d}$ = $\coupsEChL_{\rm d'}$ & $\coupsSM_{\rm d}$ = $\coupsSM_{\rm d'}$ \\ \hline
    $H\to\gamma\gamma$ & $-2e^2 g \mw a$ & $-2e^2 g\frac{1}{2} \mw$ \\
    $H\to\gamma Z$ & $2e g^2 \frac{\sinw^2}{\cosw} \mw a$ & $2e g^2 \frac{\sinw^2}{2\cosw} \mw$ 
    \end{tabular}
\end{center}
The EChL amplitudes of the diagrams type (e), in terms of the Feynman rules, are:
\bear
&&\hspace{-3mm}i\amp_{\rm e}^{\mu\nu}=\coupsEChL_{\rm e}\int_p \prop{\alpha}{\gamma}{p}\propG{p-k_1}\propG{p-k_1-k_2}(p-k_1-k_2)^\alpha g^{\gamma\mu} (-2p+2k_1+k_2)^\nu \nn\\
&&\hspace{-3mm}i\amp_{\rm e'}^{\mu\nu}= i\amp_{\rm e}^{\mu\nu}\vert_{k_1\leftrightarrow k_2;\,\mu\leftrightarrow\nu;\,\coupsEChL_{\rm e}\rightarrow\coupsEChL_{\rm e'}}\,.
\eear
The corresponding SM amplitudes differ on the $HW\pi$ vertex: 
\bear
&&\hspace{-3mm}i\ampSM_{\rm e}^{\mu\nu}=\coupsSM_{\rm e}\int_p \prop{\alpha}{\gamma}{p}\propG{p-k_1}\propG{p-k_1-k_2}(p-2k_1-2k_2)^\alpha g^{\gamma\mu} (-2p+2k_1+k_2)^\nu \nn\\
&&\hspace{-3mm}i\ampSM_{\rm e'}^{\mu\nu}=i\ampSM_{\rm e}^{\mu\nu}\vert_{k_1\leftrightarrow k_2;\,\mu\leftrightarrow\nu;\,\coupsSM_{\rm e}\rightarrow\coupsSM_{\rm e'}}\,, 
\eear
with coupling factors equal to:
\begin{center}
    \begin{tabular}{c||c|c||c|c}
    & $\coupsEChL_{\rm e}$ & $\coupsEChL_{\rm e'}$ & $\coupsSM_{\rm e}$ & $\coupsSM_{\rm e'}$ \\ \hline
    $H\to\gamma\gamma$ & $-2e^2 g \mw a$ & $-2e^2 g \mw a$ & $-2e^2 g\frac{1}{2} \mw$ & $-2e^2 g\frac{1}{2} \mw$ \\
    $H\to\gamma Z$ & $-2e g^2 \frac{\cosw^2-\sinw^2}{2\cosw} \mw a$ & $2e g^2 \frac{\sinw^2}{\cosw} \mw a$ & $-2e g^2 \frac{\cosw^2-\sinw^2}{4\cosw} \mw$ & $2e g^2 \frac{\sinw^2}{2\cosw} \mw$
    \end{tabular}
\end{center}
The EChL amplitude of the diagram type (f), in terms of the Feynman rules, is:
\be
i\amp_{\rm f}^{\mu\nu}=\coupsEChL_{\rm f}\int_p \prop{\alpha}{\gamma}{p}\propG{p-k_1}\prop{\delta}{\beta}{p-k_1-k_2} g^{\alpha\beta}g^{\gamma\mu}g^{\delta\nu} 
\ee
with coupling factors equal to:
\begin{center}
    \begin{tabular}{c||c}
    & $\coupsEChL_{\rm f}$ \\ \hline
    $H\to\gamma\gamma$ & $-2e^2 g \mw^3 a$ \\
    $H\to\gamma Z$ & $2e g^2 \frac{\sinw^2}{\cosw} \mw^3 a$
    \end{tabular}
\end{center}
The corresponding SM amplitude to this diagram is the same except for the strength of the $HWW$ vertex (where we must set $a=1$), as it can be seen from \tabref{relevant-FR1}:
\be
\ampSM_{\rm f}^{\mu\nu}=\amp_{\rm f}^{\mu\nu}\vert_{a=1}  \,.
\ee
The EChL amplitude of the diagram type (g), in terms of the Feynman rules, is:
\be
i\amp_{\rm g}^{\mu\nu}=\coupsEChL_{\rm g}\int_p \propGp\prop{\lambda}{\rho}{p-k_1}\propG{p-k_1-k_2} p\cdot (p-k_1-k_2)g^{\lambda\mu}g^{\rho\nu} 
\ee
The corresponding SM amplitude differ on the $H\pi\pi$ vertex:
\be
i\ampSM_{\rm g}^{\mu\nu}=\coupsSM_{\rm g}\int_p \propGp\prop{\lambda}{\rho}{p-k_1}\propG{p-k_1-k_2} g^{\lambda\mu}g^{\rho\nu} 
\ee
with coupling factors equal to:
\begin{center}
    \begin{tabular}{c||c||c}
    & $\coupsEChL_{\rm g}$ & $\coupsSM_{\rm g}$ \\ \hline
    $H\to\gamma\gamma$ & $2e^2 g \mw a$ & $-2e^2 g\frac{1}{2} \mw\mh^2$ \\
    $H\to\gamma Z$ & $-2e g^2 \frac{\sinw^2}{\cosw} \mw a$ & $2e g^2 \frac{\sinw^2}{2\cosw} \mw\mh^2$ 
    \end{tabular}
\end{center}
The EChL amplitudes of the diagrams type (h), in terms of the Feynman rules, are:
\bear
&&\hspace{-3mm}\quad i\amp_{\rm h}^{\mu\nu}= \nn\\
&&\hspace{-3mm}\coupsEChL_{\rm h}\int_p \propGp\propG{p-k_1}\propG{p-k_1-k_2} p\cdot (p-k_1-k_2)(-2p+k_1)^\mu (-2p+2k_1+k_2)^\nu \nn\\
&&\hspace{-3mm}\quad i\amp_{\rm h'}^{\mu\nu}= i\amp_{\rm h}^{\mu\nu}\vert_{k_1\leftrightarrow k_2;\,\mu\leftrightarrow\nu;\,\coupsEChL_{\rm h}\rightarrow\coupsEChL_{\rm h'}}\,, 
\eear
The corresponding SM amplitudes differ on the $H\pi\pi$ vertex: 
\bear
&&i\ampSM_{\rm h}^{\mu\nu} = \coupsSM_{\rm h}\int_p \propGp\propG{p-k_1}\propG{p-k_1-k_2} (-2p+k_1)^\mu (-2p+2k_1+k_2)^\nu \nn\\
&&i\ampSM_{\rm h'}^{\mu\nu} = i\ampSM_{\rm h}^{\mu\nu}\vert_{k_1\leftrightarrow k_2;\,\mu\leftrightarrow\nu;\,\coupsSM_{\rm h}\rightarrow\coupsSM_{\rm h'}}\,, 
\eear
with coupling factors equal to:
\begin{center}
    \begin{tabular}{c||c||c}
    & $\coupsEChL_{\rm h}$ = $\coupsEChL_{\rm h'}$ & $\coupsSM_{\rm h}$ = $\coupsSM_{\rm h'}$ \\ \hline
    $H\to\gamma\gamma$ & $-e^2 g \frac{1}{\mw} a$ & $e^2 g \frac{\mh^2}{2\mw}$ \\
    $H\to\gamma Z$ & $-e g^2 \frac{\cosw^2-\sinw^2}{2\cosw} \frac{1}{\mw} a$ & $e g^2 \frac{\cosw^2-\sinw^2}{4\cosw} \frac{\mh^2}{\mw}$ 
    \end{tabular}
\end{center}
The EChL amplitude of the diagram type (i), in terms of the Feynman rules, is:
\be
i\amp_{\rm i}^{\mu\nu}= \coupsEChL_{\rm i}\int_p \propGp\propG{p-k_1-k_2} p\cdot (p-k_1-k_2)g^{\mu\nu}.
\ee
The corresponding SM amplitudes differ on the $H\pi\pi$ vertex: 
\be
\ampSM_{\rm i}^{\mu\nu}= \coupsSM_{\rm i}\int_p \propGp\propG{p-k_1-k_2} g^{\mu\nu}
\ee
with coupling factors equal to:
\begin{center}
    \begin{tabular}{c||c||c}
    & $\coupsEChL_{\rm i}$ & $\coupsSM_{\rm i}$ \\ \hline
    $H\to\gamma\gamma$ & $e^2 g \frac{2}{\mw} a$ & $e^2 g \frac{\mh^2}{\mw}$ \\
    $H\to\gamma Z$ & $e g^2 \frac{\cosw^2-\sinw^2}{\cosw} \frac{1}{\mw} a$ & $e g^2 \frac{\cosw^2-\sinw^2}{2\cosw} \frac{\mh^2}{\mw}$
    \end{tabular}
\end{center}
The EChL amplitudes of the diagrams type (j) are vanishing due to the abscense of the intercation vertex of a Higgs boson with two ghosts: 
\be
i\ampEChL_{\rm j}^{\mu\nu} =i\ampEChL_{\rm j'}^{\mu\nu} = 0 \,.
\ee
On the other hand, the SM amplitudes are:
\bear
&&i\ampSM_{\rm j}^{\mu\nu} = \coupsSM_{\rm j}\int_p \propGp\propG{p-k_1}\propG{p-k_1-k_2} (p-k_1)^\mu (p-k_1-k_2)^\nu \nn\\
&&i\ampSM_{\rm j'}^{\mu\nu} = i\ampSM_{\rm j}^{\mu\nu}\vert_{k_1\leftrightarrow k_2;\,\mu\leftrightarrow\nu;\,\coupsSM_{\rm j}\rightarrow\coupsSM_{\rm j'}}\,, 
\eear
with coupling factors equal to:
\begin{center}
    \begin{tabular}{c||c}
    & $\coupsSM_{\rm j}$ = $\coupsSM_{\rm j'}$ \\ \hline
    $H\to\gamma\gamma$ & $(-2)e^2 g\frac{1}{2} \xi\mw$ \\
    $H\to\gamma Z$ & $(-2)e g^2 \frac{\cosw}{2} \xi\mw$
    \end{tabular}
\end{center}
The EChL amplitudes of the diagrams type (k) and (l), in terms of the FRs, are:
\bear
&&i\amp_{\rm k}^{\mu\nu}=\coupsEChL_{\rm k}\int_p \propGp g^{\mu\nu} \nn\\
&&i\amp_{\rm l}^{\mu\nu}=\coupsEChL_{\rm l}\int_p \propGp\propG{p-k_1} (2p-k_1)^\mu(2p-k_1)^\nu \nn\\
&&i\amp_{\rm l}^{\mu\nu}=i\amp_{\rm l}^{\mu\nu}\vert_{k_1\leftrightarrow k_2;\,\mu\leftrightarrow\nu;\,\coupsEChL_{\rm l}\rightarrow\coupsEChL_{\rm l'}}\,, 
\eear
with coupling factors equal to:
\begin{center}
    \begin{tabular}{c||c||c}
    & $\coupsEChL_{\rm k}$ & $\coupsEChL_{\rm l}$ = $\coupsEChL_{\rm l'}$ \\ \hline
    $H\to\gamma\gamma$ & $-e^2 g \frac{2}{\mw} a$ & $e^2 g \frac{1}{\mw} a$ \\
    $H\to\gamma Z$ & $-e g^2 \frac{\cosw^2-\sinw^2}{\cosw} \frac{1}{\mw} a$ & $e g^2 \frac{\cosw^2-\sinw^2}{2\cosw} \frac{1}{\mw} a$
    \end{tabular}
\end{center}
On the other hand, the corresponding SM amplitudes are vanishing since there are not the multiparticle vertices $H\gamma\pi\pi$, $HZ\pi\pi$, $H\gamma\gamma\pi\pi$ nor $H\gamma Z\pi\pi$.

Finally, the fermionic loop contributions in the EChL are taken as in the SM and are extracted from \cite{HHG} for both decays. We provide explicitly the special functions with our conventions and the definitions of the $f$ and $g$ functions of \eqref{fpaper} and \eqref{gpaper}:
\bear
A_{\gamma\gamma}^{\rm F}(x) &=& -2x(1+(1-x)f(x)) \,, \nn\\
A_{\gamma Z}^{\rm F}(x,y) &=& I_1(x,y)-I_2(x,y) \,,
\label{Amedios}
\eear
where the auxiliary functions are:
\bear
I_1(x,y) &=& \frac{x y}{2(x-y)} +\frac{x^2 y^2}{2(x-y)^2}(f(x)-f(y)) +\frac{x^2 y}{(x-y)^2}(g(x)-g(y)) \,, \nn\\
I_2(x,y) &=& -\frac{x y}{2(x-y)}(f(x)-f(y)) \,.
\eear

%
\bibliographystyle{unsrt}

\end{document}